\documentclass[journal]{IEEEtran}
\usepackage{amsmath,amsfonts}
\usepackage{amssymb}
\usepackage{algorithmic}
\usepackage{algorithm}
\usepackage{array}
\usepackage{bm}
\usepackage{booktabs}
\usepackage{cite}
\usepackage[colorlinks, linkcolor=red, citecolor=blue]{hyperref}
\usepackage{color}
\usepackage{xcolor}
\usepackage{esint}
\usepackage[mathscr]{eucal}
\usepackage{makecell}
\usepackage{multicol}
\usepackage{overpic}
\usepackage[caption=false,font=small,labelfont=rm,textfont=rm]{subfig}
\usepackage{stfloats}
\usepackage{subfig,graphicx}
\usepackage{threeparttable}
\usepackage{textcomp}
\usepackage{url}
\usepackage{verbatim}
\makeatletter

\newcommand{\Rmnum}[1]{\expandafter\@slowromancap\romannumeral #1@}
\makeatother

\begin{document}

%\title{Pixel‑Based Reconfigurable Beamforming Network Emulating Physical Movement in FAS: Design, Scaling Methodology, and Experimental Validation}
\title{Pixel‑Based Reconfigurable Beamforming Network Emulating Physical Movement in FAS}

\author{Jichen Zhang,~\IEEEmembership{Student Member,~IEEE}, Junhui Rao,~\IEEEmembership{Member,~IEEE}, Tianqu Kang,~\IEEEmembership{Student Member,~IEEE},\\Zhaoyang Ming,~\IEEEmembership{Student Member,~IEEE}, Yijun Chen,~\IEEEmembership{Student Member,~IEEE}, Alikhan Umirbayev,~\IEEEmembership{Student Member,~IEEE}, Chi-Yuk Chiu,~\IEEEmembership{Senior Member,~IEEE}, Ross Murch,~\IEEEmembership{Fellow,~IEEE}\\

\thanks{Manuscript received 01 May 2026; revised 01 June 2026; accepted 01 July 2026. Date of publication 01 August 2026; date of current version 30 June 2026. This work was supported by the Hong Kong Research Grants Council Collaborative Research Fund under Grant AoE/E-601/22-R. (\it{Corresponding author: Junhui Rao.})}
\thanks{Jichen Zhang, Junhui Rao, Tianqu Kang, Zhaoyang Ming, Yijun Chen, Alikhan Umirbayev, and Chi-Yuk Chiu are with the Department of Electronic and Computer Engineering, the Hong Kong University of Science and Technology, Hong Kong (e-mail: jzhangiq@connect.ust.hk, jraoaa@connect.ust.hk).}
\thanks{Ross Murch is with the Department of Electronic and Computer Engineering and the Institute for Advanced Study (IAS), the Hong Kong University of Science and Technology, Hong Kong (e-mail: eermurch@ust.hk).}}

%\markboth{IEEE TRANSACTIONS ON MICROWAVE THEORY AND TECHNIQUES}%


\maketitle

\begin{abstract}
	The concept of Fluid Antenna Systems (FAS) has emerged as an attractive new system technology for use in sixth-generation (6G) wireless systems. However most FAS implementations rely on mechanical antenna movement and are thus are too slow to be useful. In this paper, a novel pixel-based reconfigurable beamforming network (PRBFN) is used to emulate movement in Fluid Antenna Systems (FASs). Using the insight that changing an antenna's physical position is equivalent to changing radiation patterns that satisfy the desired pattern correlation, the PRBFN is used to control the excitation current vectors of a multi-port antenna, thereby governing the pattern correlation. Key novelties of our work involve the selection of current vectors, and the methodology for scaling the PRBFN to realize large-aperture FAS. Results are provided for our PRBFN combined with an FAS (denoted as an PRBFN-FAS) when the equivalent physical movement is set to 1.5 wavelengths. Measurements demonstrate that the PRBFN-FAS provides the desired spatial correlation, including the Bessel function relation from Clarke's model across a 5\% bandwidth, satisfying FAS requirements. System-level experiments confirm the viability of the PRBFN-FAS in communication scenarios.
\end{abstract}


\section{Introduction}

\IEEEPARstart{D}{riven} by the growth in wireless data and the promise of new technology, such as Integrated Sensing and Communications (ISAC) and Artificial Intelligence (AI), the research community is now focusing on sixth generation (6G) wireless communication systems. Among the core requirements of 6G networks \cite{6G_0,6G_1,6G_2}, ultra-reliable, low-latency, and stable connectivity remain a core target, and cannot be delivered by current wireless techniques. While Multiple-Input Multiple-Output (MIMO) systems can be extended further, these require large antenna arrays and numerous RF chains, introducing significant pressure on signal processing, channel estimation, energy consumption and cost \cite{MIMO_review}.

\begin{figure}[!t]
	\centering
	\vspace{-0.1cm}
	\begin{overpic}[width=0.85\linewidth]{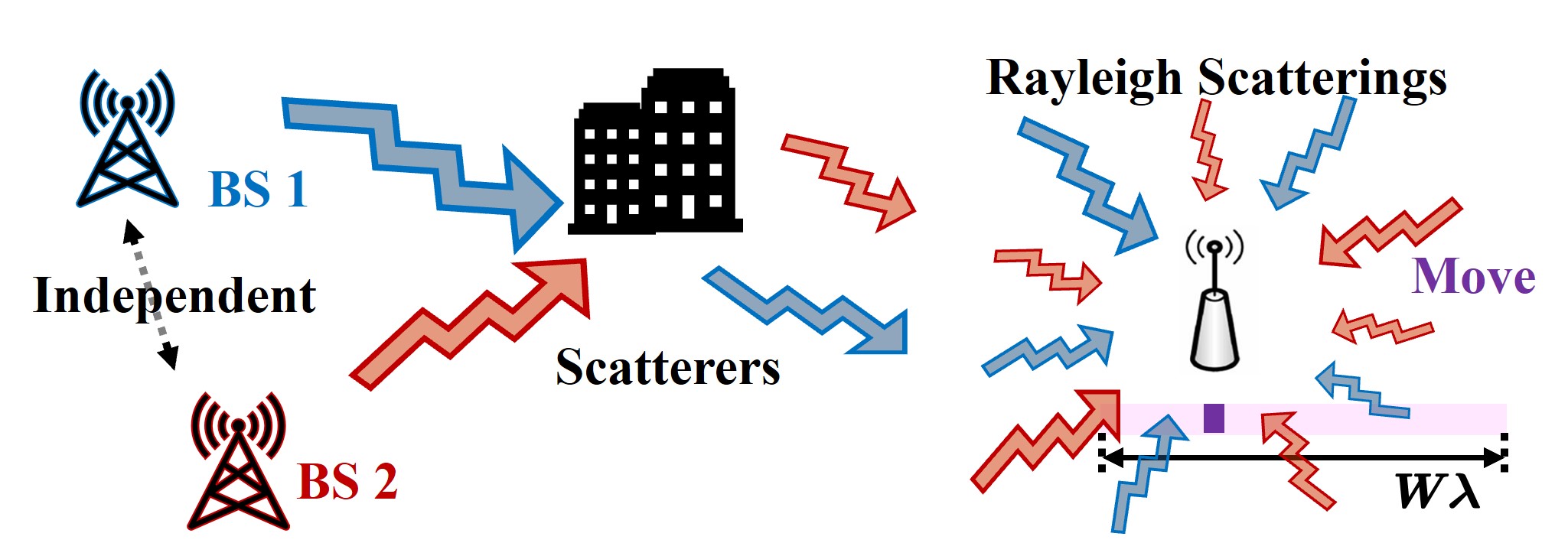}
				\put(0,1){(a)} 
	\end{overpic}
	\begin{overpic}[width=0.85\linewidth]{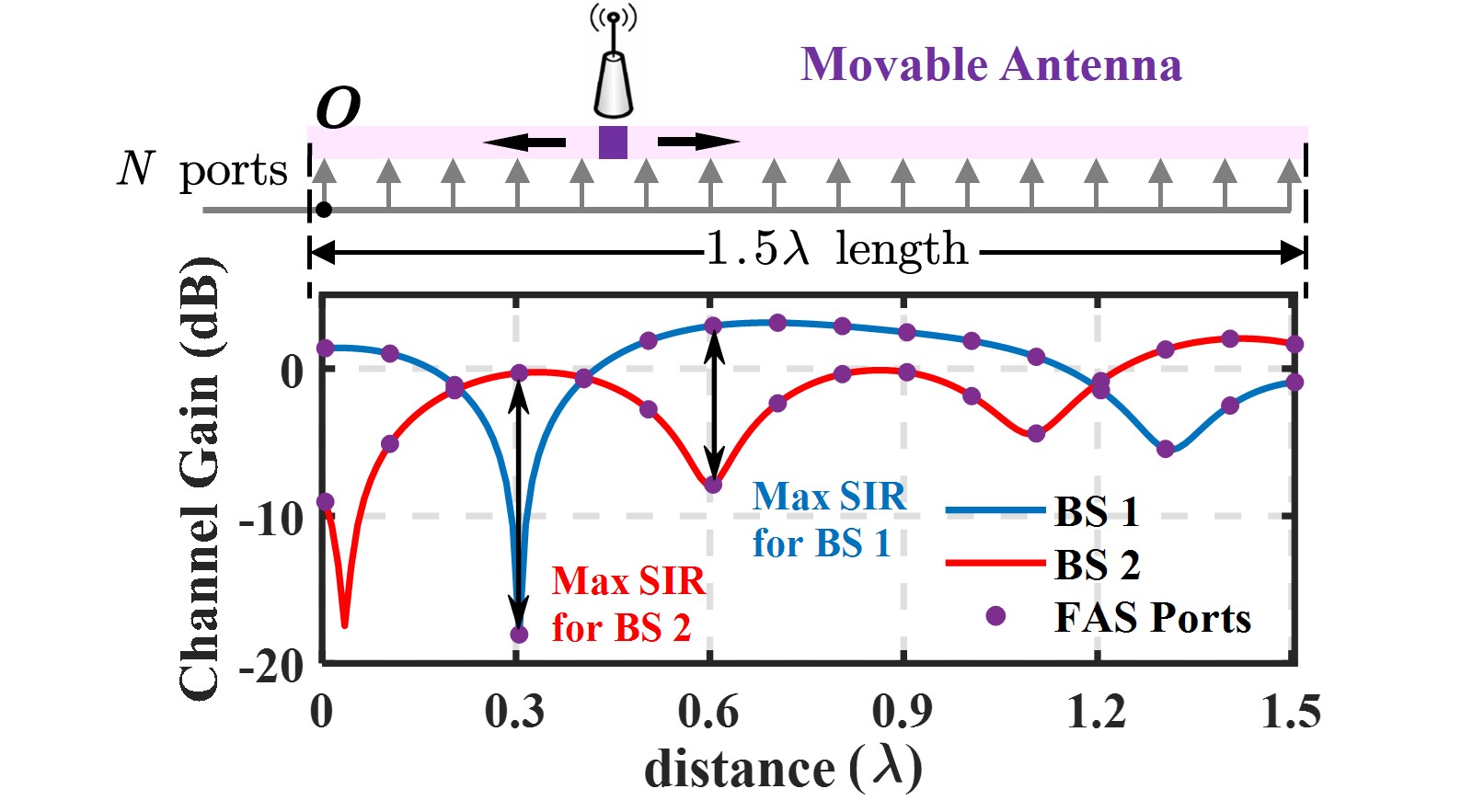}
		\put(0,1){(b)} 
	\end{overpic}
	\vspace{-0.3cm}
	\caption{FAS illustration. (a) Signals from two basestations are independently scattered by the environment, creating rich scattering at the receiver FAS, which can be moved through $N$ port positions over a $W\lambda$ linear space. (b) Example of the two BS signals received by the FAS over the linear space 1.5$\lambda$, where it can be observed both signals fade independently. The optimal FAS position can be selected from the $N$ port positions to boost wireless performance (e.g. max SIR points for BS 1 or 2).}
	\label{FAS_Scheme}
\end{figure}

In this context, the Fluid Antenna System (FAS) has emerged as a potential technology for 6G \cite{FAS,FAS_review,BruceLee}. Unlike fixed-location antennas in MIMO, FAS can move a radiator across a physical space with fine resolution, more effectively exploiting the spatial dimension without increasing the number of RF chains compared with MIMO. As depicted in Fig. \ref{FAS_Scheme}(a) and (b), the antenna in FAS dynamically selects one of $N$ possible positions ($N$ FAS ports) over a $W\lambda$ linear space ($\lambda$ is the wavelength) in a rich scattering environment \cite{FAS}. Leveraging the peaks and nulls of the signal envelope from independent signals from base stations (BSs) across all FAS locations, significant enhancements to communication system performance can be achieved \cite{FAS,FAS_review,BruceLee}. For instance, as given in Fig. \ref{FAS_Scheme}(b), FAS can maximize Signal-to-Interference Ratio (SIR) to facilitate multi-user access without intricate signal processing \cite{MIMO_signal_1,MIMO_signal_2}, known as Fluid Antenna Multiple Access (FAMA) \cite{FAMA}. 

%It is worth noting that the term $''$fluid$''$ in FAS is metaphorical \cite{FAS_review,BruceLee}. FAS refers to any wireless system including reconfigurable antennas with flexible position or shape to optimize performance. It does not require antennas made of fluid \cite{BruceLee} and any reconfigurable technology can work \cite{FAS_review, BruceLee}.

\begin{figure}[!t]
	\centering
%	\vspace{-0.3cm}
	\begin{overpic}[width=0.97\linewidth]{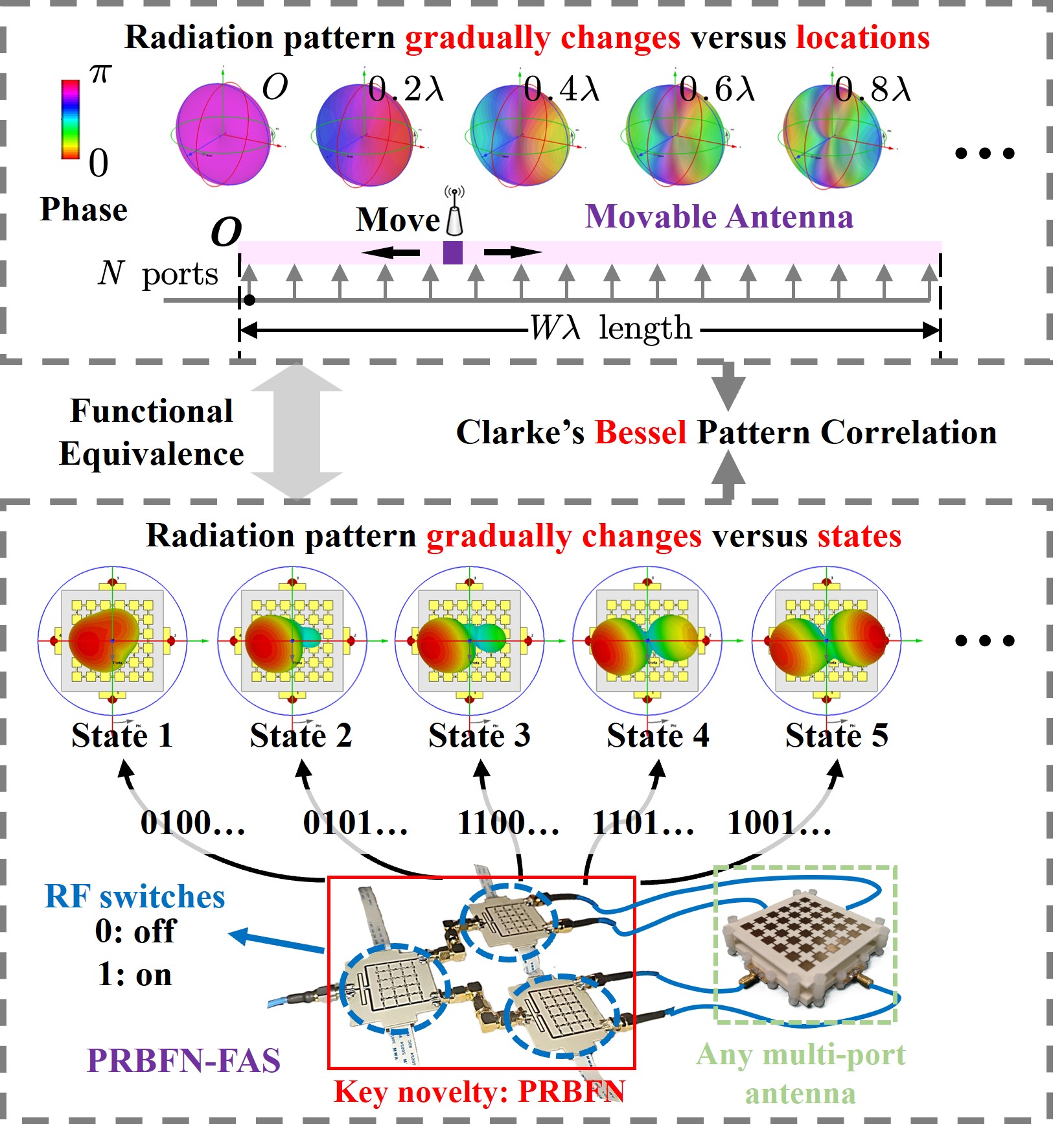}
	\end{overpic}
	\vspace{-0.3cm}
	\caption{Small scale physical movement of an antenna produces only phase change in its radiation pattern relative to the origin $O$. When signals with rich scattering are received by this antenna, at different locations, fading occurs as shown in Fig. \ref{FAS_Scheme}(b). To mimic this movement, the proposed PRBFN produces feeding currents that form radiation patterns with the same pattern correlation characteristics as produced by the movement. In particular, $N$ PRNFN-FAS states can emulate $N$ FAS port positions and is equivalent to it.}
	\label{FAS_Core}
\end{figure}

Extensive system simulations have demonstrated that FAS can significantly enhance communication performance and strongly support cutting-edge technologies \cite{ISAC1,ISAC2,RIS}. However, when the spotlight turns from algorithms to hardware, the literature thins drastically. Generally, the few FAS prototypes reported so far can be divided into three categories. The first category is based on mechanical or physical-movement-based reconfigurable antennas, whose movement is controlled by motors \cite{MMA-FAS}, pumps \cite{liquid1,liquid2,liquid3}, or electrowetting \cite{droplet}. The operating principles of these antennas are straightforward. However, their FAS port switching speed is too slow for rapid wireless channel dynamics, limiting their FAS potential \cite{s-FAMA}. As a second category, meta-fluid antennas \cite{Meta-fluid} or reconfigurable holographic surfaces (RHSs) \cite{RHS}, create diode-controlled slots across a waveguide as FAS ports. With great scalability and compatibility with PCBs, meta-fluid antennas or RHSs achieve microsecond switching speeds that can timely track fading wireless channels. However, the required $\lambda/2$ slot size inherently limits FAS port density ($N/W$), resulting in sparse spatial sampling which reduces spatial diversity and ultimately compromises the performance of FAS \cite{FAS,FAMA}.
 
In the third category, pixel-based reconfigurable antennas (PRAs) can also emulate physical movement in FAS \cite{PRA-FAS}. The PRA employs sub-wavelength metallic pixels interconnected by RF switches, enabling dynamic control of the current distribution and thus radiation patterns. Through optimized pixel connections, PRA states can generate the desired FAS spatial correlation, including the Bessel function result obtained from Clarke's model, effectively emulating movement. With microsecond switching speed and high port density, PRAs fully release the spatial diversity of FAS, and recent studies have explored their applications in communication systems \cite{PRA_est,REMAA,PRA_iotj}. However, the scalability of PRAs is limited by the $\lambda/2$ antenna size, which constrains achievable $W$ values. The non-linearity of PIN diodes also restricts PRAs to low power levels ($<$15 dBm) \cite{pixel0}, preventing their deployments at the transmitter (Tx).

To address the challenges described, an FAS based on beamforming networks (BFNs) offers a new possibility. Switching BFN states allows the combining of the orthogonal radiation components with specific ratios to control pattern correlation, mimicking FAS movement while maintaining $\mu \text{s}$ switching speed. In addition, the BFN can be positioned ahead of amplifiers to keep reconfigurable devices, including PIN diodes or varactors \cite{BFN1,BFN2,BFN3,BFN4,BFN5}, within their linear operating region. This eliminates nonlinear distortion, and permits Tx-side deployment. Moreover, the BFN is scalable, thus it can achieve large FAS size $W$, breaking the size limitation of PRA-FAS.

In this paper, a pixel-based reconfigurable beamforming network (PRBFN) is proposed for FAS applications in Fig. \ref{FAS_Core}. When combined with the FAS, we denote it as an PRBFN-FAS. The key novelties and distinctive features are summarized in the following. 

\subsubsection{Novel implementation of FAS using beamforming} 
By mapping each FAS port position to predefined excitation currents of PRBFN, port position selection in FAS is shown to be equivalent to beamforming in PRBFN-FAS, as shown in Fig. \ref{FAS_Core}. Based on our PRBFN-FAS design, we provide a beamforming-based framework for FAS implementation. 

\subsubsection{High reconfigurability with high port density}
PRBFN-FAS can support numerous ($N$) states, enabling the high port density ($N/W$) required for FAS. Reconfigurable pixels enable precise control of the required excitation beamforming currents, providing the required FAS correlation \cite{pixel_RIS,pixel_re1,pixel1, pixel_switch}. 

\subsubsection{Cascaded topology with high scalability} 
Our design for the PRBFN-FAS adopts a cascaded topology composed of identical unit cells. FAS with large $W$ can be achieved by cascading multiple unit cells, breaking the limits in PRAs.

\subsubsection{Enhanced linearity for Tx applications}
PIN diode nonlinearity under high power \cite{pixel0} prevents previous high-speed FAS designs from Tx deployment \cite{Meta-fluid, RHS, PRA-FAS}. The PRBFN‑FAS overcomes this by placing the PRBFN before the amplifiers, ensuring small‑signal operation of reconfigurable components.

\subsubsection{Stable correlation across sufficient bandwidth}
PRA operates in the resonant region, making broadband correlation difficult, e.g. 2\% in \cite{PRA-FAS}. In contrast, PRBFN‑FAS controls correlation using excitation current vectors via PRBFN, and a stable correlation of over 5\% bandwidth is achieved, making it suitable for practical applications. 

\subsubsection{Experimental validation}
A prototype of the proposed PRBFN‑FAS with FAS parameters of $W = 1.5$, $N = 18$ is designed, fabricated, and measured. Testbed experiments verify its effectiveness under practical communication scenarios.

%\subsubsection{Facilitate FAS channel estimation}
%The difficulty of channel estimation presents a significant challenge for FAS \cite{FAS_est1,FAS_est2,FAS_est3,FAS_est4}. One conventional method is performing channel measurements at a limited number of observation ports, and extrapolating the results to all $N$ ports \cite{FAS_est4}. The PRBFN-FAS architecture, combined with one compact MIMO antenna, allows the estimation process to be performed more easily.

The remainder of the paper is organized as follows. Section \Rmnum{2} analyzes how physical movement in FAS can be emulated by beamforming. Section \Rmnum{3} presents the design methodology of the PRBFN-FAS. In Section \Rmnum{4}, measurement results for an PRBFN-FAS with FAS parameters of $W = 1.5, N = 18$ are described, along with the prototype details.   System-level experiments of the PRBFN-FAS in a real scattering environment are reported. In Section \Rmnum{5}, further discussion is provided along with comparisons with previous related work. Finally, we summarize and draw conclusions in Section \Rmnum{6}.

\section{Design Analysis}

\subsection{Background: From Spatial to Pattern Correlation}

To understand how beamforming can emulate movement in FAS, we must first illustrate that switching reconfigurable radiation patterns controlled by beamforming currents is functionally equivalent to switching the physical position of the radiator in conventional FAS. In the following discussions, we assume a wireless communication scenario where there is rich scattering with polarizations that are spatially uncorrelated and equally likely, and use the approach in \cite{antenna_diversity}. 

For FAS with a size of $W\lambda$ and number of port positions, $N$, we assume that the far-field radiation pattern at the $n$-th FAS port position can be written as $\mathbf{e}_n(\mathbf{\Omega})= [e_{\theta,n}(\mathbf{\Omega}),e_{\phi,n}(\mathbf{\Omega})]^\mathrm{T}$, where $\mathbf{\Omega}=(\theta,\phi)$. All $N$ radiation patterns, from the $N$ port positions, share a common coordinate origin $O$ and the voltage at the $n$-th port is denoted as
\begin{equation}
	\label{eqn2-1}
	g_n = a\iint{\mathbf{e}_n(\mathbf{\Omega})\cdot\mathbf{h}(\mathbf{\Omega})}\mathrm{d}\mathbf{\Omega},
\end{equation}
where $a$ is a proportionality constant \cite{1969Antenna}, and $\mathbf{h}(\mathbf{\Omega})=[h_{\theta}(\mathbf{\Omega}),h_{\phi}(\mathbf{\Omega})]^\mathrm{T}$ represents the incident field from the scattering environments. 

The correlation coefficient between the $i$-th and $j$-th FAS port, $C_{i,j}$, is written
\begin{equation}
	\label{eqn2-2}
	C_{i,j} = \frac{\mathcal{E}[g_i^{}g_j^*]}{\sqrt{\mathcal{E}[|g_i|^2]\mathcal{E}[|g_j|^2]}}=\frac{\text{Cov}(g_i,g_j)}{\sqrt{\mathcal{E}[|g_i|^2]\mathcal{E}[|g_j|^2]}},
\end{equation}
where $\mathcal{E}[\cdot]$ is expectation and $g_n \sim \mathcal{CN}(0,\sigma_n^2)$ due to the random scattering. Using (\ref{eqn2-1}), the numerator of (\ref{eqn2-2}) can be expanded as
\begin{equation}
	\label{eqn2-2-1}
		\text{Cov}(g_i,g_j)=a^2\mathcal{E}\left[ \iint{\mathbf{e}_i^{}(\mathbf{\Omega})\cdot\mathbf{e}_j^*(\mathbf{\Omega})\cdot S(\mathbf{\Omega})} \mathrm{d}\mathbf{\Omega} \right],
\end{equation}
where $S(\mathbf{\Omega}) = \mathcal{E}\left[\iint{  \mathbf{h}(\mathbf{\Omega}) \cdot\mathbf{h}^*(\mathbf{\Omega})}\mathrm{d}\mathbf{\Omega}\right]$ is the power angular spectrum (PAS) over 3D space. Therefore, the correlation $C_{i,j}$ can be expressed in terms of radiation patterns, as
\begin{equation}
	\label{eqn2-5-a}
	C_{i,j}=\frac{\iint{\mathbf{e}_i\cdot\mathbf{e}_j^*\cdot S\text{d}\bf{\Omega}} }{\sqrt{ \iint{|\mathbf{e}_i|^2\cdot S\text{d}\bf{\Omega}}} \sqrt{\iint{|\mathbf{e}_j|^2\cdot S\text{d}\bf{\Omega}} }},
\end{equation}
where $\bm{(\Omega)}$ is omitted for brevity, and $S(\mathbf{\Omega})=S_0\mathbf{U}_2$ in which $\mathbf{U}_2$ is the $2\times2$ identity matrix due to the two independent polarization components over 3D space \cite{antenna_diversity}. 

For conventional FAS in Fig. \ref{FAS_Scheme}(a), switching the location of the activated port alters the radiation pattern. Specifically, all $N$ ports share identical amplitude patterns with $|\mathbf{e}_i(\mathbf{\Omega})|= |\mathbf{e}_j(\mathbf{\Omega})|$, while their phases can be related by
\begin{equation}
	\label{eqn2-3}
%	\mathbf{e}_j(\mathbf{\Omega}) = \mathbf{e}_i(\mathbf{\Omega})\cdot e^{jkd_{i,j}\cos\phi\sin\theta} = \mathbf{e}_i(\mathbf{\Omega})\cdot e^{j\Phi_{i,j}(\bm{\Omega})},
	\mathbf{e}_j(\mathbf{\Omega}) = \mathbf{e}_i(\mathbf{\Omega})\cdot e^{j\Phi_{i,j}(\bm{\Omega})},
\end{equation}
where $\Phi_{i,j}(\bm{\Omega})=kd_{i,j}\cos\phi\sin\theta$ and $d_{i,j}=\frac{|i-j|W\lambda}{N-1}$ is the distance between the $i$-th and $j$-th FAS ports. Using (\ref{eqn2-2}) to (\ref{eqn2-3}) and considering PAS with isotropic patterns in the $x-y$ plane only, the $C_{i,j}$ (also the spatial correlation) becomes
\begin{equation}
	\hspace{-0.2cm}
	\label{eqn2-4}
	C_{i,j}=\iint{e^{-j\Phi_{i,j}(\bm{\Omega})}\delta(\theta -\frac{\pi}{2} )}\mathrm{d}\mathbf{\Omega}=J_0\left( \frac{2\pi|i-j|W}{N-1}\right),
\end{equation}
where $S(\mathbf{\Omega})=\delta(\theta -\frac{\pi}{2})$ \cite{antenna_diversity}, and $J_0(\cdot)$ is the Bessel function of the first kind, order zero \cite{FAS_new_model}, agreeing with the correlation relationship of Clarke's model in \cite{FAS}.

It should also be observed that the phase of the correlation coefficient $C_{i,j}$ is not critical for evaluating FAS performance, since key metrics, such as channel capacity \cite{FAS} and multiplexing gain \cite{FAMA}, solely depend on the magnitudes of FAS port voltages. Therefore, the absolute value is introduced in (\ref{eqn2-5-a}), as 
\begin{equation}
	\label{eqn2-5}
	C_{i,j}=\left|\frac{\iint{\mathbf{e}_i\cdot\mathbf{e}_j^*\cdot S\text{d}\bf{\Omega}} }{\sqrt{ \iint{|\mathbf{e}_i|^2\cdot S\text{d}\bf{\Omega}}} \sqrt{\iint{|\mathbf{e}_j|^2\cdot S\text{d}\bf{\Omega}} }}\right|,
\end{equation}
and similarly for conventional FAS (\ref{eqn2-4}) becomes
\begin{equation}
	\hspace{-0.2cm}
	\label{eqn2-4-a}
	C_{i,j}=\left| J_0\left( \frac{2\pi|i-j|W}{N-1}\right) \right|,
\end{equation}
which serves as the final target expression for FAS correlation. 

From equations (\ref{eqn2-5}) and (\ref{eqn2-4-a}), it can be observed that carefully selecting reconfigurable patterns that satisfy the spatial correlation given in (\ref{eqn2-4-a}), can emulate spatially distributed FAS ports \cite{PRA-FAS}. This approach effectively converts spatial correlation between FAS ports into pattern correlation between reconfigurable radiation patterns as depicted in Fig. \ref{FAS_Core}. PRA \cite{PRA-FAS} exploited this approach by utilizing reconfigurable antennas to provide FAS capable antenna systems. However, an obvious alternative approach is also possible that is based on conventional beamforming. Therefore, in this paper, a PRBFN is proposed to find appropriate excitation currents to feed a multi-port antenna that can provide $N$ reconfigurable radiation patterns meeting FAS spatial correlation requirements.

%In the remainder of this work, the development is for FAS transmitter configurations. We select this configuration because the pixel-based approach to FAS \cite{PRA-FAS} suffers from non-linearity when the transmit power is more than 15 dBm \cite{pixel0} due to the use of active components, while our novel PRBFN avoids that issue. The PRBFN approach can also be configured as a receiver (Rx) straightforwardly and that is described at the end of the paper. 

\subsection{Correlation Control using Beamforming Network}

\begin{figure}[!t]
	\centering
	%	\vspace{-0.2cm}
	\begin{overpic}[width=1\linewidth]{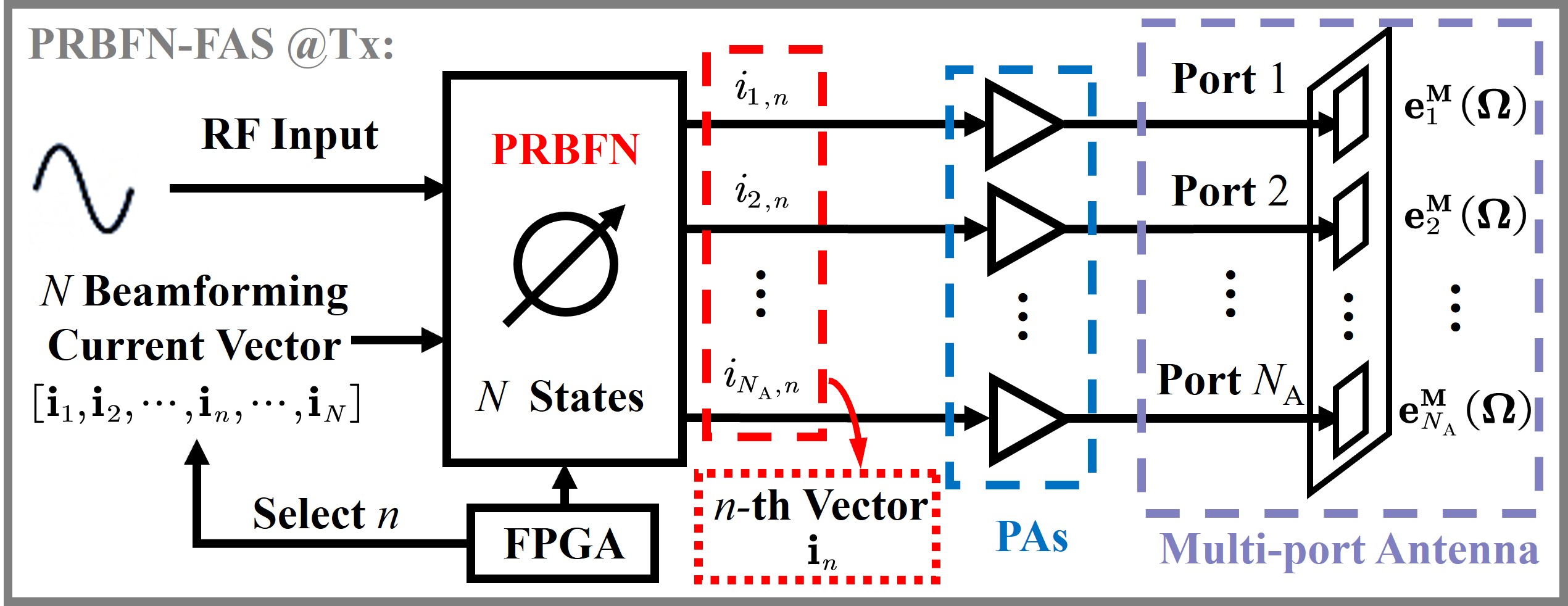}
		\put(1,1){(a)} 
	\end{overpic}
%	\hfil
	\vfil
%	\vspace{0.0cm}
	\begin{overpic}[width=1\linewidth]{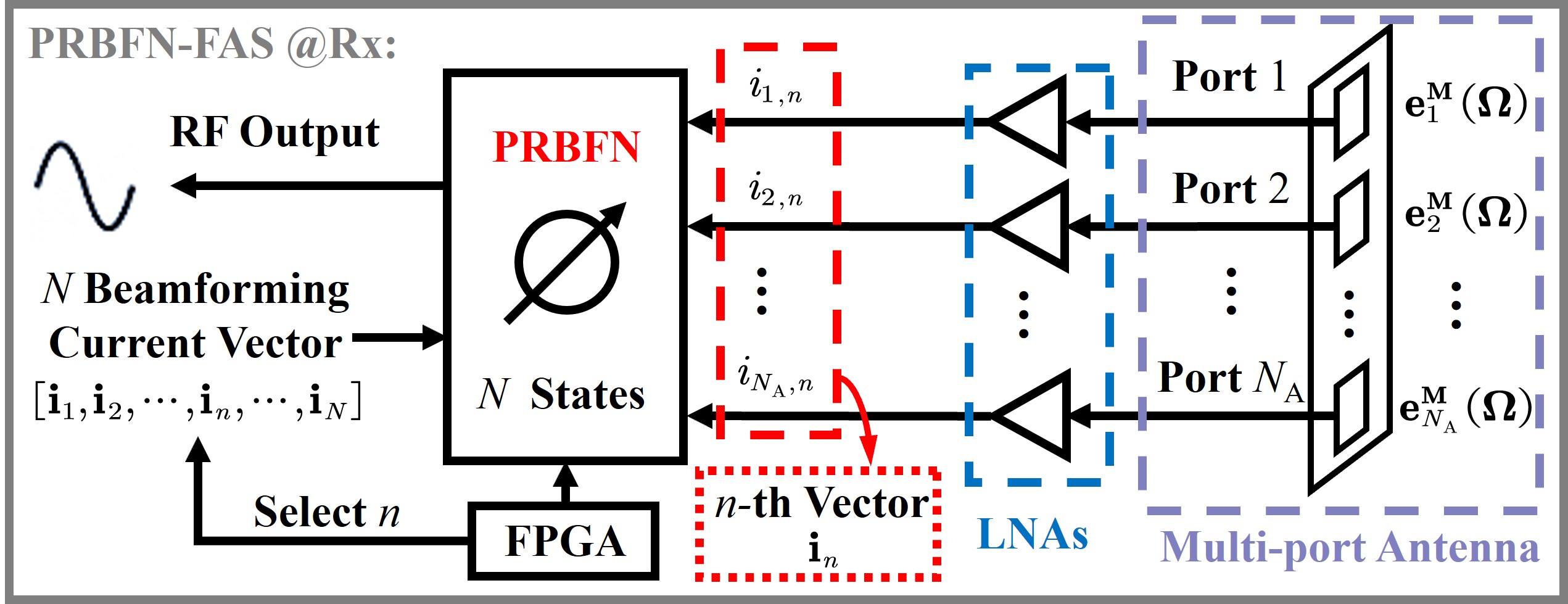}
		\put(1,1){(b)} 
	\end{overpic}
%	\hfil
%	\includegraphics[width=1\linewidth]{Figures/Architecture.jpg}
	\vspace{-0.6cm}
	\caption{PRBFN‑FAS architecture at (a) Tx and (b) Rx. Both architectures include an PRBFN with $N$ reconfigurable states. For Tx,  PAs are used after the PRBFN and before the $N_A$ antennas so that the PBFN can operate at low signal power. For Rx, LNAs after the $N_A$ antennas and before the PRBFN are used in order to maintain noise figure performance.}
	\label{FAS_Architecture}
\end{figure}

The proposed beamforming architecture for FAS, denoted as PRBFN‑FAS, is shown in Fig. \ref{FAS_Architecture}(a) and (b). The PRBFN generates $N$ reconfigurable beamforming current vectors (aligning with the number of FAS ports), each feeding a highly-isolated multi-port antenna with $N_\text{A}$ ports. Power amplifiers (PAs) are placed after PRBFN at Tx to avoid the linearity issue, and low-noise amplifiers (LNAs) are placed ahead of PRBFN at receiver (Rx) to compensate the insertion loss.

%Bidirectional amplifiers (BDAs) \cite{BDA1,BDA2} are placed after the PRBFN to avoid the linearity issues in Tx and compensate for insertion loss in receiver (Rx).

For the $n$-th PRBFN state, the output excitation current vector can be written as $\mathbf{i}_n = [i_{1,n},i_{2,n},\cdots,i_{N_\text{A},n}]^\mathrm{T}\in \mathbb{C}^{N_\text{A}\times1}$. The radiation pattern matrix of the multi-port antenna is written as $\mathbf{E_M} = [\mathbf{e}^\mathbf{M}_1(\mathbf{\Omega}),\mathbf{e}^\mathbf{M}_2(\mathbf{\Omega}),\cdots, \mathbf{e}^\mathbf{M}_{N_\text{A}}(\mathbf{\Omega})]$, where $\mathbf{e}^\mathbf{M}_m(\mathbf{\Omega}),\ m=1,2,\cdots,N_\text{A}$ is the radiation pattern of the $m$-th antenna port excited by unit current with all other ports well matched. Therefore, the $n$-th reconfigurable beamforming radiation pattern $\mathbf{e}_n(\mathbf{\Omega})$ is
\begin{equation}
	\label{eqn2-6}
	\mathbf{e}_n(\mathbf{\Omega}) = \sum^{N_\text{A}}_{m=1}{\mathbf{e}^\mathbf{M}_m(\mathbf{\Omega})i_{m,n}} = \mathbf{E_M}\mathbf{i}_n.
\end{equation}
The excitation current vectors of all $N$ states can be collected into a beamforming current matrix $\mathbf{B}$, expressed as
\begin{equation}
	\label{eqn2-7}
	\mathbf{B} = [\mathbf{i}_1,\mathbf{i}_2,\cdots, \mathbf{i}_N]\in\mathbb{C}^{N_\text{A}\times N}.
\end{equation} 
Thus $N$ reconfigurable beamforming radiation patterns fed by $N$ PRBFN states, written as $\mathbf{E}(\mathbf{\Omega})$, are  
\begin{equation}
	\label{eqn2-8}
	\mathbf{E} = [\mathbf{e}_1,\mathbf{e}_2,\cdots,\mathbf{e}_N] = \mathbf{E_M}\mathbf{B},
\end{equation}
where we omit $\mathbf{(\Omega)}$ for brevity.

The correlation relationship between the $N$ beamforming radiation patterns is \cite{PCDM}
\begin{eqnarray}
	\label{eqn2-9}
	\begin{split}
		\mathbf{C} &= \left| \iint{\mathbf{E}^\mathrm{H}\mathbf{E}S}\mathrm{d}\mathbf{\Omega}\right| =\left|\iint{\mathbf{B}^\mathrm{H}\mathbf{E_M}^\mathrm{H}\mathbf{E_M}\mathbf{B}S}\mathrm{d}\mathbf{\Omega}\right|\\
		&= \left|\mathbf{B}^\mathrm{H}\left(\iint{\mathbf{E_M}^\mathrm{H}\mathbf{E_M}S}\mathrm{d}\mathbf{\Omega}\right)\mathbf{B}\right|=\left|\mathbf{B}^\mathrm{H}\mathbf{K_M}\mathbf{B}\right|,
	\end{split}
\end{eqnarray}
where the denominator for normalization, as presented in (\ref{eqn2-5}), is omitted for brevity. Notably, $\mathbf{K_M}\in\mathbb{C}^{N_\text{A}\times N_\text{A}}$, the pattern correlation of the multi-port antenna, is determined exclusively by its $N_\text{A}$ patterns, independent of the PRBFN. Given sufficiently high orthogonality between the antenna ports, $\mathbf{K_M}$ can be approximated by an identity matrix $\mathbf{U}_{N_\text{A}}$. Consequently, the correlation relationship $\mathbf{C}\in\mathbb{C}^{N\times N}$ in (\ref{eqn2-9}) of the $N$ PRBFN-FAS radiation patterns can be simplified to
\begin{equation}
	\label{eqn2-10}
	\mathbf{C} = \left|\mathbf{B}^\mathrm{H}\mathbf{B}\right|.
\end{equation}
When all antenna ports are well-matched, the $N$ excitation current vectors collected in $\mathbf{B}$ are determined exclusively by the PRBFN. This characteristic decouples the PRBFN design process from the subsequent design of the multi-port antenna, allowing these two parts to be developed independently.

As a result, the PRBFN must be carefully designed to generate $N$ reconfigurable states such that $\mathbf{B}$ yields a correlation $\mathbf{C}$ satisfying the desired FAS correlation (\ref{eqn2-4-a}). Meanwhile, the multi-port antenna with $N_\text{A}$ ports must exhibit excellent port matching, low mutual coupling, and high orthogonality to ensure that $\mathbf{K_M}$ approximates an identity matrix. In following sections, we will discuss the detailed designs separately.

\section{PRBFN Design Methodology}

The PRBFN design methodology consists of four steps: (1) specifying FAS parameters $W$ and $N$; (2) optimizing $N$ beamforming current vectors to achieve the desired correlation, guiding the following PRBFN design; (3) proposing a typology with cascaded unit cells and the scaling methodology of the PRBFN; (4) detailing the unit cell design.

\subsection{Select Appropriate FAS Parameters}

As a method to realize FAS, the proposed PRBFN-FAS adopts standard FAS parameters, including the number of FAS ports $N$, the size $W\lambda$ and thus the target correlation. To realize its full potential, appropriate parameter selection is essential. The target correlation is denoted as 
$\mathbf{C}_\text{obj}$ \cite{FAS_new_model}, and the $(i,j)$-th entry of $\mathbf{C}_\text{obj}$ is defined by (\ref{eqn2-4-a}), as
\begin{equation}
	\label{eqn2-11}
	[\mathbf{C}_\text{obj}]_{i,j}= \left|J_0\left(\frac{2\pi|i-j|W}{N-1} \right)\right|.
\end{equation}

Another key parameter is the FAS port density, denoted as $N/W$. Using previous results in \cite{FAS,FAMA}, the rate of increase in channel capacity and multiplexing gain significantly decreases as the FAS port density satisfies $N/W \geq 10$. Considering the increasing design complexity with higher port density, we maintain $N/W = 10$ as a sufficient condition in the following PRBFN-FAS designs. 

\subsection{Find Optimal Beamforming Current Vectors $\hat{\mathbf{B}}$}

For a given $W$, we set $N = 10W$ to meet the FAS port density requirement. Substituting $\mathbf{C}_\text{obj}$ into (\ref{eqn2-10}) yields a closed-form solution via singular value decomposition (SVD) for $\mathbf{B}\in\mathbb{C}^{N_\text{A}\times N}$ when $N = N_\text{A}$. However, $N$ is typically much larger than $N_\text{A}$, requiring a minimum least‑squares formulation via SVD. Since FAS disregards the phase of the correlation matrix (\ref{eqn2-10}), we target the correlation of $|\mathbf{B}^\mathrm{H}\mathbf{B}|$, which leads to non-uniqueness, implying multiple feasible solutions. Thus, we adopt numerical optimization to find the optimal beamforming current matrix (denoted as $\hat{\mathbf{B}}$), such that $|\hat{\mathbf{B}}^\mathrm{H}\hat{\mathbf{B}}|$ can best approximate $\mathbf{C}_\text{obj}$.

To minimize the deviation between the target correlation matrices, we formulate the following optimization problem for a system with $N_A$ antenna ports
\begin{equation}
	\label{eqn3-0-1}
	\begin{split}
		&\mathop{\min}\limits_{\mathbf{B}} f_{N_\text{A}}(\mathbf{B}) = \Vert |\mathbf{B}^\mathrm{H}\mathbf{B}|- \mathbf{C}_\text{obj} \Vert_F^2\\
		&\text{s.t.} \ \Vert \mathbf{i}_n \Vert^2_2=1, \ \forall n = 1,2,\cdots,N,
	\end{split}
\end{equation}
where $\Vert\cdot\Vert_F$ denotes the Frobenius norm, and $\mathbf{i}_n$ is the $n$-th vector of $\mathbf{B}$ in (\ref{eqn2-7}), following the power constraint. The optimal solution of (\ref{eqn3-0-1}) is the desired PRBFN output 
\begin{equation}
\label{eqn2-11-1}
\hat{\mathbf{B}}=[\hat{\mathbf{i}}_1,\hat{\mathbf{i}}_2,\cdots,\hat{\mathbf{i}}_N]\in\mathbb{C}^{N_\text{A}\times N}
\end{equation}
where $\hat{\mathbf{i}}_n = [\hat{i}_{1,n},\hat{i}_{2,n},\cdots, \hat{i}_{N_\text{A},n}]^\mathrm{T}\in \mathbb{C}^{N_\text{A}\times1}$. The subscript indicates that $f_{N_\text{A}}(\mathbf{B})$ also depends on $N_\text{A}$, and $N_\text{A}$ is typically set before optimization (\ref{eqn3-0-1}).

Given that the variable $\mathbf{B}$ is complex, and the objective function contains absolute values, solving this optimization problem presents significant challenges. To address this, we employ the projected gradient descent (PGD) method and incorporate modifications based on $\mathbb{CR}$-calculus \cite{CR}, enabling effective application to this complex optimization problem. Using Wirtinger calculus \cite{Wirtinger}, i.e., treating $\mathbf{B}$ and its conjugate $\mathbf{B}^\mathrm{H}$ as independent variables, the gradient of $f_{N_\text{A}}(\mathbf{B})$ with respect to $\mathbf{B}$ is
\begin{equation}
	\label{eqn3-0-5}
	\nabla_{\mathbf{B}}f_{N_\text{A}} = 2\mathbf{B}\left[\left( |\mathbf{B}^\mathrm{H} \mathbf{B}| -\mathbf{C}_\text{obj} \right) \circ \mathrm{sgn}\left(\mathbf{B}^\mathrm{H} \mathbf{B} \right) \right],
\end{equation}
where $\circ$ denotes the Hadamard product, and $\mathrm{sgn}(\cdot)$ is the element-wise sign function. Under the unit-norm column constraint on $\mathbf{B}$, The PGD optimization iteratively performs gradient descent steps followed by projection onto the feasible set. The pseudo code is provided in Algorithm \ref{Optim_target}.

\begin{algorithm}[!t]
	\renewcommand{\algorithmicrequire}{\textbf{Input:}}
	\renewcommand{\algorithmicensure}{\textbf{Output:}}
	\caption{PGD for Optimal $\hat{\mathbf{B}}$}
	\label{Optim_target}
	\begin{algorithmic}[1]
		\REQUIRE Target Bessel correlation $\mathbf{C}_\text{obj} \in \mathbb{R}^{N\times N}$,\\ 
		\hspace{0.4cm} Antenna ports $N_\text{A}$, step size $\eta>0$, tolerance $\varepsilon>0$,\\ 
		\hspace{0.4cm} $\mathbf{B}^{(0)}=[\mathbf{i}_1^{(0)},\dots,\mathbf{i}_N^{(0)}] \in \mathbb{C}^{N_\text{A}\times N}$ with $\|\mathbf{i}_n^{(0)}\|_2^2=1$.
		
		\FOR{$k=1$ to $\mathrm{maxIter}$}
		\STATE \textbf{Gradient:} Compute $\nabla_{\mathbf{B}}f_{N_\text{A}}$ from (\ref{eqn3-0-5}) using $\mathbf{B}^{(k-1)}$.
		\STATE \textbf{Descent:} $\mathbf{B}^{(k)} = \mathbf{B}^{(k-1)} - \eta \cdot \nabla_{\mathbf{B}}f_{N_\text{A}}$.
		\STATE \textbf{Projection:} Following the constraint in (\ref{eqn3-0-1})
		\FOR{$n=1,\dots,N$}
		\STATE $s = \|\mathbf{i}_n^{(k)}\|_2$
			\IF{$s \neq 0$} \STATE $\mathbf{i}_n^{(k)} \leftarrow \mathbf{i}_n^{(k)} / s$ 
			\ELSE \STATE $\mathbf{i}_n^{(k)} \leftarrow \text{random unit vector}$ \ENDIF
		\ENDFOR
		\STATE \textbf{Convergence Check:} Compute $f_{N_\text{A}}(\mathbf{B}^{(k)})$ by (\ref{eqn3-0-1}).	\IF{$f_{N_\text{A}}(\mathbf{B}^{(k)}) < \varepsilon$} \STATE \textbf{Output} $\hat{\mathbf{B}} = \mathbf{B}^{(k)}$ and terminate. \ENDIF
		\ENDFOR
		\ENSURE $\hat{\mathbf{B}}\in\mathbb{C}^{N_\text{A}\times N}$ with $\hat{\mathbf{C}}=|\hat{\mathbf{B}}^\mathrm{H}\hat{\mathbf{B}}|\approx \mathbf{C}_\text{obj}$.
	\end{algorithmic}
\end{algorithm}

\begin{figure}[!t]
	\vspace{-0.35cm}
	\centering
	\includegraphics[width=0.85\linewidth]{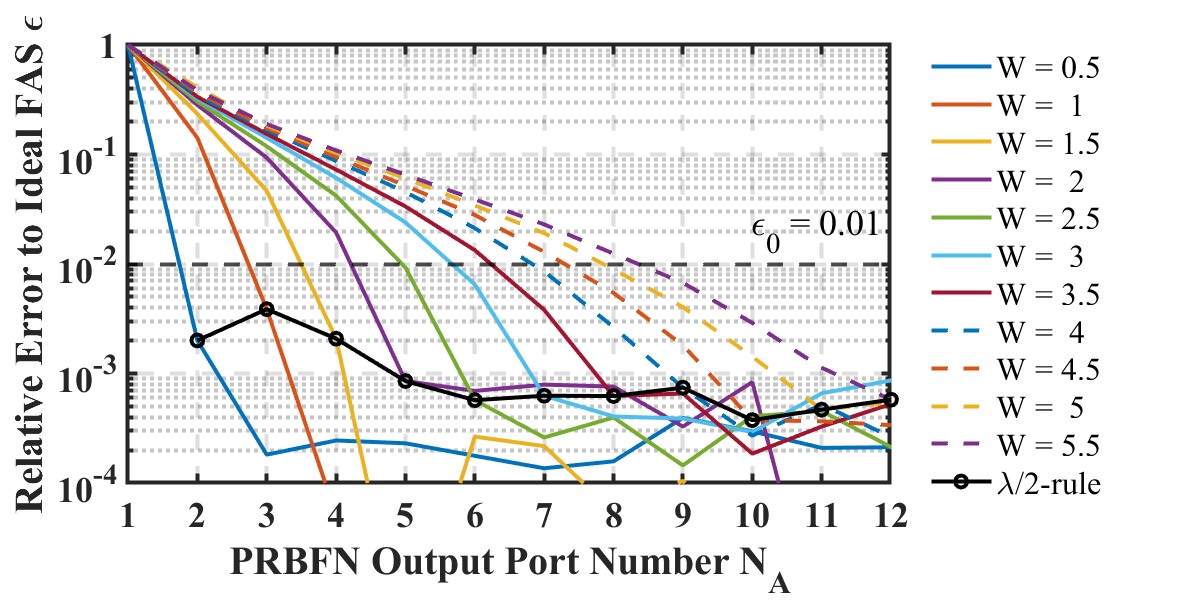}
	\vspace{-0.35cm}
	\caption{The relative error $\epsilon(N_A)$ of optimized correlation result $|\hat{\mathbf{B}}^\mathrm{H}\hat{\mathbf{B}}|$ against target Bessel $\mathbf{C}_\text{obj}$ versus PRBFN output port number $N_\text{A}$, with the FAS port density of $N/W=10$.}
	\label{Parameters}
\end{figure}

For a set of given $W$ and $N$, Algorithm \ref{Optim_target} can be applied to find the optimum $\hat{\mathbf{B}}$ with different $N_\text{A}$. To determine a suitable but minimal $N_\text{A}$, the relative error $\epsilon(N_\text{A})$ is denoted by 
\begin{equation}
	\label{eqn3-0-6}
	\epsilon(N_\text{A})=\frac{f_{N_\text{A}}(\hat{\mathbf{B}})}{f_{1}(\hat{\mathbf{B}'})}=\frac{\Vert |\hat{\mathbf{B}}^\mathrm{H}\hat{\mathbf{B}}|- \mathbf{C}_\text{obj} \Vert_F^2}{\Vert \mathbf{U}_N-\mathbf{C}_\text{obj}\Vert^2_F},
\end{equation}
where the denominator for normalization is the error $f_{N_\text{A}}$ achieved by setting $N_\text{A}=1$. In this case, $|(\hat{\mathbf{B}'})^\mathrm{H}\hat{\mathbf{B}'}|=\mathbf{U}_N$ is the all-one matrix since there is only one antenna. Assuming FAS port density as $N/W=10$, the $\epsilon(N_\text{A})$ with various $W$ are plotted in Fig. \ref{Parameters}. The relative error decreases monotonically with increasing $N_\text{A}$, and fluctuation occurs below $10^{-3}$ due to numerical limits. Setting a threshold $\epsilon_0 = 0.01$, we select the minimal $N_\text{A}$ that satisfies $\epsilon(N_\text{A}) < \epsilon_0$.

For any given $W$, a simple but effective way to determine a suitable $N_\text{A}$ is to ensure the antenna separation exceeds $\lambda/2$, referred to here as the $\lambda/2-$rule, where $N_A$ satisfies
\begin{equation}
	\label{eqn3-0-7}
	N_\text{A} \geq \left[ W/0.5 \right]+1,
\end{equation}
in which $[\cdot]$ denotes the floor function. In Fig. \ref{Parameters}, using this minimal $N_\text{A}$ in (\ref{eqn3-0-7}), it can be observed that $\epsilon(N_\text{A})$ falls below $\epsilon_0$ with various $W$, ensuring that $\mathbf{C}_\text{obj}$ is approached as closely as possible. Therefore, the number of antennas $N_A$ required in our beamforming FAS can be determined by (\ref{eqn3-0-7}).

\begin{figure}[!t]
	\vspace{-0.3cm}
	\centering
	\begin{overpic}[width=0.49\linewidth]{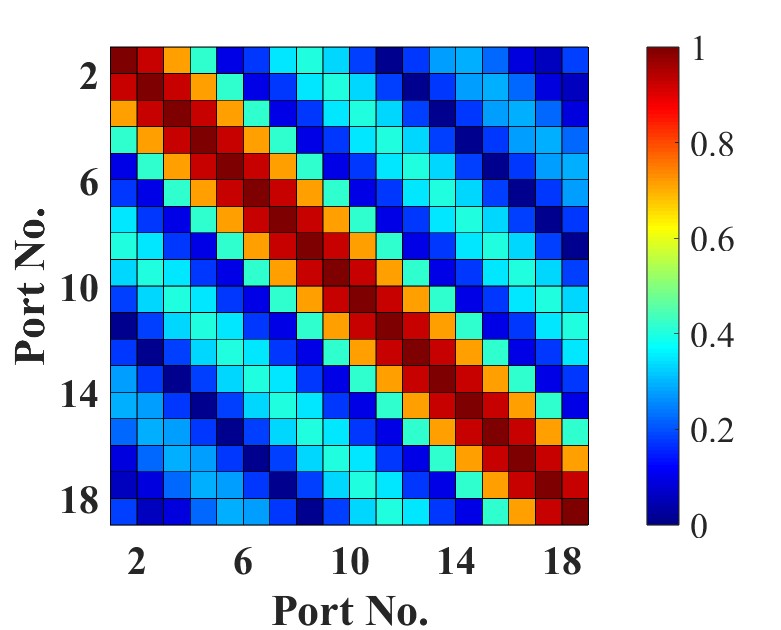}
		\put(0,3){(a)}
	\end{overpic}
	\hfil
	\begin{overpic}[width=0.49\linewidth]{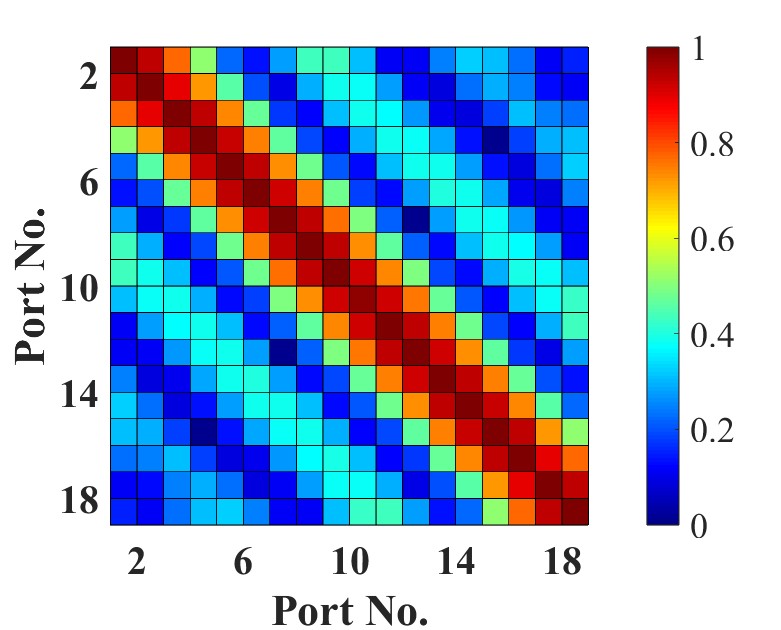}
		\put(0,3){(b)}
	\end{overpic}
	\vspace{-0.7cm}
	\caption{Correlation between states. (a) Ideal Bessel correlation matrix $\mathbf{C}_\text{obj}$ with the FAS parameters of $W=1.5, N=18, N_\text{A}=4$. (b) Optimized correlation $|\hat{\mathbf{B}}^\mathrm{H}\hat{\mathbf{B}}|$ with a relative error of $\epsilon(N_\text{A})=0.001$. }
	\label{Correlation_target}
\end{figure}

To validate the proposed optimization method above, we present a design example with $W=1.5$, $N=18$, $N_\text{A}=4$. Fig. \ref{Correlation_target}(a) shows the ideal Bessel correlation $\mathbf{C}_\text{obj}$, and Fig. \ref{Correlation_target}(b) shows the optimized correlation $\hat{\mathbf{C}} = |\hat{\mathbf{B}}^\mathrm{H}\hat{\mathbf{B}}|$ from Algorithm \ref{Optim_target}. The close matching confirms the effectiveness of the proposed PGD optimization.

Furthermore, the PGD optimization can generate multiple viable beamforming currents $\hat{\mathbf{B}}$. Given that the PGD in Algorithm \ref{Optim_target} is an efficient linear optimization process, we randomly generate 30 initial $\mathbf{B}^{(0)}$ using a uniform distribution with unit constraint for each column, and select the $\hat{\mathbf{B}}$ with the smallest phase differences across the $N_\text{A}$ output ports among all available results. This strategy helps minimize the circuit complexity and physical size of the PRBFN in the subsequent implementation, thereby reducing overall design difficulty.

\subsection{Cascaded Topology of the PRBFN}

\begin{figure}[!t]
	\centering
	\vspace{-0.1cm}
	\includegraphics[width=1\linewidth]{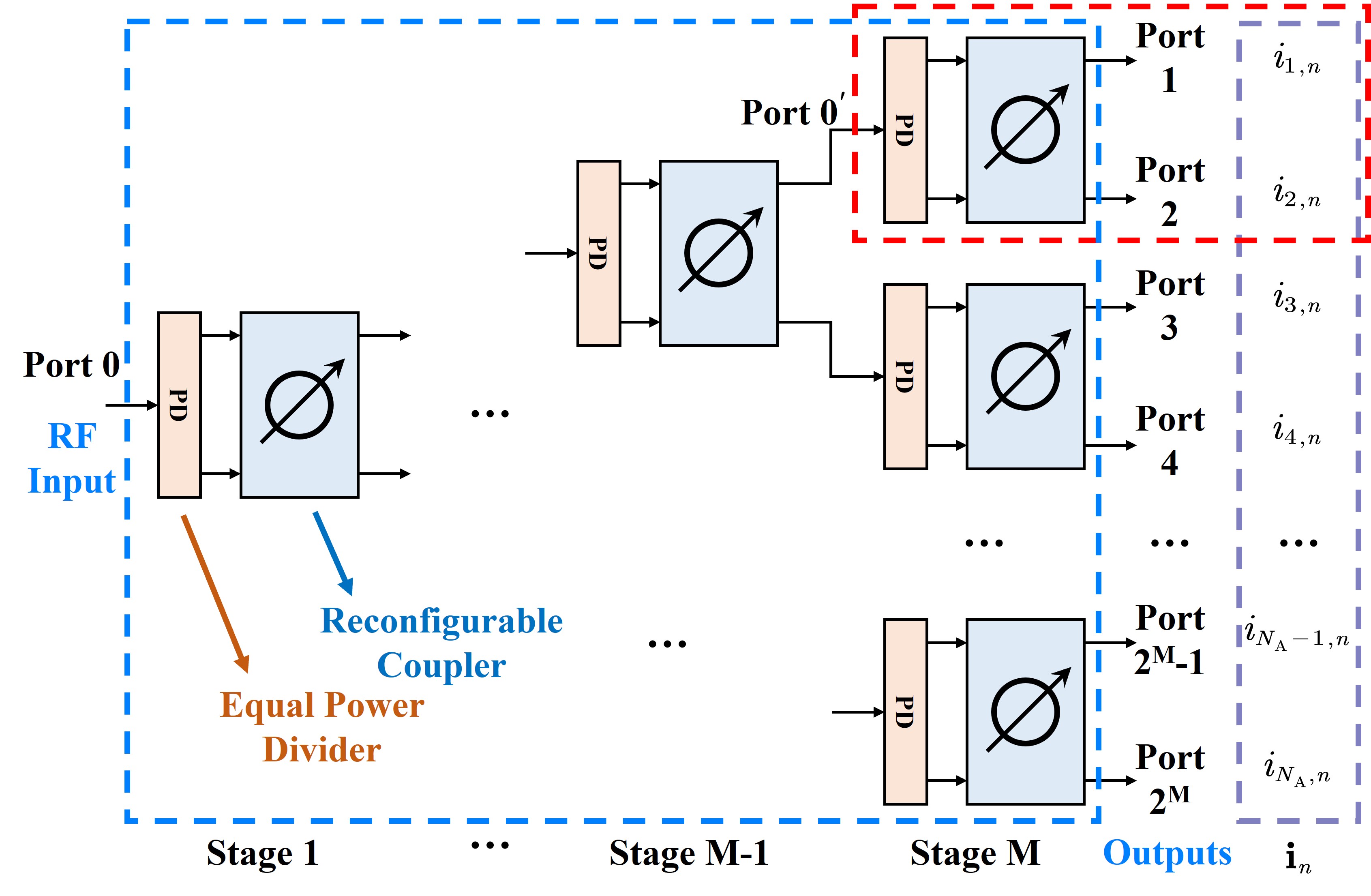}
	\vspace{-0.75cm}
	\caption{The $M$-stage cascaded topology of the proposed PRBFN, with single input (single RF chain of FAS) and $N_\text{A}=2^M$ output ports. The $n$-th reconfigurable state $\mathbf{i}_n$ of $\mathbf{B}$ is presented. }
	\label{PRBFN_Cascade}
\end{figure}

In this subsection, we describe the third step in our design process, i.e. constructing the PRBFN using a cascaded topology, as shown in Fig. \ref{PRBFN_Cascade}. The single RF chain in FAS acts as RF input, which is scaled by cascaded unit cells to support $N_\text{A}$ outputs. Each unit cell comprises an equal power divider (PD) and a reconfigurable coupler, allowing the precise control over both the amplitude and phase of the $N$ output currents at $N_\text{A}$ ports. In this subsection, insertion loss is initially neglected for methodological simplicity, and can be compensated later by amplifiers as shown in 
Fig. \ref{FAS_Architecture}.

For clarity, the unit highlighted in the red box in  is selected as an example. This unit cell can be identified by its stage ($M$-th stage) and its vertical position within that stage (the $1^\text{st}$ in the $M$-th stage), and we therefore denote the currents at its output for the $n$-th state as $\mathbf{i}^{1}_{M,n} = \left[	i_{1,n}	i_{2,n}\right]^\text{T}$. Other unit cells at various locations follow the same rule. In addition, for the topology in Fig. \ref{PRBFN_Cascade}, the number of PRBFN output ports $N_\text{A}$ must be a power-of-two, and thus we select $N_\text{A}=2^M$ while following formula in (\ref{eqn3-0-7}). Therefore, for the $m$-th ($m\leq M$) stage, there are a total of $2^m$ unit cells.
 
The output current vector of the $k$-th ($k\leq 2^{(m-1)}$) cell in the $m$-th stage can be denoted as $\mathbf{i}^k_{m,n}\in\mathbb{C}^{2\times1}$ corresponding to its $n$-th reconfigurable state. Under good matching conditions, the output currents of the $n$-th state can be written as
\begin{equation}
	\label{eqn3-8}
	\mathbf{i}_n = \mathbf{H}_{M,n}\mathbf{H}_{M-1,n}\cdots\mathbf{H}_{2,n}\mathbf{H}_{1,n},
\end{equation}
where $\mathbf{i}_n = [i_{1,n},i_{2,n},\cdots,i_{N_\text{A},n}]^\mathrm{T} \in \mathbb{C}^{2^M\times1}$, and $\mathbf{H}_{m,n} = \mathrm{blkdiag}(\mathbf{i}^1_{m,n},\mathbf{i}^2_{m,n},\cdots,\mathbf{i}^{2^{(m-1)}}_{m,n})\in\mathbb{C}^{2^m \times 2^{(m-1)}}$ is the transmission response of the overall $m$-th PRBFN stage. To construct the overall architecture of the PRBFN, the output beamforming currents $\mathbf{B}$ for all $N$ states must be obtained. This requires solving for all $\mathbf{H}_{m,n}$ in equation (\ref{eqn3-8}).

Based on the optimal $\hat{\mathbf{B}}$ obtained from the optimization algorithm described previously, the required output current from the PRBFN, for each state, is known. Using this we aim to optimize each individual module such that the overall output currents $\mathbf{B}$ of the PRBFN in (\ref{eqn2-7}) closely approximates the optimal $\hat{\mathbf{B}}$ in (\ref{eqn2-11-1}) across all $N$ reconfigurable states. The PRBFN design follows a backward iterative procedure, starting with the unit cells at the final ($M$-th) stage and progressing stage-by-stage until reaching the first stage. Following the lossless and matching assumption, the PRBFN design procedure consists of iterative calculations, whose general design process can be summarized as follows:

\subsubsection{Design $2^{(M-1)}$ unit cells for the $M$-th PRBFN stage} According to $\hat{\mathbf{i}}_n$ from $\hat{\mathbf{B}}$, we design each unit cell of the $M$-th stage in PRBFN separately. To illustrate the design process, the unit cell highlighted in the red box in Fig. \ref{PRBFN_Cascade} is again taken as an example. In the $n$-th of $N$ reconfigurable states, the unit cell should satisfy the following conditions
\begin{equation}
	\label{eqn3-9}
	\begin{split}
%		&\mathbf{J}^1_{M,n}=\mathop{\arg\min}\limits_{\mathbf{J}^1_{M,n}} \\
		&|i_{1,n}| = \frac{|\hat{i}_{1,n}|}{\Vert\hat{\mathbf{i}}_{M,n}^1 \Vert_2},\ |i_{2,n}| = \frac{|\hat{i}_{2,n}|}{\Vert\hat{\mathbf{i}}_{M,n}^1 \Vert_2},\\
		&\angle i_{1,n}-\angle i_{2,n} = \angle \hat{i}_{1,n}-\angle \hat{i}_{2,n},\\
	\end{split}
\end{equation}
where $\hat{\mathbf{i}}_{M,n}^1=[\hat{i}_{1,n},\hat{i}_{2,n}]^\mathrm{T}$ are the first two terms in the optimal current vector $\hat{\mathbf{i}_n}$ (see (\ref{eqn2-11-1})). Similarly, the same constraints apply uniformly to all the other units within the $M$-th PRBFN stage, thus the transmission matrix $\mathbf{H}_{M,n}=\mathrm{blkdiag} (\mathbf{i}^1_{M,n},\mathbf{i}^2_{M,n},\cdots,\mathbf{i}^{2^{(M-1)}}_{M,n})$ of the $M$-th PRBFN stage can be derived.

\subsubsection{Iterative calculations to obtain the PRBFN} With the known $\mathbf{H}_{M,n}$ derived above, the conjugate transpose of $\mathbf{H}_{M,n}$ is taken on both sides of equation (\ref{eqn3-8}), yielding
\begin{equation}
	\label{eqn3-10}
		\mathbf{H}_{M,n}^\mathrm{H}\mathbf{i}_n = \left(\mathbf{H}_{M,n}^\mathrm{H}\mathbf{H}_{M,n}\right) \mathbf{H}_{M-1,n}\cdots\mathbf{H}_{2,n}\mathbf{H}_{1,n}
\end{equation}
where the Gram term $\left(\mathbf{H}_{M,n}^\mathrm{H}\mathbf{H}_{M,n}\right)$ can be expanded as
\begin{equation}
	\label{eqn3-11}
	\begin{bmatrix}
		\Vert\mathbf{i}^{1}_{M,n}\Vert_2^2 & 0 & \cdots & 0\\
		0 & \Vert\mathbf{i}^{2}_{M,n}\Vert_2^2	&\cdots & 0\\
		\cdots & \cdots & \cdots & \cdots\\
		0 & 0 & \cdots & \Vert\mathbf{i}^{2^{(M-1)}}_{M,n}\Vert_2^2
	\end{bmatrix} = \mathbf{U}_{2^{(M-1)}},
\end{equation}
where $\mathbf{U}_{2^{(M-1)}} \in \mathbb{R}^{2^{(M-1)}\times 2^{(M-1)}}$ is the identity matrix. Therefore, (\ref{eqn3-10}) can be updated using
\begin{equation}
	\label{eqn3-12}
		\mathbf{i'}_n =\mathbf{H}_{M-1,n}\cdots\mathbf{H}_{2,n}\mathbf{H}_{1,n},
\end{equation}
where $\mathbf{i'}_n = \mathbf{H}_{M,n}^\mathrm{H}\mathbf{i}_n$. The problem thus reduces to the same form as equation (\ref{eqn3-8}). Using the same approach as in the previous step, all unit cells of the $(M-1)$-th stage can be solved. Similarly, the entire PRBFN can be systematically synthesized through this iterative backward solution.

\subsection{Design of Pixel-Based Reconfigurable Unit Cell}

\begin{figure}[!t]
	\vspace{-0.1cm}
	\centering
	\includegraphics[width=1\linewidth]{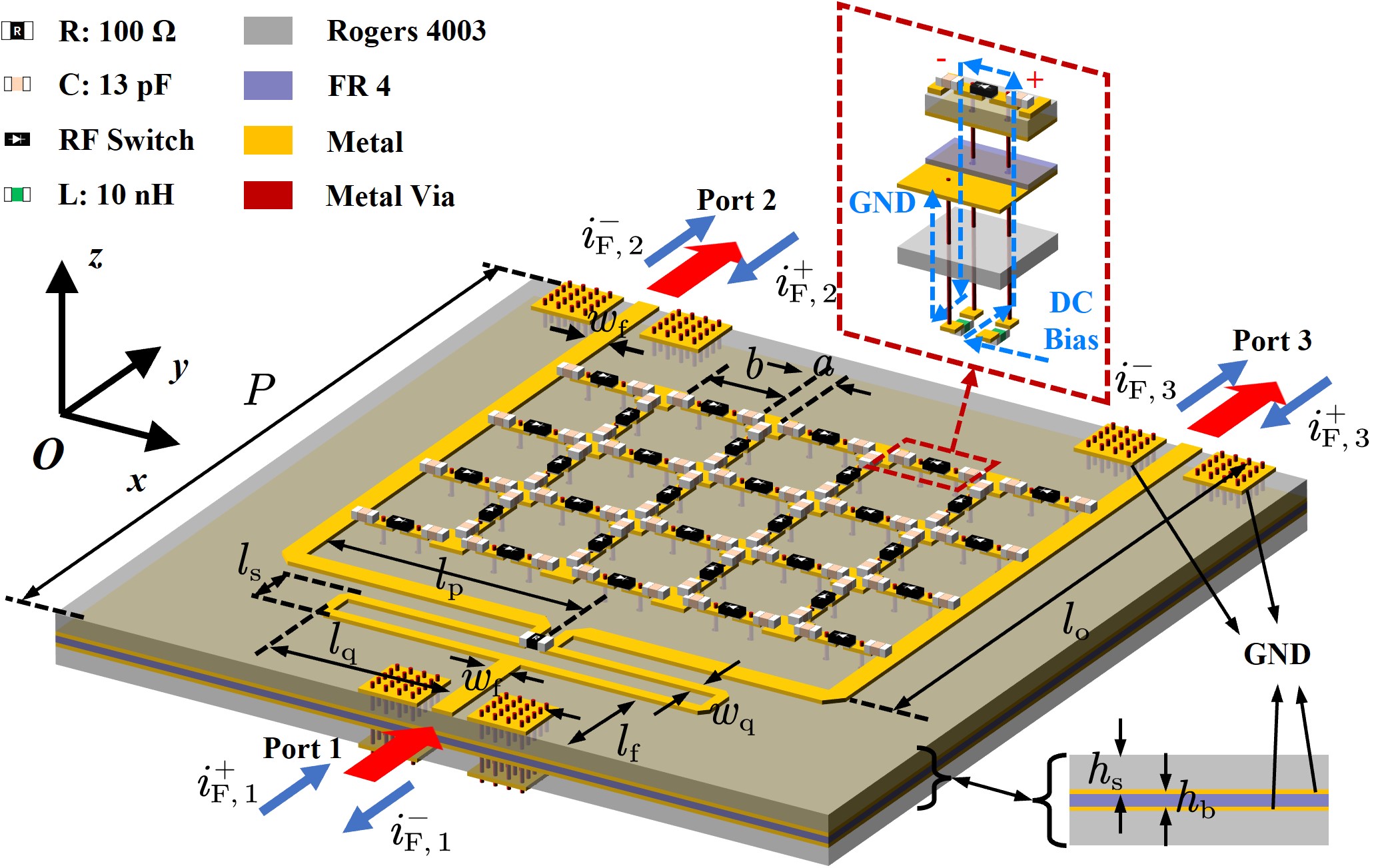}
	\vspace{-0.7cm}
	\caption{The configuration of the proposed pixel-based reconfigurable unit cell in the cascaded PRBFN topology. Dimensions: $P=42$, $h_\text{s}=0.508$, $h_\text{b}=0.1$, $w_\text{f}=1.1$, $w_\text{q}=0.6$, $l_\text{f}=6$, $l_\text{s}=2.5$, $l_\text{q}=10$, $l_\text{p}=13.6$, $l_\text{o}=31$, $a=1.8$, $b=4$ (Unit: mm).}
	\label{PRBFN_Unit_Cell}
\end{figure}

In the final step, the design of each unit cell is devised. The configuration of the unit cell in our proposed PRBFN is illustrated in Fig. \ref{PRBFN_Unit_Cell}, consisting of an equal PD and a reconfigurable coupler based on a pixel structure. The unit is fabricated using two PCB layers bonded with an intermediate FR4 dielectric substrate. The top PCB layer regulates the amplitude and phase of the RF signal by altering pixel connections via RF switches, enabling multiple reconfigurable states. The bottom PCB layer hosts the control DC circuit for these RF switches. A detailed view of a single RF switch control structure is provided in the red box of Fig. \ref{PRBFN_Unit_Cell}. To ensure signal integrity, the bottom layer employs inductors \cite{coilcraft} to isolate RF from the control DC signals, while the upper layer uses capacitors \cite{murata} to block DC signals from the RF pathways, thereby facilitating independent control of all independent RF switches by MCUs or FPGAs.

\begin{figure}[!t]
	\vspace{-0.1cm}
	\centering
	\begin{overpic}[width=1\linewidth]{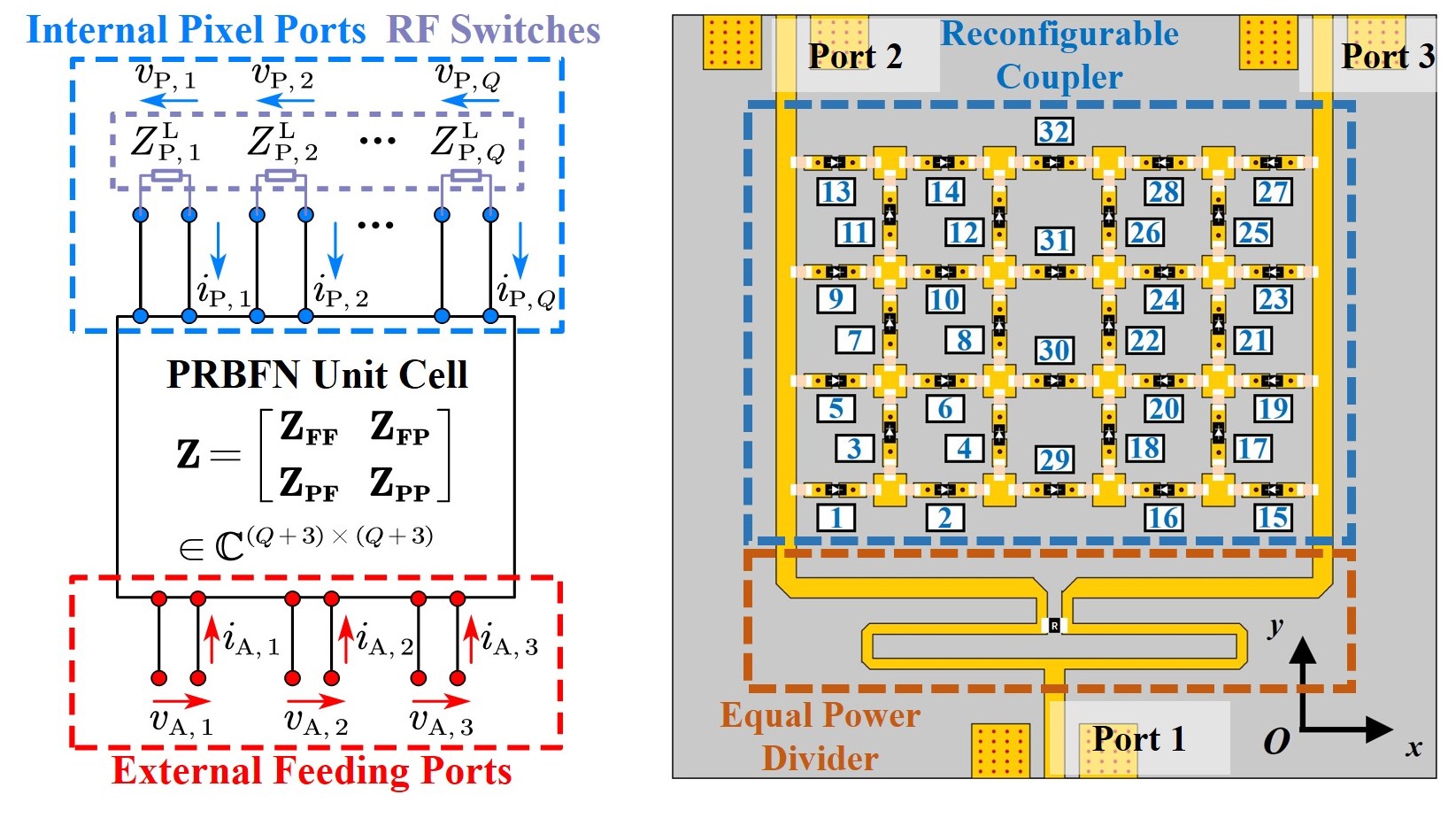}
		\put(0,2){(a)} \put(41,2){(b)}
	\end{overpic}
	\vspace{-0.8cm}
	\caption{Unit cell model and structure. (a) The equivalent circuit model of the unit cell. (b) Top view of the unit cell, with $Q=32$ internal pixel connections and their port number.}
	\label{PRBFN_ECM}
\end{figure}

As shown in Fig. \ref{PRBFN_ECM}(a), the unit cell can be modeled as a multi‑port circuit with three external feeding ports and $Q=32$ internal pixel ports, numbered in Fig. \ref{PRBFN_ECM}(b). Optimizing the internal pixel connections yields $N$ reconfigurable output currents $\mathbf{B}$ that approximate the desired $\hat{\mathbf{B}}$ from Section \Rmnum{3}, thereby satisfying the FAS specifications. All internal ports are terminated with RF switches, diodes SMP1345‑079LF, as loads.  The state of each diode can be expressed using a binary vector $\bm{x}\in\{0,1\}^Q$, with $''$0$''$ and $''$1$''$ representing off and on state, respectively. The corresponding diagonal load impedance matrix is written as $\mathbf{Z^L}=\mathrm{diag}(Z^\mathrm{L}_1,Z^\mathrm{L}_2,\cdots,Z^\mathrm{L}_Q)$.

The voltages and currents of all ports can be related by
\begin{equation}
	\label{5-1}
	\begin{bmatrix}
		\mathbf{v_F} \\ \mathbf{v_P}
	\end{bmatrix}
	= \mathbf{Z}
	\begin{bmatrix}
		\mathbf{i_F} \\ \mathbf{i_P}
	\end{bmatrix}
	= \begin{bmatrix}
		\mathbf{Z_{FF}} & \mathbf{Z_{FP}} \\ 
		\mathbf{Z_{PF}} & \mathbf{Z_{PP}}
	\end{bmatrix}
	\begin{bmatrix}
		\mathbf{i_F} \\ \mathbf{i_P}
	\end{bmatrix}, 
\end{equation}
where $\mathbf{Z_{FF}}\in\mathbb{C}^{3\times3}$ is the impedance matrix between external feeding ports, $\mathbf{Z_{PF}}\in\mathbb{C}^{Q\times3}$ is the mutual impedance between the source feeding ports and each pixel port, $\mathbf{Z_{FP}}$ is the transpose of $\mathbf{Z_{PF}}$, and $\mathbf{Z_{PP}}\in\mathbb{C}^{Q\times Q}$ is the impedance matrix between internal pixel ports. $\mathbf{i_F}\in\mathbb{C}^{3\times1}$, $\mathbf{i_P}\in\mathbb{C}^{Q\times1}$ are the current vectors of feeding ports and pixel ports, respectively, and $\mathbf{v_F}$ and $\mathbf{v_P}$ are voltage vectors. The Z-matrix of the three feeding ports, denoted as $\mathbf{Z_{PR}}$, is expressed by \cite{IMPM}
\begin{equation}
	\label{5-2}
	\mathbf{Z_{PR}}(\bm{x}) = \mathbf{Z_{FF}}-\mathbf{Z_{FP}}\left[\mathbf{Z_{PP}}+\mathbf{Z^L}(\bm{x})\right]^{-1}\mathbf{Z_{PF}},
\end{equation}
from which the S-parameters $\mathbf{S_{PR}}(\bm{x})\in\mathbb{C}^{3\times3}$ of one unit cell with corresponding switch states can be derived. 

\begin{figure}[!t]
	\vspace{-0.05cm}
	\centering
	\begin{overpic}[width=0.498\linewidth]{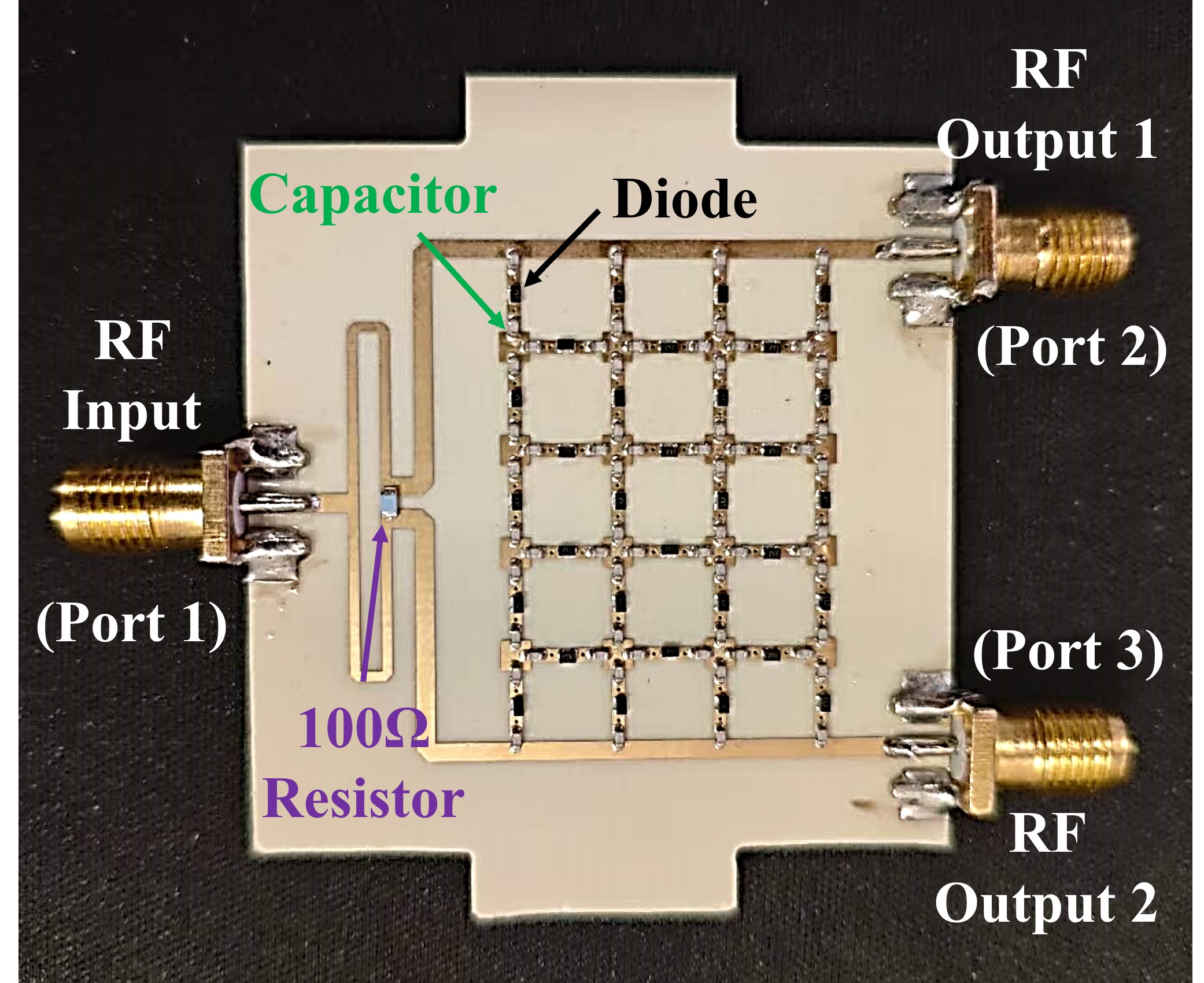}
		\put(1,3){\textcolor{white}{(a)}}
	\end{overpic}
	\hfil
	\begin{overpic}[width=0.488\linewidth]{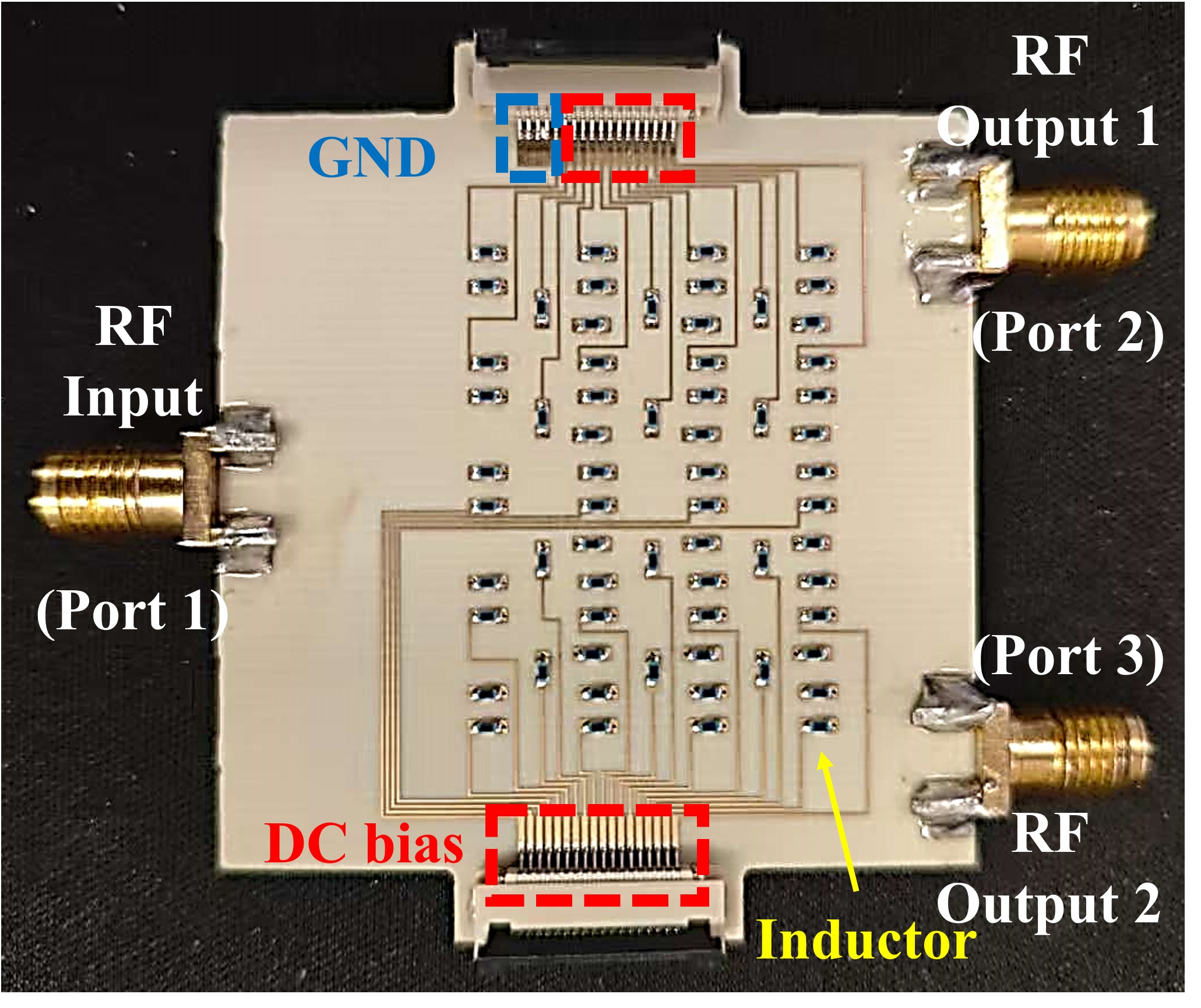}
		\put(1,3){\textcolor{white}{(b)}}
	\end{overpic}
	%	\hfil
	\vspace{-0.6cm}
	\caption{(a) Top view and (b) bottom view of the PRBFN unit cell.}
	\label{PRBFN_Unit_Cell_Prototype}
\end{figure}

Our objective is to align the output beamforming currents $\mathbf{i}^k_{m,n}$ of each unit cell with the target $\hat{\mathbf{i}}^k_{m,n}$ in both amplitudes and phases. For each unit cell, reflected currents are negligible under well‑matched conditions, enabling $i_{\text{F},1}^{-},i_{\text{F},2}^{+},i_{\text{F},3}^{+}\approx0$, while the outgoing currents $i_{\text{F},1}^{+},i_{\text{F},2}^{-},i_{\text{F},3}^{-}$ dominate. Thus the relationship between the outgoing currents at the external feeding ports can be described as 
\begin{equation}
	\label{5-3}
	i_{\text{F},2}^{-}=S_{21}\cdot i_{\text{F},1}^{+},\quad i_{\text{F},3}^{-}=S_{31}\cdot i_{\text{F},1}^{+},
\end{equation}
where $S_{21}$ and $S_{31}$ are the $(2,1)$-th and $(3,1)$-th term of $\mathbf{S_{PR}}$. Therefore, S-parameters can be directly used as optimization goals instead of each output $\mathbf{i}^{k}_{m,n}$ in the PRBFN topology. In subsequent optimization steps, the S-parameters are deployed instead of current-based conditions for simplicity.

Following the steps in last subsection, the pixel connections of each unit cell are optimized separately. For ease of explanation, the unit cell highlighted by the red box in Fig. \ref{PRBFN_Cascade} is taken as an instance again, whose pixel connections of the $n$-th state, denoted as $\bm{x}_n$, need to be optimized. Collecting all $N$ vectors $\bm{x}_n, \forall n=1,2,\cdots,N$ into set $\mathcal{X}$, the optimization problem for the PRBFN unit cell can be expressed as
\begin{equation}
	\label{eqn5-4}
	\begin{split}
		\mathop{\min}\limits_{\bm{x}_n\in\mathcal{X}}& \mathop{\max}\limits_{f_s}\left\{ \sum^N_{n=1}{ \left[ c_1 G_1(\bm{x}_n,f_s)+ c_2 G_2(\bm{x}_n,f_s) \right] } \right\},\\
		\text{s.t.} \ & (C1): \max\{S_{kk,n}(\bm{x}_n,f_s)\} < t_\text{s}, \ k=1,2,3,\\	
		&(C2): \max\{S_{23,n}(\bm{x}_n,f_s)\} < t_\text{m},\\	
		&(C3): \max\{1-G_3(\bm{x}_n,f_s)\} < t_\text{loss}, 
	\end{split} 
\end{equation}
where $c_1, c_2$ are weights, $(C1)$ is the constraint of reflection coefficients, $(C2)$ is the constraint of mutual couplings, $(C3)$ is the constraint of energy loss. $t_\text{s}=-$10 dB and $t_\text{m}=-$20 dB are the thresholds for impedance matching and mutual coupling, and $t_\text{loss}=-$3 dB is the threshold for insertion loss. $G_1(\bm{x}_n,f_s)$, $G_2(\bm{x}_n,f_s)$ measure the amplitude and phase match to the desired $\hat{\mathbf{i}}^1_{M,n}$, respectively, and $G_3(\bm{x}_n,f_s)$ expresses the total transmission power, which can be expanded by
\begin{equation}
	\label{eqn5-5}
	\begin{split}
		&G_1 = \left\Vert \frac{[|S_{21,n}|,|S_{31,n}|]^\mathrm{T}}{\sqrt{G_3}}-\frac{\left[|\hat{i}_{1,n}|,|\hat{i}_{2,n}|\right]^\mathrm{T}}{\Vert\hat{\mathbf{i}}^1_{M,n} \Vert_2} \right\Vert_2^2,\\
		&G_2 = \left|\left(\angle S_{21,n} -\angle S_{31,n} \right)- \left( \angle \hat{i}_{1,n}-\angle \hat{i}_{2,n}\right)\right|^2,\\
		&G_3 = \left|S_{21,n}\right|^2+\left|S_{31,n}\right|^2,	\\
	\end{split} 
\end{equation}
where $(\bm{x}_n,f_s)$ is omitted for simplicity. Following the optimization (\ref{eqn5-4}), the pixel connections of each unit cell in the PRBFN across $N$ reconfigurable states can be confirmed.

Given the symmetry of the target correlation matrix in Fig. \ref{Correlation_target}, the target $\hat{\mathbf{B}}$ can also be set to exhibit this symmetry. This property allows us to design only half of the reconfigurable states in the PRBFN unit. For instance, state 1 and state $N$ are symmetric, i.e, the outputs at ports 2 and 3 in state 1 correspond to those at ports 3 and 2 in state $N$, respectively. This symmetry significantly reduces the optimization burden. 

Using the four steps described above, each PRBFN unit cell can be designed, and a prototype of a unit cell is shown Fig. \ref{PRBFN_Unit_Cell_Prototype}(a) and (b). Using these unit cells, the entire PRBFN-FAS can be prototyped, as described in the next section.  

\section{PRBFN-FAS Prototype and Experimental Results}

To verify the feasibility of the proposed design methodology, we implement a 4‑port PRBFN‑FAS that emulates a conventional FAS with $N=18$ ports over a $1.5\lambda$ linear space ($W=1.5$). We first provide details of the PRBFN prototype, and the experimental results followed by brief details of the required 4-port antenna. Measurements for the integrated 4‑port PRBFN‑FAS are then provided. Finally, the performance of our entire proposed PRBFN-FAS in practical wireless communication scenarios is shown.

\subsection{PRBFN Design}

\begin{figure}[!t]
	\vspace{-0.cm}
	\centering
	\begin{overpic}[width=0.505\linewidth]{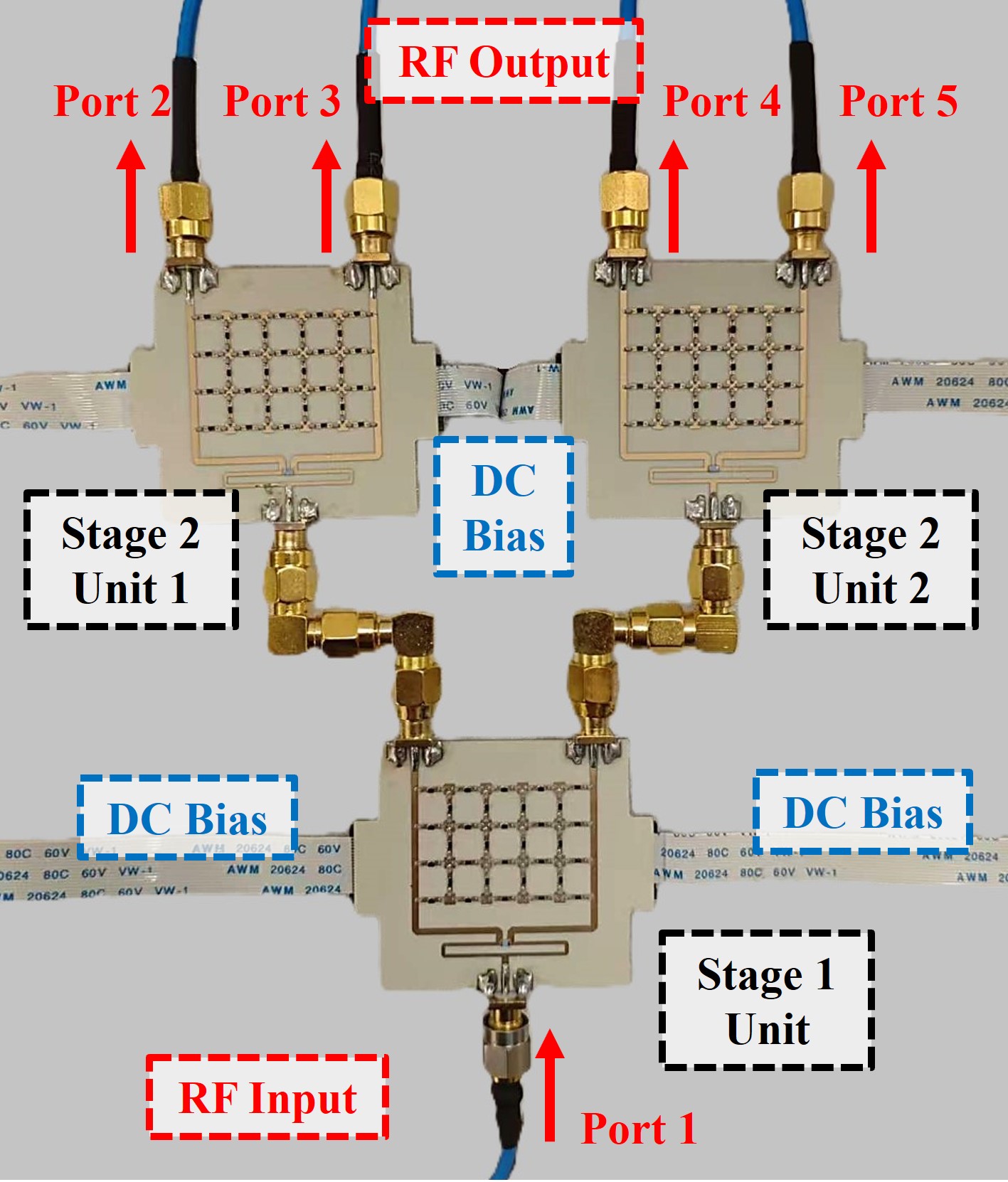}
		\put(0.5,2){(a)}
	\end{overpic}
	\hfil
	\begin{overpic}[width=0.482\linewidth]{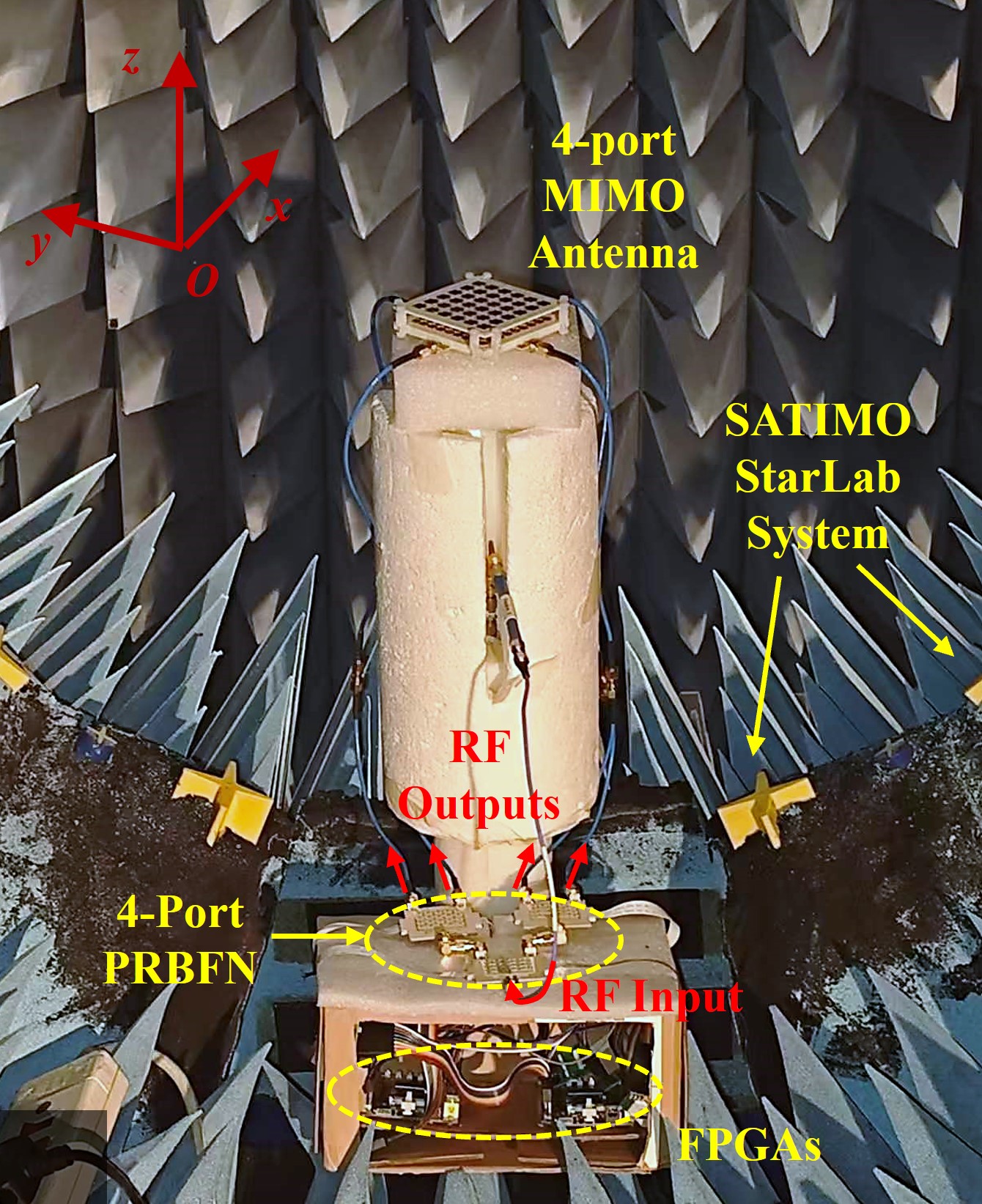}
		\put(0.5,2){\textcolor{white}{(b)}}
	\end{overpic}
	\vspace{-0.65cm}
	\caption{(a) The prototype of the 4-port PRBFN, including 3 unit cells. (b) The measurement setup of the PRBFN-FAS.}
	\label{PRBFN_4Ports_Setup}
\end{figure}

 As depicted in Fig. \ref{PRBFN_4Ports_Setup}(a), the PRBFN prototype has $N_\text{A}=4$ outputs. Using the optimal beamforming currents $\hat{\mathbf{B}}$ from Section \Rmnum{3}, the pixel connections of the three unit cells across $N=18$ reconfigurable states are optimized. Leveraging the symmetry of the target correlation matrix $\hat{\mathbf{C}}= |\hat{\mathbf{B}}^\mathrm{H}\hat{\mathbf{B}}|$ in Fig. \ref{Correlation_target}(b), the Unit in Stage 1 need only support $N/2$ states, while the remaining $N/2$ are obtained by mirroring the diode configurations. Unit 1 and Unit 2 in Stage 2 both support all $N$ states and have axis symmetry. For instance, The output 1 and 2 of Unit 1 in PRBFN State 1 correspond to the output 2 and 1 of Unit 2 in PRBFN State $N$.

\begin{figure}[!t]
	\centering
	\vspace{-0.05cm}
	\begin{overpic}[width=0.6\linewidth]{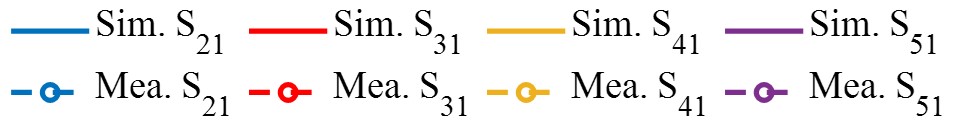}
	\end{overpic}
	\vspace{0.01cm}\\
	\begin{overpic}[width=0.325\linewidth]{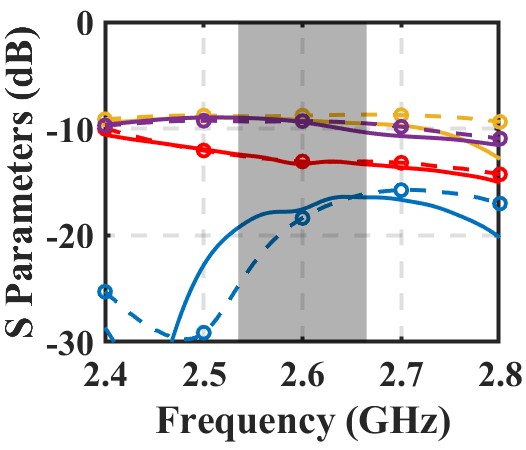}
		\put(0,1){(a)}
	\end{overpic}
	\hfil
	\begin{overpic}[width=0.325\linewidth]{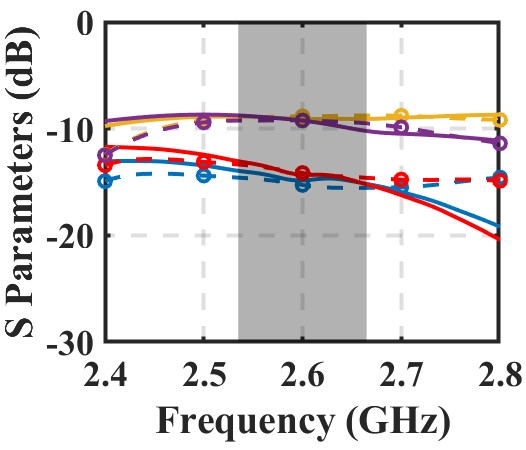}
		\put(0,1){(b)}
	\end{overpic}
	\hfil
	\begin{overpic}[width=0.325\linewidth]{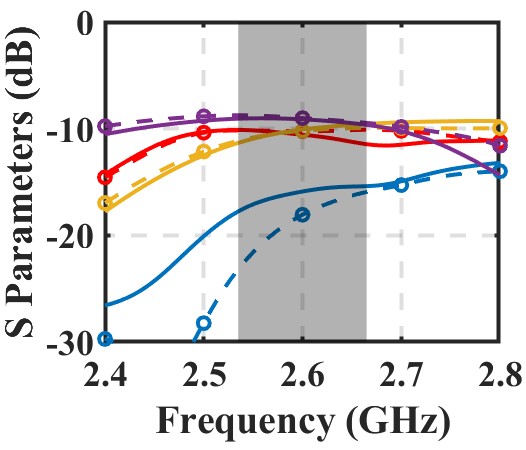}
		\put(0,1){(c)}
	\end{overpic}
	\vfil
	\vspace{0.02cm}
	\begin{overpic}[width=0.325\linewidth]{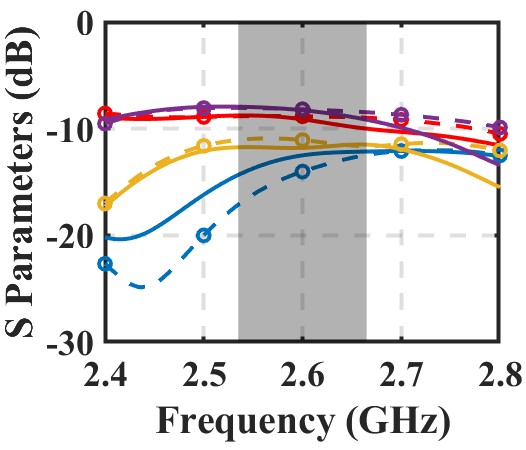}
		\put(0,1){(d)}
	\end{overpic}
	\hfil
	\begin{overpic}[width=0.325\linewidth]{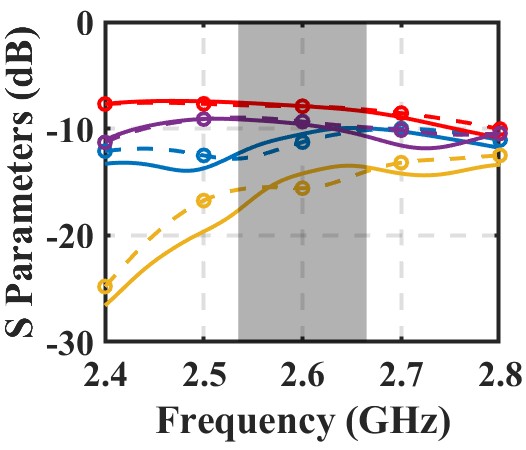}
		\put(0,1){(e)}
	\end{overpic}
	\hfil
	\begin{overpic}[width=0.325\linewidth]{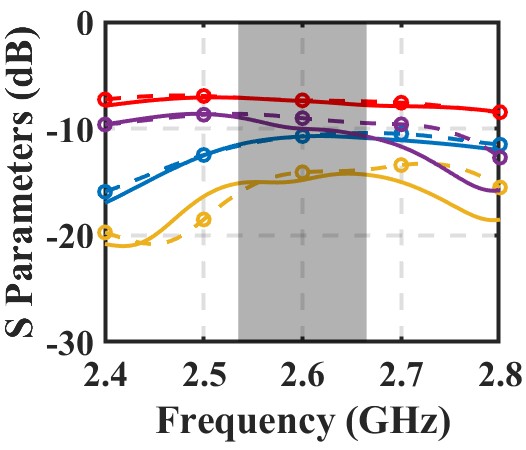}
		\put(0,1){(f)}
	\end{overpic}
	\vfil
	\vspace{0.02cm}
	\begin{overpic}[width=0.325\linewidth]{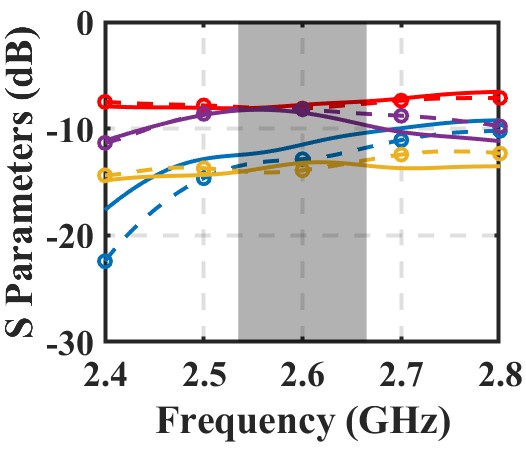}
		\put(0,1){(g)}
	\end{overpic}
	\hfil
	\begin{overpic}[width=0.325\linewidth]{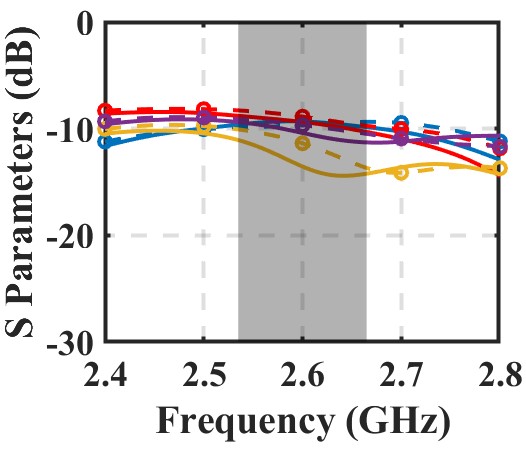}
		\put(0,1){(h)}
	\end{overpic}
	\hfil
	\begin{overpic}[width=0.325\linewidth]{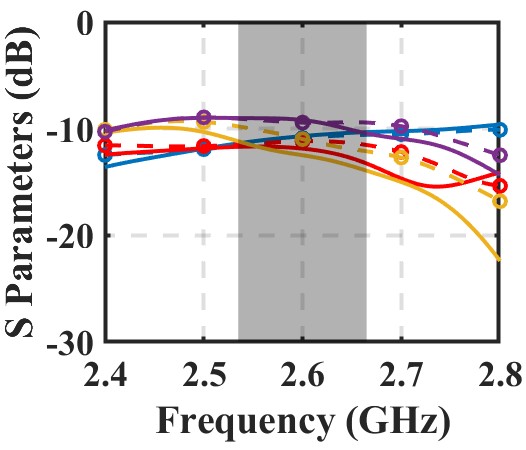}
		\put(0,1){(i)}
	\end{overpic}\\
	\vspace{-0.2cm}
	\caption{The $S_{21},S_{31},S_{41},S_{51}$ of the proposed 4-port PRBFN (a)-(i) State 1 to 9. States 18 to 10 mirror States 1 to 9 with the four output ports interchanged. (Shadow area: center 5\% bandwidth, 130 MHz at 2.6 GHz.)}
	\label{PRBFN_4Ports_Tran}
\end{figure}

\begin{figure}[!t]
	\centering
	\vspace{-0.3cm}
	\begin{overpic}[width=0.49\linewidth]{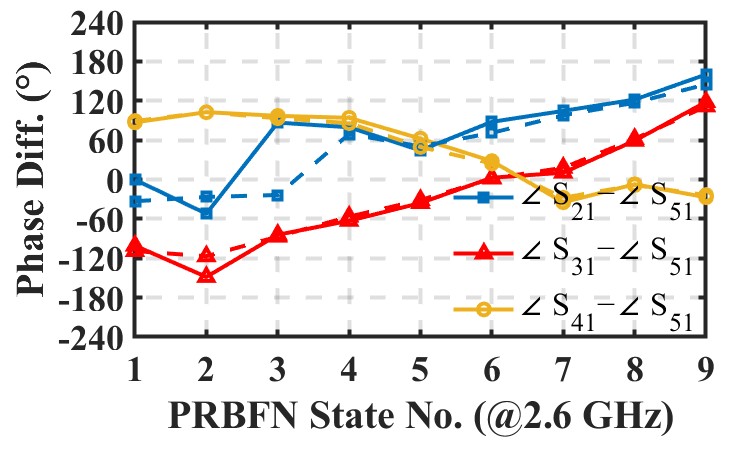}
		\put(0,1){(a)}
	\end{overpic}
	\hfil
	\begin{overpic}[width=0.49\linewidth]{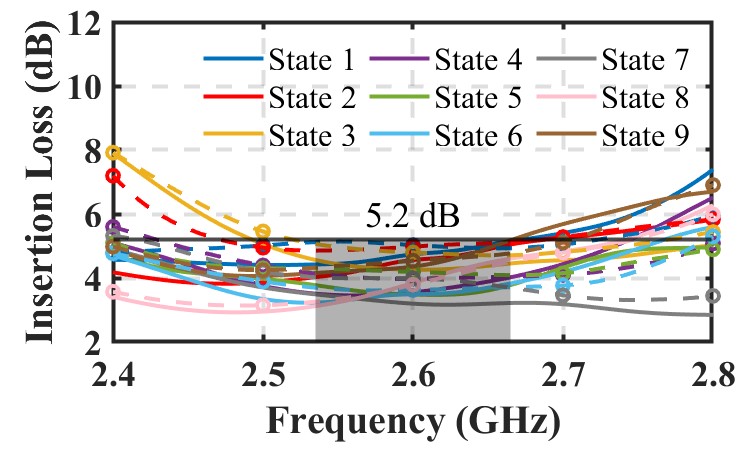}
		\put(0,1){(b)}
	\end{overpic}
	\vspace{-0.65cm}
	\caption{(a) The phase difference among 4 transmission outputs versus all PRBFN-FAS states at 2.6 GHz. (b) Insertion loss of all $N$ reconfigurable PRBFN-FAS states. (Solid line: Sim. Dashed line: Mea.)}
	\label{PRBFN_4Ports_Phase_Loss}
\end{figure}

\begin{figure}[!t]
	\vspace{-0.3cm}
	\centering
	\begin{overpic}[width=0.49\linewidth]{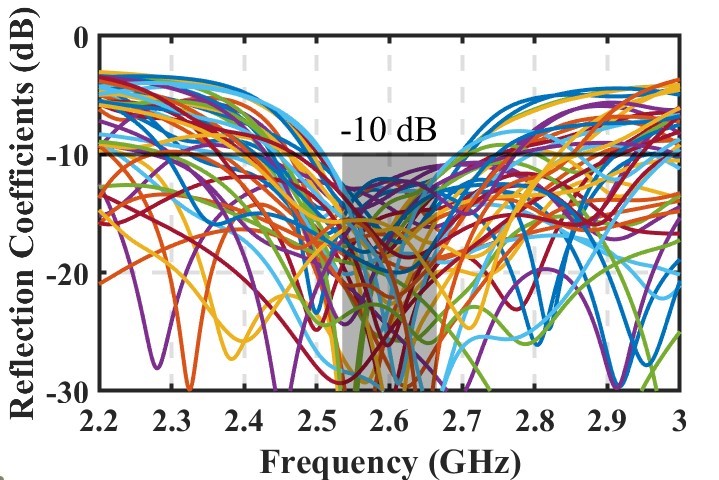}
		\put(0,1){(a)}
	\end{overpic}
	\hfil
	\begin{overpic}[width=0.49\linewidth]{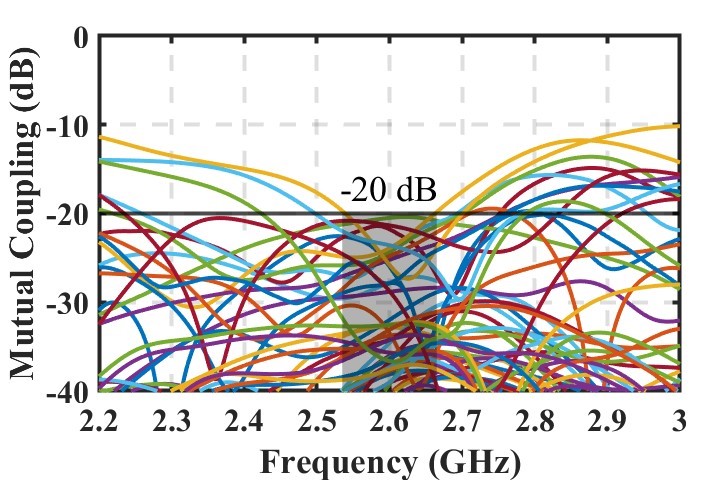}
		\put(0,1){(b)}
	\end{overpic}
	\vspace{-0.65cm}
	\caption{Measured (a) reflection coefficients ($S_{11},S_{22},S_{33},S_{44},S_{55}$), and (b) mutual couplings ($S_{23},S_{24},S_{25},S_{34},S_{35},S_{45}$) of the 4-port PRBFN across all $N=18$ states. (Shadow area: center 5\% bandwidth at 2.6 GHz.)}
	\label{PRBFN_4Ports_Refl}
\end{figure}

Fig. \ref{PRBFN_4Ports_Tran}(a) to (i) presents the simulated and measured transmission coefficients of the first $N/2=9$ states, including $S_{21},S_{31},S_{41},S_{51}$, respectively. Given the mirror symmetry of the PRBFN, the remaining states (State 18 to 10) mirror the behavior of State 1 to 9, with identical S-parameters except for a reversal among Port 2 to 5 in Fig. \ref{PRBFN_4Ports_Setup}(a). Both simulation and measurement results show strong agreement in magnitude across the central 130 MHz, i.e., 5\% bandwidth at 2.6 GHz, covering the n41 band \cite{3gpp}. 

Besides amplitudes, the phase of the PRBFN output is also critical, as the correlation between states depends primarily on phase differences across output ports. With $S_{51}$ magnitudes stable across all $N$ states in Fig. \ref{PRBFN_4Ports_Tran}, we select $S_{51}$ as a reference. Fig. \ref{PRBFN_4Ports_Phase_Loss}(a) shows that most measured phase differences agree well with simulation at 2.6 GHz. Outputs with large phase deviations, such as $S_{21}$ in State 3, have negligible impact on overall correlation due to their very small magnitudes. 

Fig. \ref{PRBFN_4Ports_Phase_Loss}(b) shows the total transmission loss of the 4‑port PRBFN. The maximum loss across all reconfigurable states by two cascaded units of each channel is below 5.2 dB, readily compensated by the following four amplifiers in Fig. \ref{FAS_Architecture}.  The reflection coefficients and mutual couplings for all five external ports across all $N=18$ states are presented in Fig. \ref{PRBFN_4Ports_Refl}(a) and (b). Over the central 5\% operating bandwidth, reflection coefficients stay below $-$10 dB and mutual coupling remains under $-$20 dB, confirming excellent impedance matching and high port isolation in the PRBFN design.

\subsection{Compact Highly Orthogonal 4-Port Antenna}

Following the theory in Section \Rmnum{2}, the multi-port antenna design required for our approach can be an arbitrary antenna exhibiting high orthogonality between feeding ports, which ensures that $\mathbf{K_M}$ in formula (\ref{eqn2-9}) closely approximates the identity matrix. This section briefly introduces the design of a compact 4-port ($N_\text{A}=4$) antenna with high orthogonality, intended for radiation in the PRBFN-FAS in Fig. \ref{FAS_Architecture}.

\begin{figure}[!t]
%	\vspace{-0.1cm}
	\centering
	\includegraphics[width=1\linewidth]{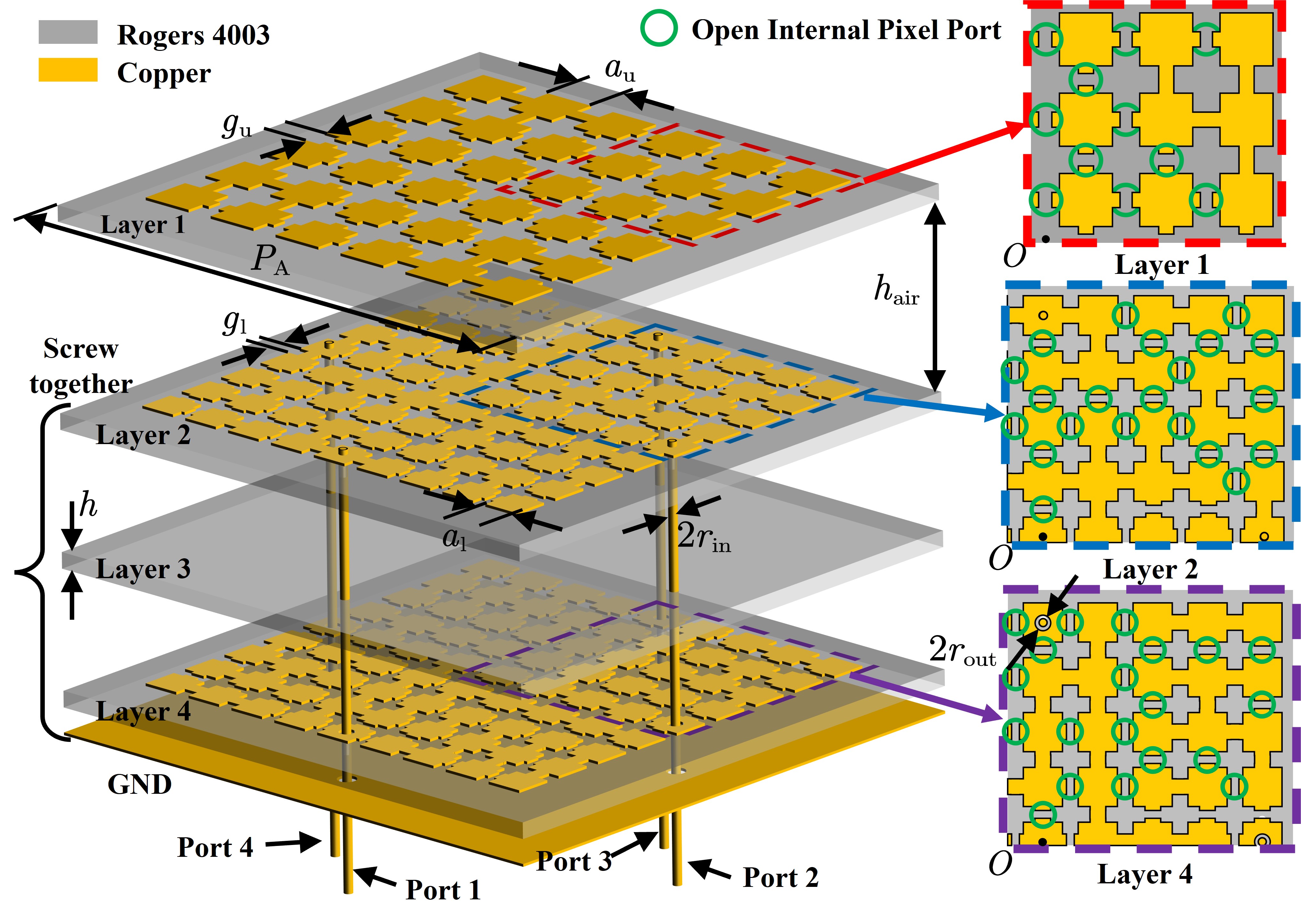}
	\vspace{-0.8cm}
	\caption{The geometry of the compact 4-port antenna with symmetry and orthogonal patterns. Dimensions: $P_\text{A}=75$, $h_\text{air}=8$, $h=1.524$, $a_\text{u}=7$, $g_\text{u}=3.6$, $a_\text{l}=5$, $g_\text{l}=2$, $r_\text{in}=0.3$, $r_\text{out}=0.6$ (Unit: mm).}
	\label{Antenna_Configuration}
\end{figure}

\begin{figure}[!t]
	\vspace{-0.3cm}
	\centering
	\begin{overpic}[width=0.48\linewidth]{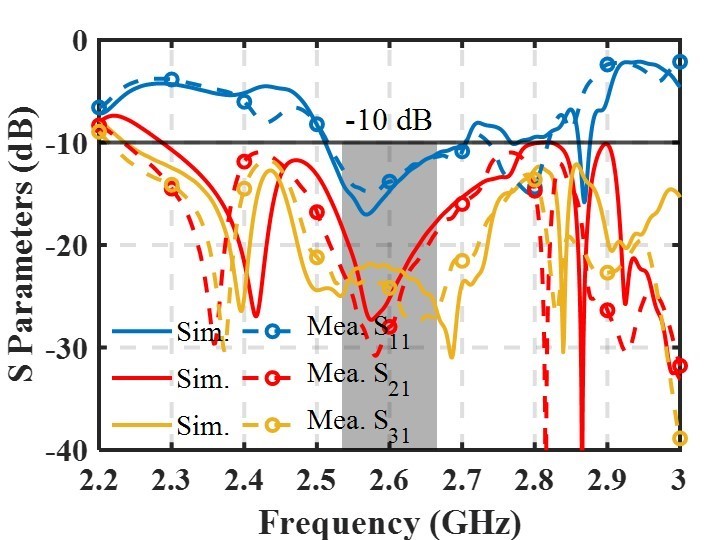}
		\put(0,3){(a)}
	\end{overpic}
	\hfil
	\begin{overpic}[width=0.505\linewidth]{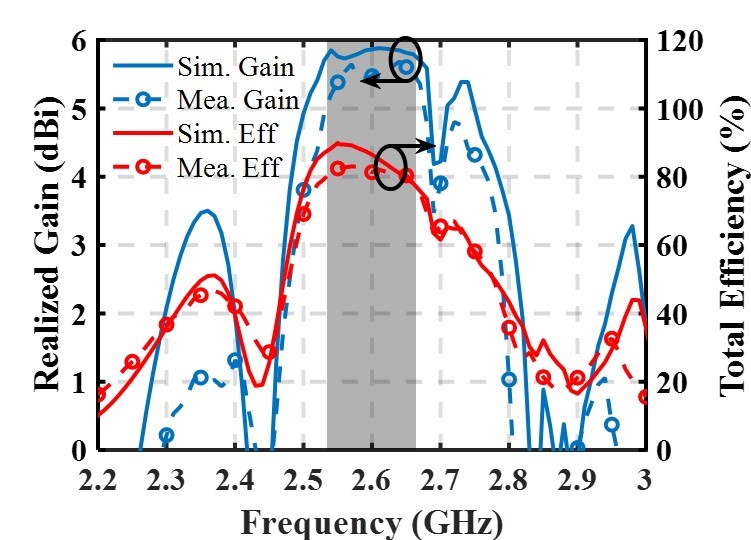}
		\put(0,3){(b)}
	\end{overpic}
	\vspace{-0.65cm}
	\caption{Simulated and measured (a) S-parameters, (b) total efficiency and broadside realized gain of the proposed MIMO antenna.}
	\label{Antenna_S_Eff_Gain}
\end{figure}

\begin{figure}[!t]
	\vspace{-0.3cm}
	\centering
	\begin{overpic}[width=0.57\linewidth]{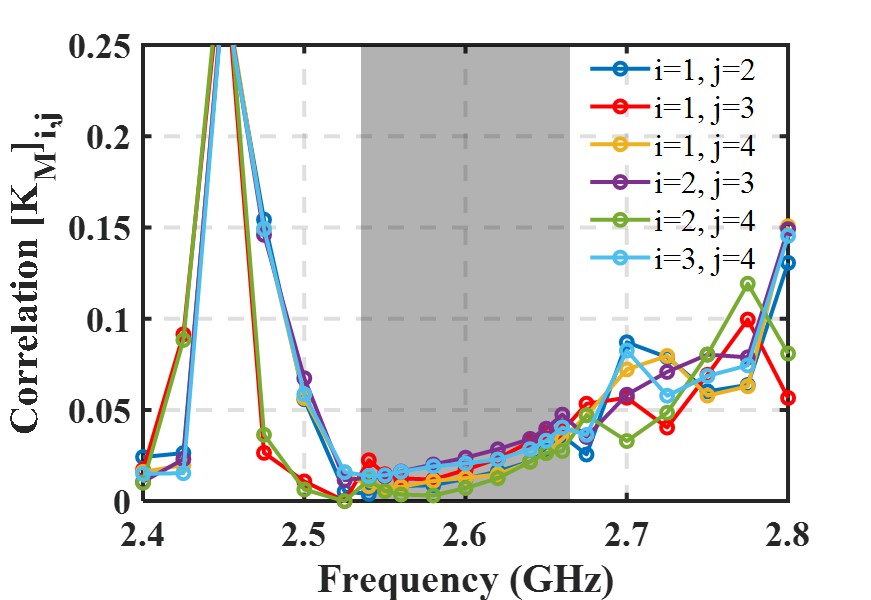}
		\put(0,1){(a)}
	\end{overpic}
	\hfil
	\begin{overpic}[width=0.41\linewidth]{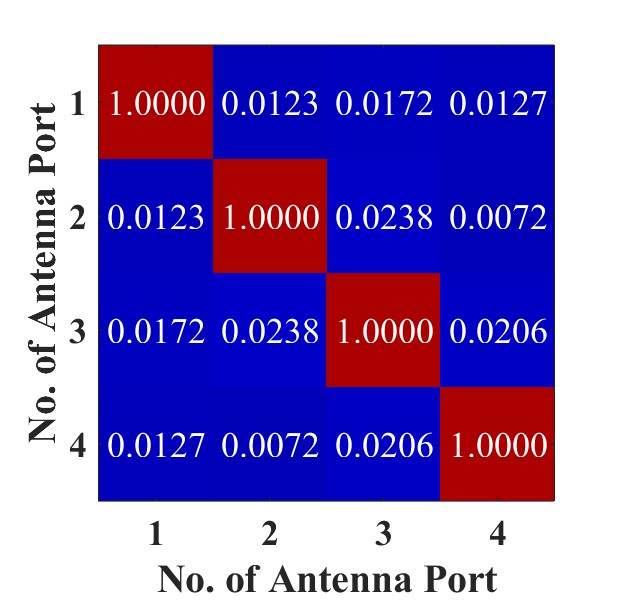}
		\put(0,1){(b)}
	\end{overpic}
	\vspace{-0.65cm}
	\caption{(a) The correlation terms, $[\mathbf{K_{M}}]_{i,j}\ \forall i\neq j$, approach zero across the operating bandwidth shown in gray. (b) Correlation matrix $\mathbf{K_{M}}$ at 2.6 GHz, approximating identity matrix $\mathbf{U}_4$.}
	\label{Antenna_Pattern}
\end{figure}

Based on the methodologies used in \cite{IMPM,duplex,MIMO3}, the geometry of the  4-port antenna is shown in Fig. \ref{Antenna_Configuration}. It is also includes pixel elements and by exploiting rotational symmetry the computation for the optimization of pixel connections is reduced \cite{IMPM}. It incorporates three metal pixel layers, with all four feeds connected to the middle layer (Layer 2). Extending the design in \cite{MIMO3}, we add an upper pixel layer (Layer 1) to enhance bandwidth and a lower pixel layer (Layer 4) to reduce couplings between the 4 feeding ports. 

Due to the rotational symmetry, S-parameters of all four ports exhibit nearly identical behavior. The simulated and measured S-parameters are presented in Fig \ref{Antenna_S_Eff_Gain}(a). It can be observed that across the 5\% fractional bandwidth at the center frequency of 2.6 GHz, all four ports are well matched, and the mutual couplings between any two ports remain below –20 dB. The realized gain and total efficiency are presented in Fig. \ref{Antenna_S_Eff_Gain}(b). Within the operating bandwidth, the realized gain exceeds 5 dBi and the radiation efficiency is consistently above 80\%. These results demonstrate that our proposed 4-port antenna exhibits excellent performance. Fig. \ref{Antenna_Pattern}(a) shows that the multi-port antenna's correlation terms $[\mathbf{K_{M}}]_{i,j}\ \forall i\neq j$ closely approximate zero across the bandwidth. Based on the measured patterns, the most critical performance metric, $\mathbf{K_M}$ at 2.6 GHz, is given in Fig. \ref{Antenna_Pattern}(b), which agree well with the identity matrix $\mathbf{U}_{N_\text{A}}$, fully satisfying the requirements of the PRBFN-FAS architecture.

\subsection{PRBFN-FAS}

\begin{figure}[!t]
	\vspace{-0.1cm}
	\centering
	\includegraphics[width=1\linewidth]{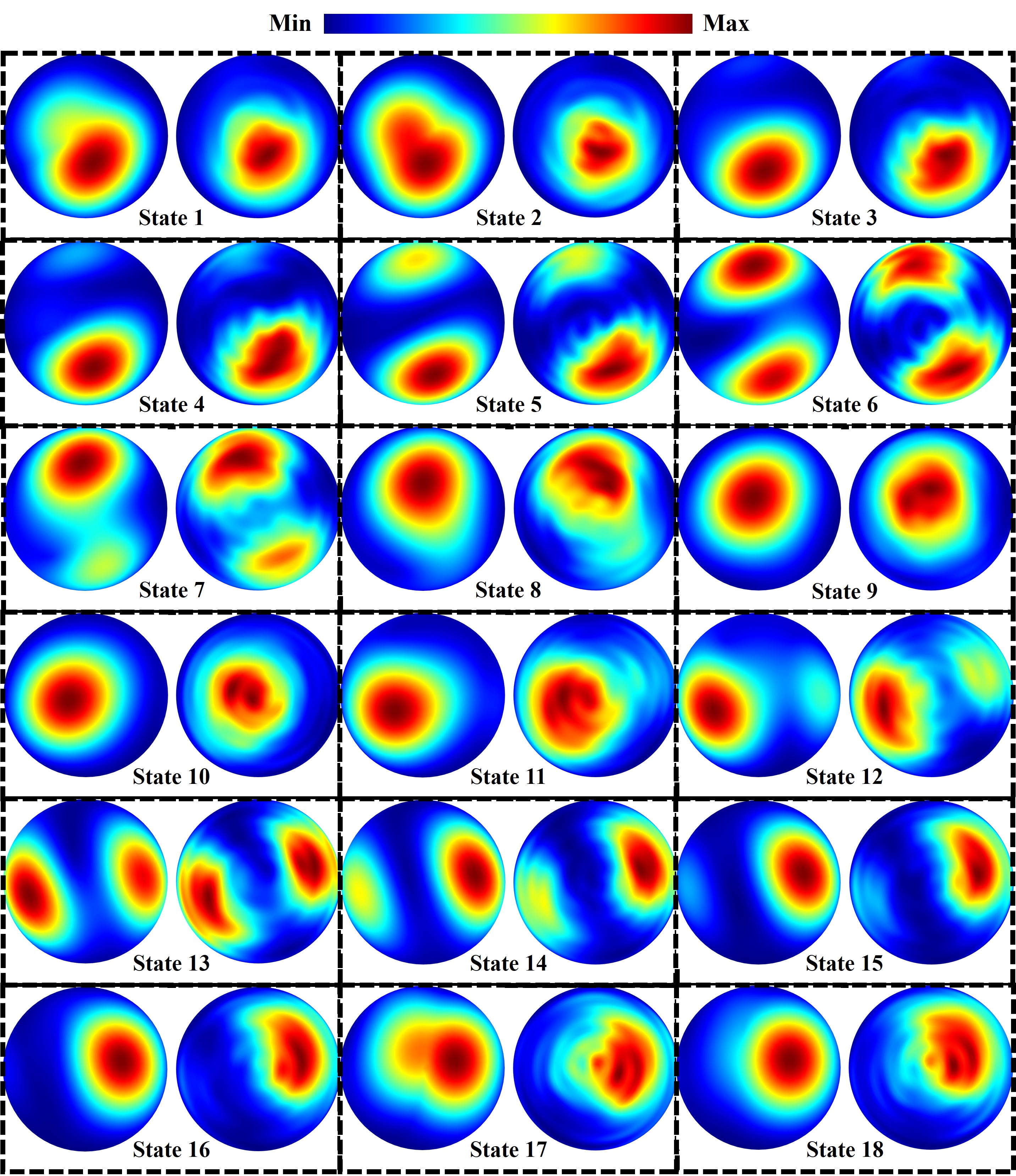}
	\vspace{-0.65cm}
	\caption{Orthographic projection of the normalized radiation patterns (linear) of $N=18$ PRBFN-FAS states. (Left: Simulation. Right: Measurement.)}
	\label{PRBFN_4Ports_Pattern}
\end{figure}

\begin{figure}[!t]
	\vspace{-0.15cm}
	\centering
	\begin{overpic}[width=0.32\linewidth]{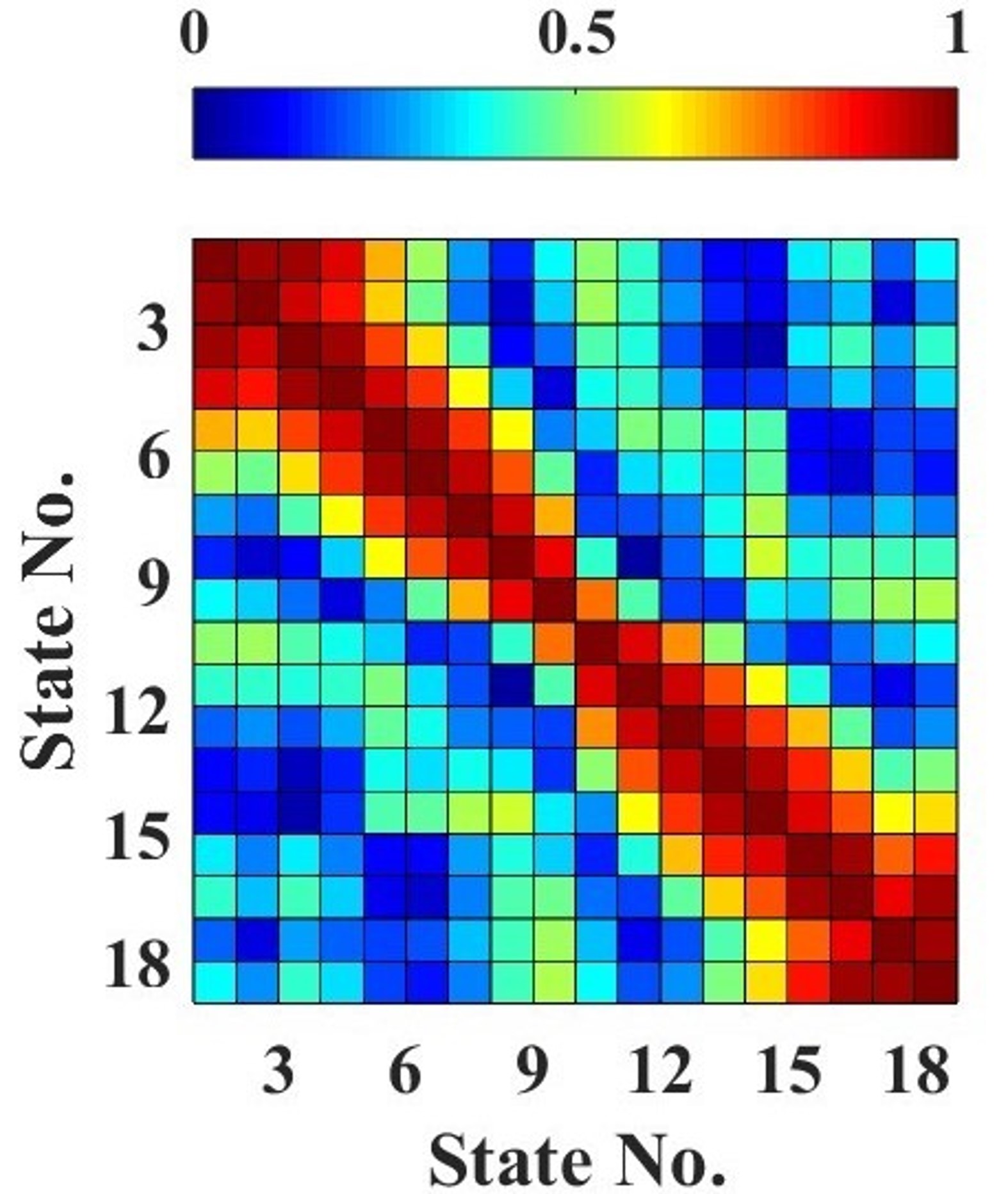}
		\put(0,2){(a)}
	\end{overpic}
	\hfil
	\begin{overpic}[width=0.32\linewidth]{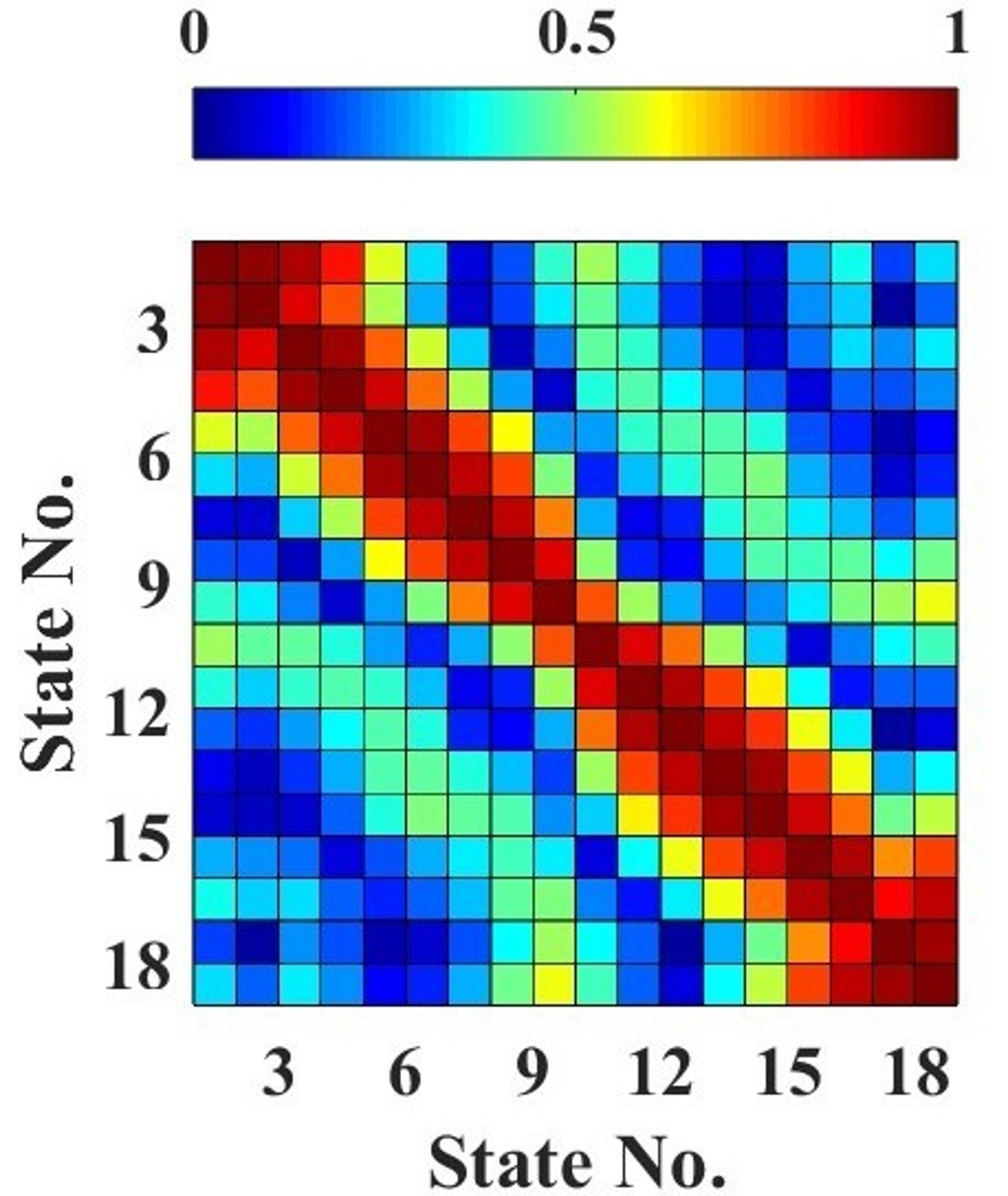}
		\put(0,2){(b)}
	\end{overpic}
	\hfil
	\begin{overpic}[width=0.32\linewidth]{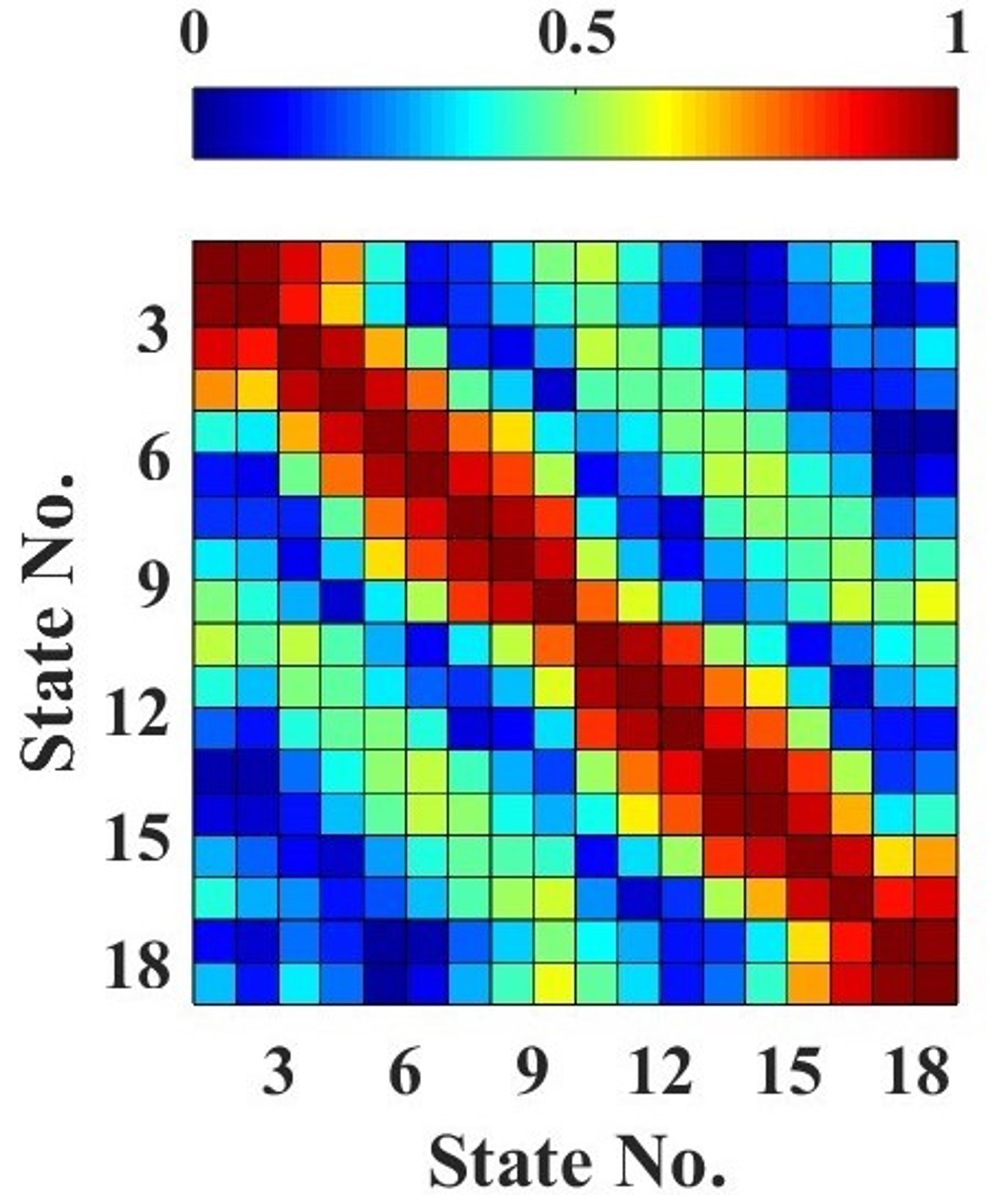}
		\put(0,2){(c)}
	\end{overpic}
	\vspace{-0.3cm}
	\caption{Measured correlation of $N$ reconfigurable states of PRBFN-FAS including the 4-port PRBFN and the proposed MIMO antenna at (a) 2.55, (b) 2.6, (c) 2.65 GHz. The relative errors are $\epsilon=0.062, 0.041, 0.035$.}
	\label{PRBFN_4Ports_Correlation}
\end{figure}

The PRBFN and multiport antenna were prototyped and integrated together to form a complete PRBFN-FAS. The test setup, shown in Fig. \ref{PRBFN_4Ports_Setup}(b), employs two FPGAs to control the reconfigurable state switching. The simulated and measured radiation patterns of all $N=18$ states are presented in Fig. \ref{PRBFN_4Ports_Pattern}. Generally, measured results agree well with the simulated ones, while the fluctuations are caused by the DC bias lines. Radiation patterns in adjacent states exhibit strong similarity, consistent with the high correlation property of $C_{i,j}\approx1$. In contrast, patterns from states far apart show significant variation, demonstrating high orthogonality with $C_{i,j}\approx0$. The measured correlation of $N=18$ beamforming radiation patterns $\mathbf{C}$ is shown in Fig. \ref{PRBFN_4Ports_Correlation}, which remains stable across the operating bandwidth. 

To quantify the deviation between $\mathbf{C}$ and $\mathbf{C}_\text{obj}$, formula (\ref{eqn3-0-6}) is modified and the relative error $\epsilon$ given as
\begin{equation}
	\label{eqn5-5-0}
	\epsilon = \frac{\Vert \mathbf{C}-\mathbf{C}_\text{obj}\Vert^2_F}{f_{1}(\hat{\mathbf{B}'})}=\frac{\Vert |\mathbf{E}^\mathrm{H}\mathbf{E}|-\mathbf{C}_\text{obj}\Vert^2_F}{\Vert \mathbf{U}_N-\mathbf{C}_\text{obj}\Vert^2_F},
\end{equation}
where $\mathbf{E}=[\mathbf{e}_1,\mathbf{e}_2,\cdots,\mathbf{e}_N]$ is $N$ normalized patterns obtained by measurements in Fig. \ref{PRBFN_4Ports_Pattern}. Across the 5\% bandwidth, $\epsilon=0.062, 0.041, 0.035$ at 2.55, 2.6, 2.65 GHz, respectively. The low relative errors demonstrate that the measured correlation results exhibit strong agreement with the ideal Bessel correlation $\mathbf{C}_\text{obj}$ and the target correlation $\hat{\mathbf{C}}$ calculated by $\hat{\mathbf{B}}$ given in Fig. \ref{Correlation_target}(a) and (b), demonstrating that our proposed PRBFN-FAS can be functionally equivalent to ideal FAS with $N=18$ ports across $1.5\lambda$ $(W=1.5)$ space.

It is worth noting that the proposed PRBFN-FAS indeed exhibits beam-scanning capabilities. However, the primary focus lies not on the specific direction of each radiation pattern, but on the relationships among all reconfigurable states. In essence, this approach can be viewed as a form of dense sampling in the radiation pattern domain, which effectively corresponds to dense spatial sampling. Thereby, the PRBFN-FAS achieves functional equivalence to a conventional FAS.

\subsection{System Experiments and Results}

Having validated the correlation performance among reconfigurable states (FAS ports) of the PRBFN‑FAS via radiation patterns, this subsection presents system‑level experiments and evaluation of the proposed 4‑port PRBFN‑FAS in practical communication systems.

\begin{figure}[!t]
	\vspace{-0.1cm}
	\centering
	\includegraphics[width=1\linewidth]{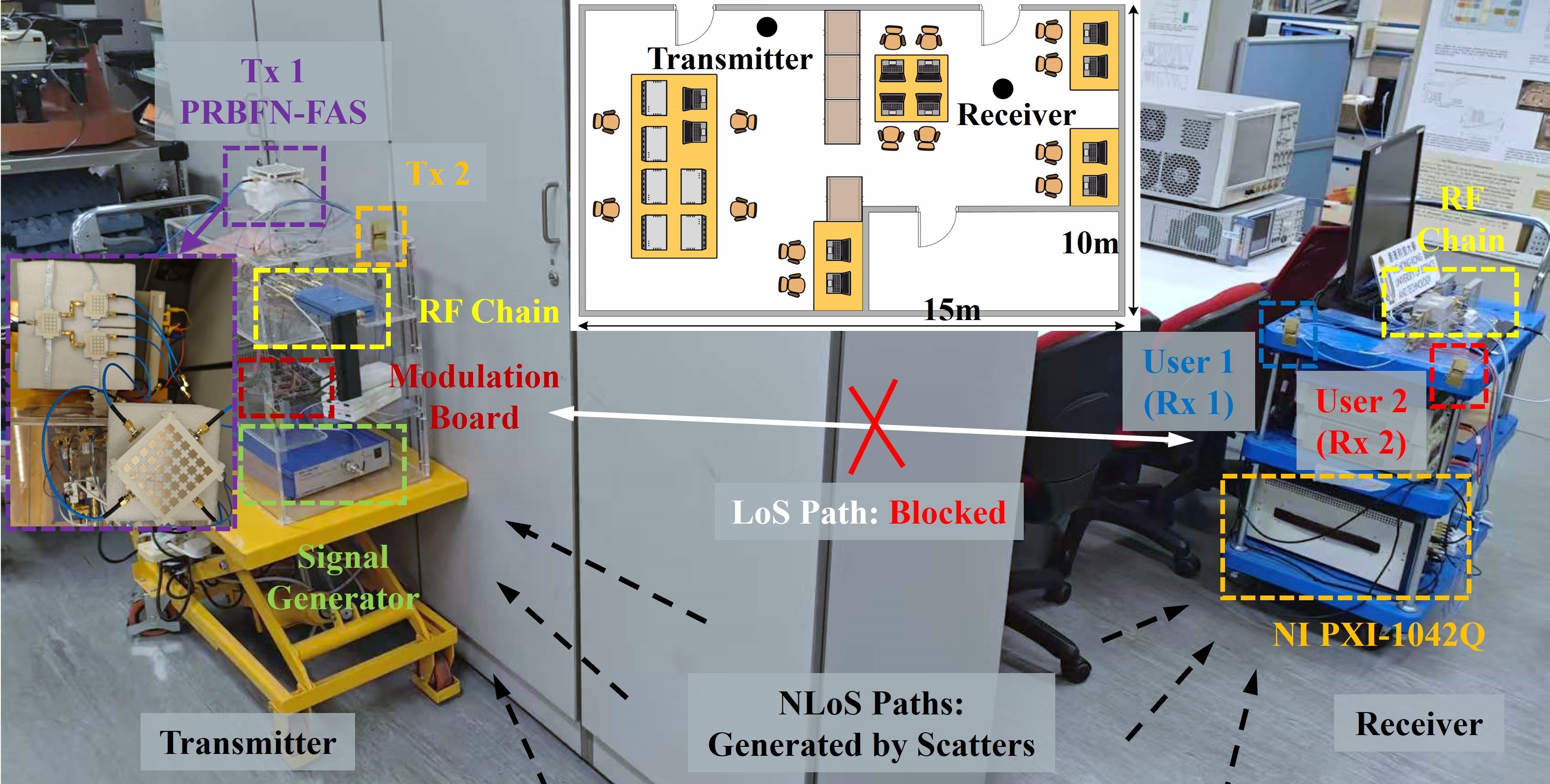}
	\vspace{-0.7cm}
	\caption{Communication system experiment setup with the PRBFN-FAS.}
	\label{System_Measurement}
\end{figure}

We measured the variations in PRBFN-FAS channels across various PRBFN-FAS states in a rich scattering environment. As shown in Fig. \ref{System_Measurement}, a MIMO testbed \cite{testbed} is utilized, which can provide measurements of $4\times4$ wireless channels every 0.01s. In the FAS channel measurement, a simplified $2\times2$ MIMO configuration is utilized. The Tx side of the test setup consists of the proposed PRBFN-FAS (Tx 1) and a reference dipole antenna (Tx 2). The dipole serves to monitor channel stability during measurements. Since it is impractical to measure all FAS port channels simultaneously, we instead sequentially scan through the $N=18$ reconfigurable states of the PRBFN-FAS while ensuring channel stability. Under stable conditions, the channels corresponding to all $N$ FAS ports can be considered quasi-static and thus approximately equivalent at any given time instant. The Rx consists of two dipoles set far apart for independence. 

The measured channel elements of the MIMO system are denoted as $h_{u,v}$ where $u$ refers to the Rx antenna index and $v$ the Tx antenna index. Therefore, with both sides equipped with dipoles, $h_{1,2}$ and $h_{2,2}$ are set as benchmark channels which are expected to remain stable. Channels $h_{1,1}, h_{2,1}$ are two FAS channels with the PRBFN-FAS at the Tx side. The PRBFN-FAS is controlled by FPGAs and set to cycle through $N=18$ reconfigurable states in order, with the switching synchronized to the sampling rate of the MIMO testbed. The channel measurements were conducted in the Wireless Communication Laboratory at the Hong Kong University of Science and Technology. The indoor environment, shown in Fig. \ref{System_Measurement}, features non-line-of-sight (NLoS) environments with the Tx and Rx separated by at least 5 meters and obstructed by cupboards blocking the line-of-sight (LoS) path.

\begin{figure}[!t]
	\vspace{-0.3cm}
	\centering
	\includegraphics[width=0.4\linewidth]{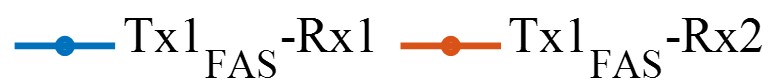}\\
	\vspace{0.02cm}
	\begin{overpic}[width=0.49\linewidth]{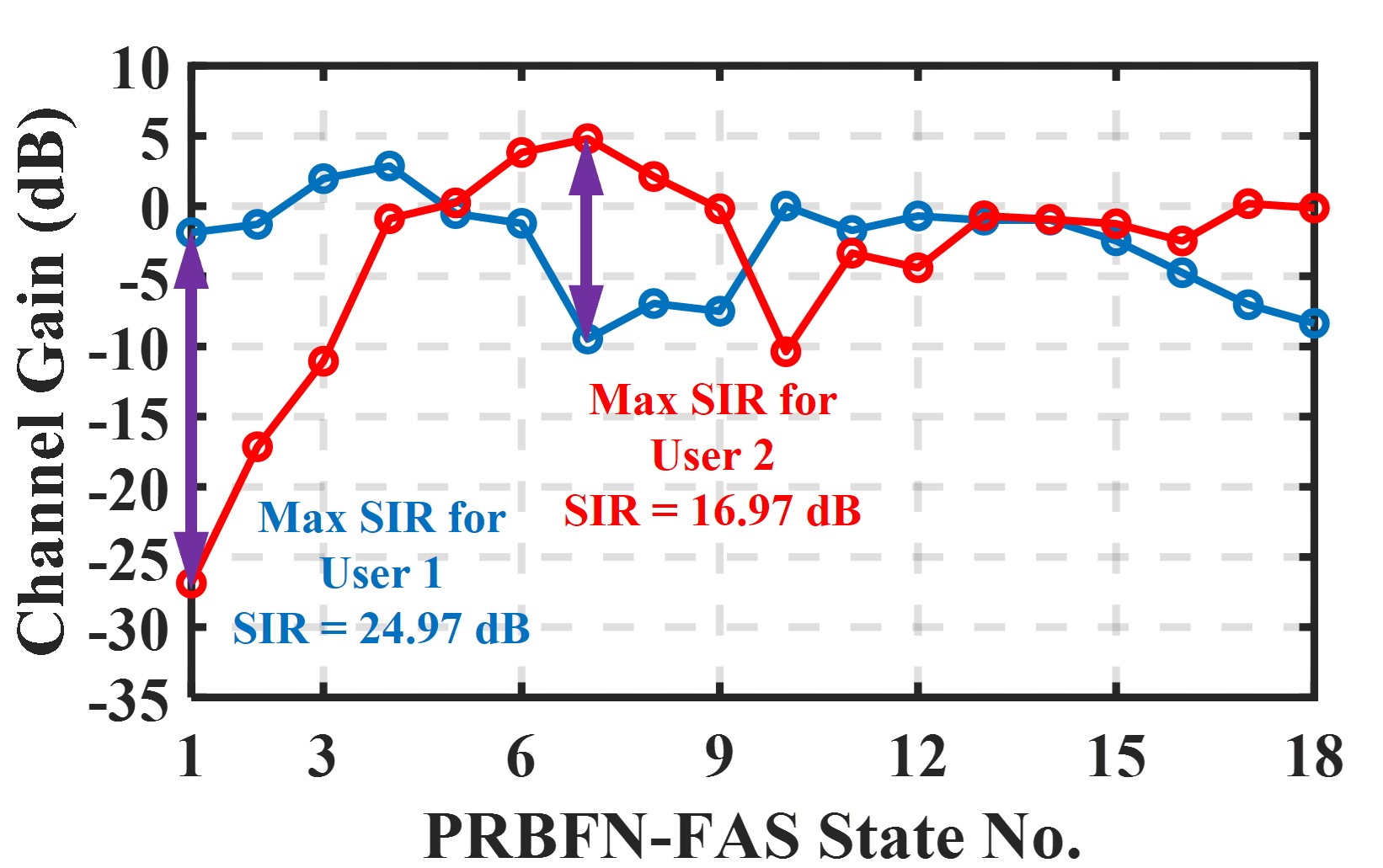}
		\put(0,2){(a)}
	\end{overpic}
	\hfil
	\begin{overpic}[width=0.49\linewidth]{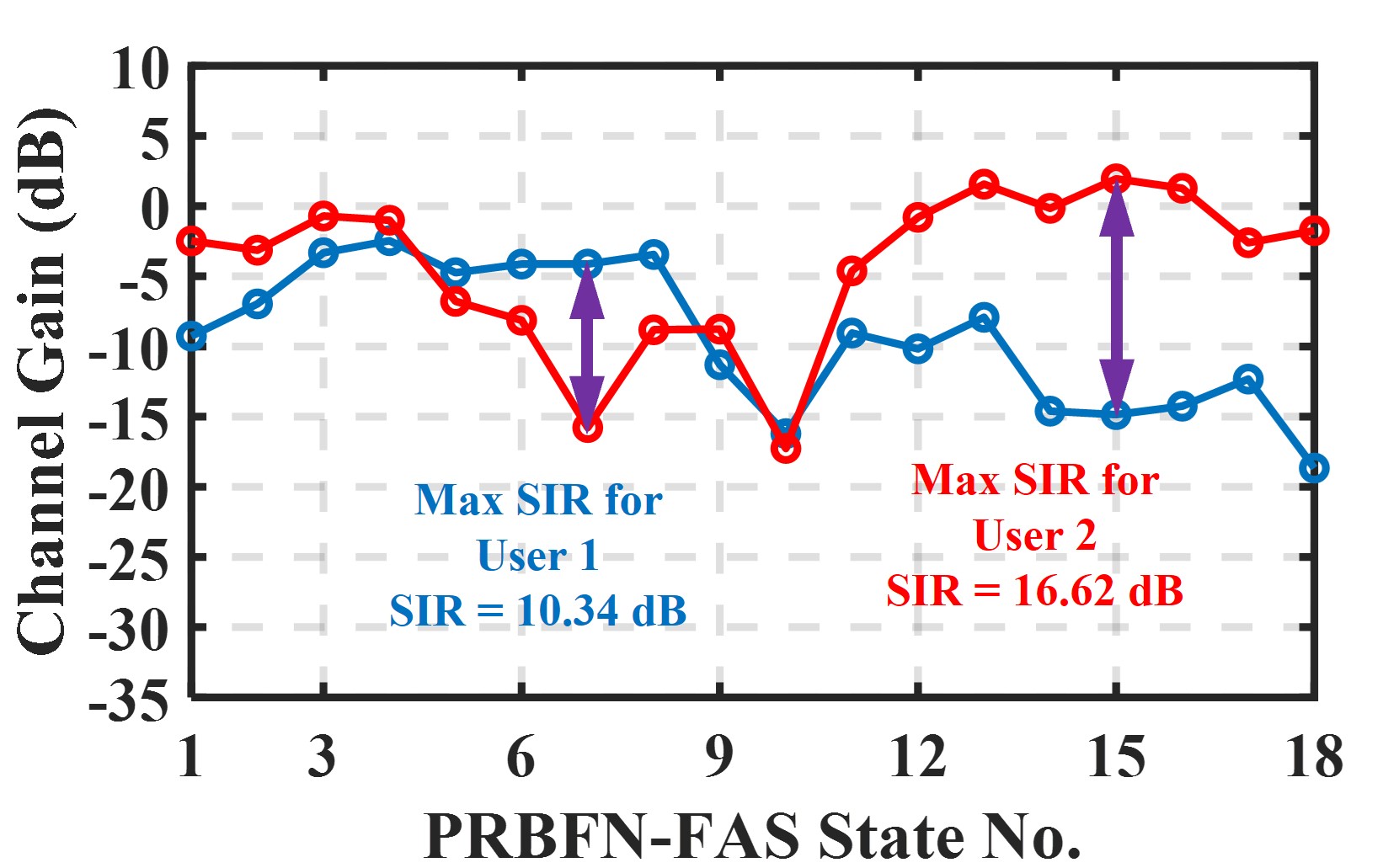}
		\put(0,2){(b)}
	\end{overpic}\\
	\vspace{0.05cm}
	\begin{overpic}[width=0.49\linewidth]{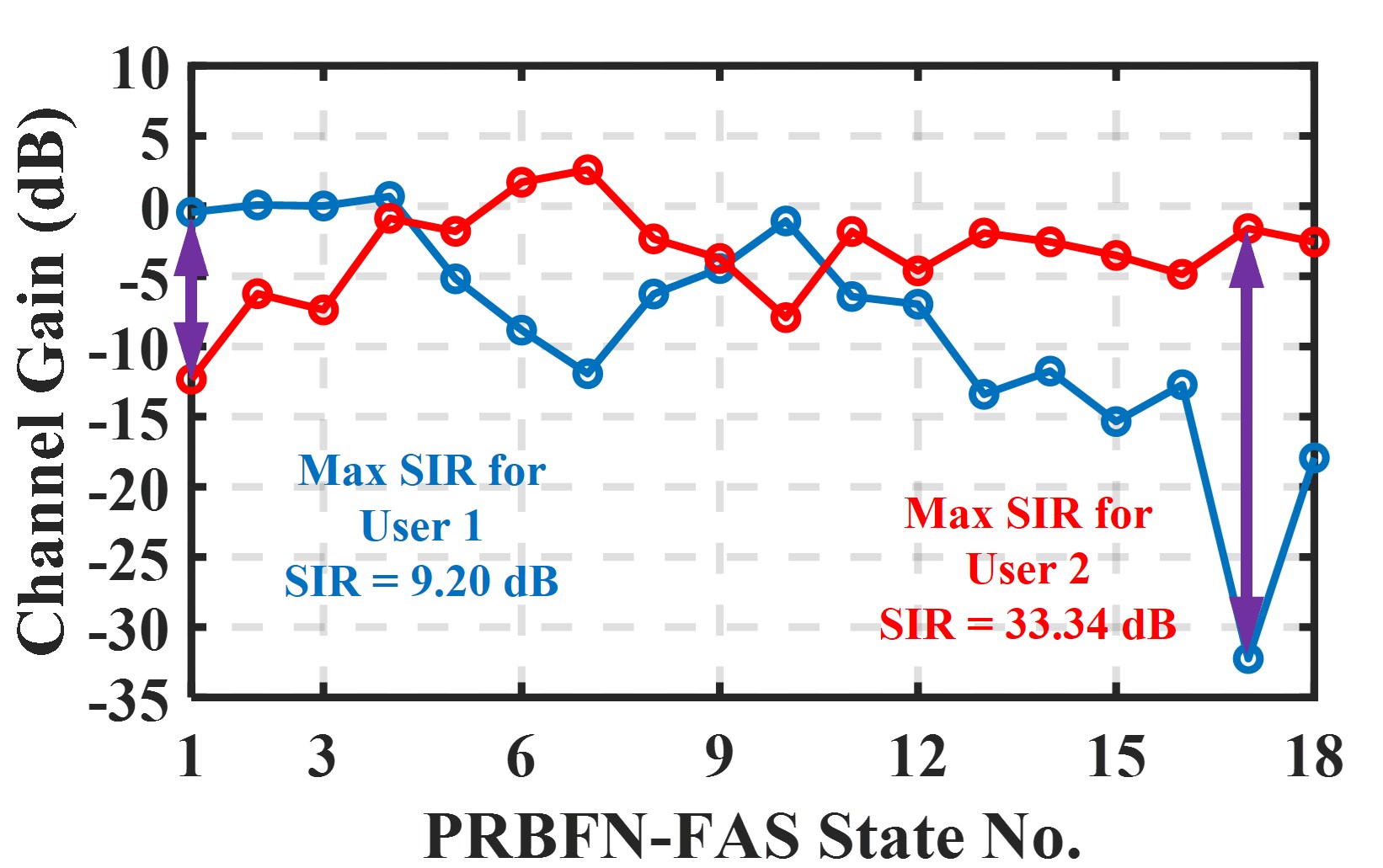}
		\put(0,2){(c)}
	\end{overpic}
	\hfil
	\begin{overpic}[width=0.49\linewidth]{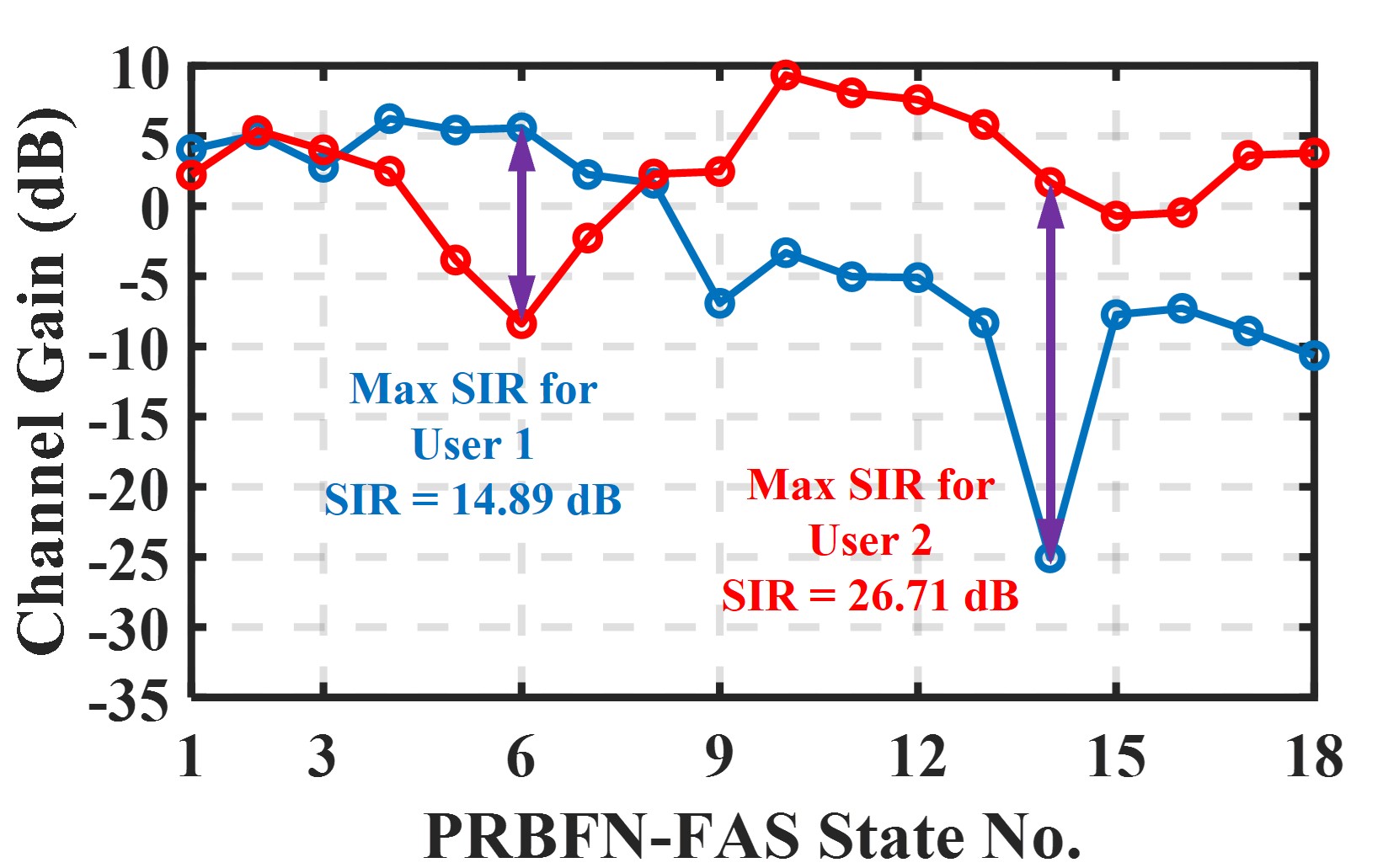}
		\put(0,2){(d)}
	\end{overpic}
	\vspace{-0.7cm}
	\caption{Measured stationary channels for $N=18$ PRBFN-FAS ports, with two users (Rx) at (a)-(d) 4 different locations in Fig. \ref{System_Measurement}. The optimal FAS ports with largest SIR are marked for each user in FAMA.}
	\label{System_Gain}
\end{figure}

Fig. \ref{System_Gain} illustrates the variation in measured channel gains of $h_{1,1}$ and $h_{2,1}$ across different PRBFN-FAS states, corresponding to various FAS ports, evaluated at four distinct locations shown in Fig. \ref{System_Measurement}. The results indicate that the PRBFN-FAS generates obvious antenna diversity at various states. Furthermore, by selecting optimal PRBFN-FAS state, over 10 dB signal strength difference of 2 users can be achieved, which supports FAMA applications. The observed variation in signal strength with port switching confirms the effectiveness of the PRBFN-FAS design in practical communication systems.

\begin{figure}[!t]
	\vspace{-0.1cm}
	\centering
	\includegraphics[width=1\linewidth]{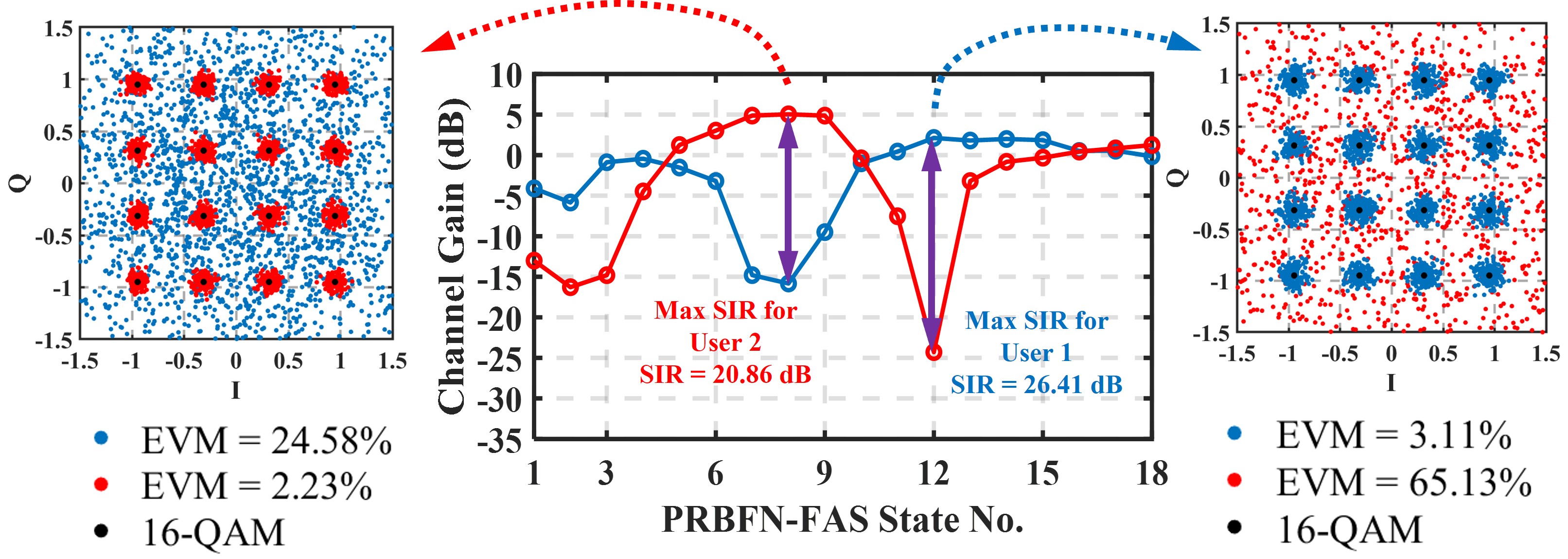}
	\vspace{-0.7cm}
	\caption{Constellation diagram under measured FAMA case. Selecting the optimal PRBFN‑FAS port for each user significantly improves EVM and thus enhances performance for multi-use access.}
	\label{System_Constellation}
\end{figure}

Deploying 16 QAM modulation under the measured channel conditions, we verify the  performance of the PRBFN‑FAS for uplink signals, as given in Fig. \ref{System_Constellation}. For User 1, selecting the 12-th PRBFN‑FAS state yields clear differentiation with a demodulation EVM of 3.11\%. Similarly, for User 2, the 8-th state achieves an EVM of 2.23\%. Simply selecting the optimal PRBFN‑FAS state enables effective multi-user differentiation, where the interference can be overcome without relying on the complex signal processing \cite{FAMA}. 

\begin{figure}[!t]
	\vspace{-0.35cm}
	\centering
	\includegraphics[width=0.6\linewidth]{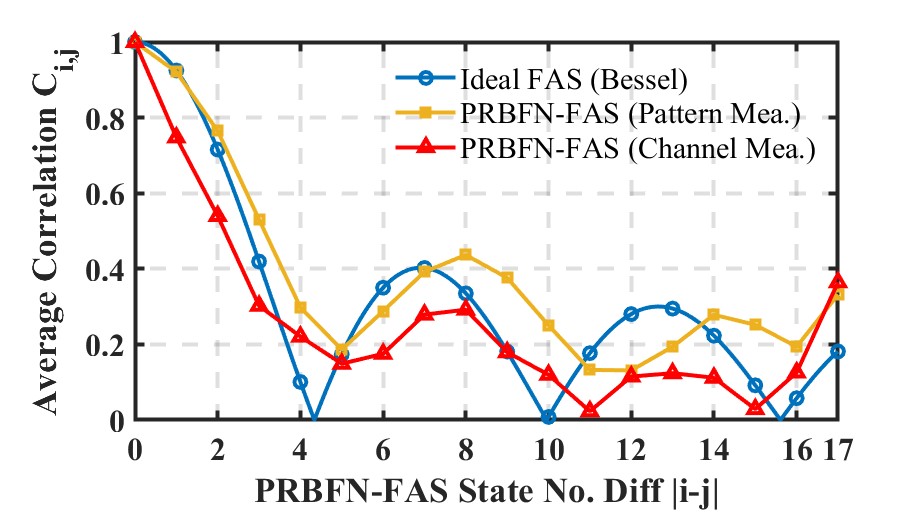}
	\vspace{-0.4cm}
	\caption{Correlation of the 4-port PRBFN-FAS reconfigurable states, including measurement results by $N$ radiation patterns (Fig. \ref{PRBFN_4Ports_Correlation}) and practical communication system in Fig. \ref{System_Measurement}.}
	\label{System_Correlation}
\end{figure}

Based on the channel measurement results, the correlation between different states of the PRBFN-FAS can be evaluated using the averaged auto-correlation coefficient derived from the channel measurements collected from $U=2$ users at $K$ various locations, which can be expressed as
\begin{equation}
	\label{eqn7-1}
	C^\text{mea}_{i} = \frac{1}{UK}\sum^K_{k=1}{ \sum^U_{u=1}{ \frac{ \mathcal{R}(i,u,k) }{\sigma^2(i,u,k)} } },
\end{equation}
where $\mathcal{R}(\cdot)$ is the auto-correlation function, and $\sigma^2$ is the standard deviation used for normalization. Collecting all FAS port numbers into a set $\mathcal{V}=\{1,2,\cdots,N\}$, these terms in (\ref{eqn7-1}) can be expanded as
\begin{equation}
	\label{eqn7-2}
	\begin{split}
		&\mathcal{R}(i,u,k) = \sum_{j,i+j\in\mathcal{V}}{ h^{(k)}_{u,1}(j)h^{(k)*}_{u,1}(i+j) },\\
		&\sigma^2(i,u,k) = \sum_{j,i+j\in\mathcal{V}}{ | h^{(k)}_{u,1}(j)| \cdot |h^{(k)}_{u,1}(i+j)|},
	\end{split}
\end{equation}
where $h^{(k)}_{u,1}(j)$ represents the measured channel between the $u$-th Rx and the PRBFN-FAS at Tx 1 with its $j$-th state at location $k$. The measured average correlation is shown in Fig. \ref{System_Correlation}, where the correlation obtained from the measured radiation patterns $\mathbf{C}$ and ideal Bessel curve $\mathbf{C}_\text{obj}$ are plotted for comparison. It can be observed that the correlation performance obtained through the communication system exhibits strong consistency with previous results. The deviations can be attributed to the non-ideal three-dimensional scattering in the actual test environment \cite{PRA-FAS}.

\begin{figure}[!t]
	\vspace{-0.1cm}
	\centering
	\begin{overpic}[width=0.48\linewidth]{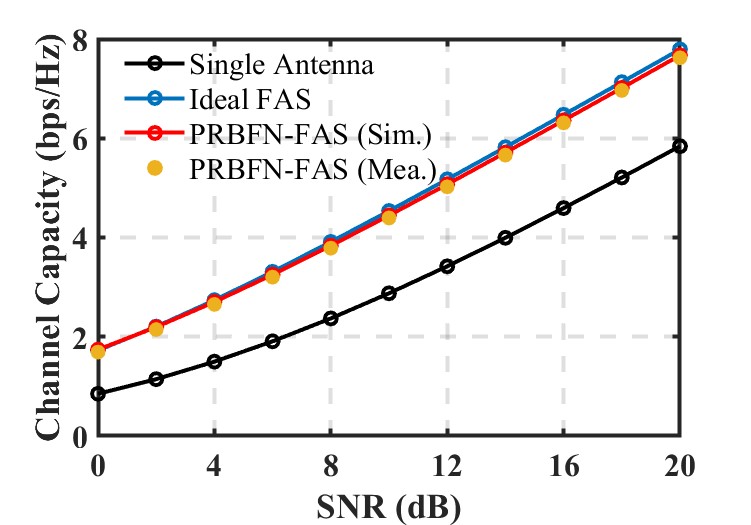}
		\put(0,2){(a)}
	\end{overpic}
%	\hfil
	\begin{overpic}[width=0.48\linewidth]{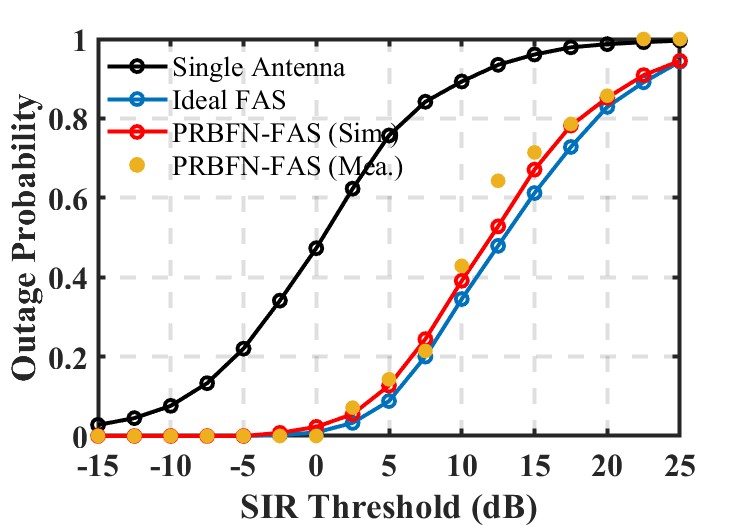}
		\put(0,2){(b)}
	\end{overpic}
	\vspace{-0.3cm}
	\caption{Measured (a) channel capacity of the PRBFN‑FAS versus SNR and (b) outage probability of the $U=2$ user system versus SIR threshold, using the measured channel in Fig. \ref{System_Gain}. (See the yellow dots.) Results are compared with a single antenna and an ideal FAS with $N=18$, $W=1.5$.}
	\label{System_Capacity_Outage}
\end{figure}

The channel capacity of the PRBFN-FAS was evaluated based on channel measurements conducted at $K = 8$ distinct locations, each involving $U = 2$ users. The maximum PRBFN‑FAS channel capacity, $C_\text{FAS}$, is achieved when the selected state yields the maximum channel gain, by
\begin{equation}
	\label{eqn7-3}
	C_\text{FAS} = \frac{1}{UK}\sum^K_{k=1}{ \sum^U_{u=1}{ \mathop{\max}\limits_n \left[\log_2\left(1+\gamma\cdot|h^{(k)}_{u,1}(n)|^2_2\right)\right]}},
\end{equation}
where $\gamma$ is the Signal-to-Noise Ratio (SNR). As shown in Fig. \ref{System_Capacity_Outage}(a), the results indicate that the data transmission performance of the PRBFN-FAS closely approaches that of ideal FAS with the same parameters.

Furthermore, the outage probability of the $U=2$ user system using the PRBFN-FAS is analyzed under FAMA, which can be expressed as 
\begin{equation}
	\label{eqn7-4}
	P_\text{out} = \text{Prob}\left(\mathop{\max}\limits_n{\{\varGamma^{(k)}_u(n)\}}< \varGamma_0 \right),
\end{equation}
where $\varGamma^{(k)}_u(n)$ is the SIR of the $u$-th user as the PRBFN-FAS selects the $n$-th state at location $k$, and $\varGamma_0$ is the SIR judgment threshold. $\varGamma_u(n)$ can be easily obtained from the channel measurement results in Fig. \ref{System_Gain}. Fig. \ref{System_Capacity_Outage}(b) plots the outage probability of the 2-user system versus the threshold $\varGamma_0$. The measured outage probability shows strong agreement with simulated PRBFN-FAS results and the ideal FAS with $N=18$ and $W=1.5$, demonstrating significant improvement against traditional fixed antenna.

\begin{figure}[!t]
	\vspace{-0.1cm}
	\centering
	\includegraphics[width=0.75\linewidth]{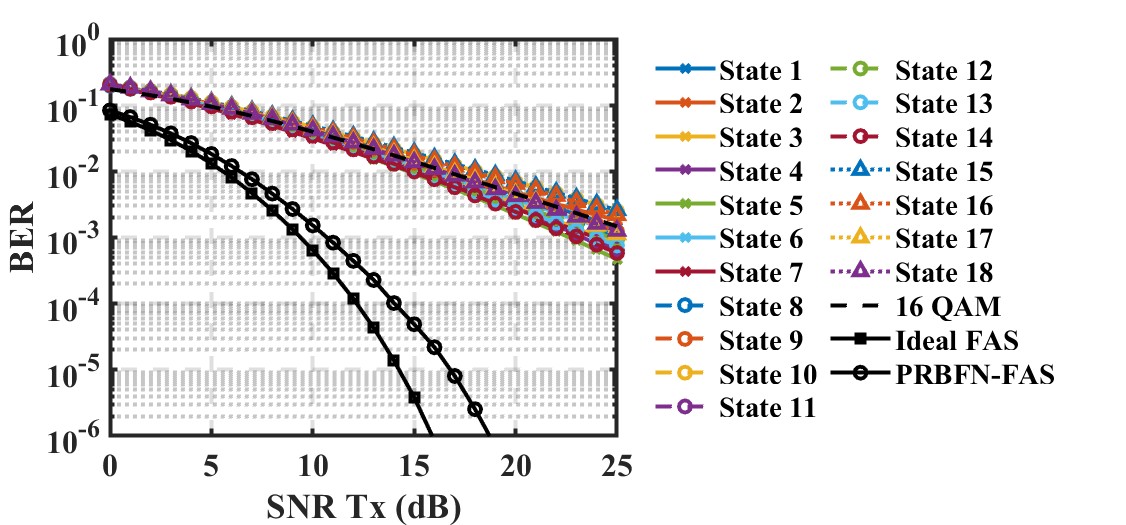}
	\vspace{-0.3cm}
	\caption{BER measurement results of PRBFN-FAS versus SNR of Tx.}
	\label{System_BER}
\end{figure}

We also provide bit error rate (BER) performance estimates by using the beamspace model \cite{PCDM} and our radiation pattern measurements in Fig. \ref{PRBFN_4Ports_Pattern}. The PRBFN‑FAS (or FAS) enhances communication system performance by improving channel gain through state switching, while leaving other RF chain components, such as modulation/demodulation, unaffected. Under these assumptions, BER performance is analyzed using 16‑QAM modulation in random scattering scenarios. As shown in Fig. \ref{System_BER}, dynamic PRBFN‑FAS state selection significantly improves BER under NLoS scattering compared to fixed‑state antennas, where we used a fixed PRBFN-FAS state instead. At high SNR, the improvement is even more pronounced, reaching one to three orders of magnitude. These results confirm that fixed antenna designs cannot handle the rapid fading NLoS wireless channels, while dynamic state selection can achieve more impressive performance \cite{PRA_exp}.  

%The channel $h^{(k)}_{u,1}(n)$ of the PRBFN-FAS can be expressed in beamspace  by
%\begin{equation}
%	\label{eqn7-5}
%	h^{(k)}_{u,1}(n) = \mathbf{e}_{\mathbf{R},u}^\mathrm{H}\mathbf{H}^{(k)}_\mathbf{v}\mathbf{e}^{}_\mathbf{T}(n),
%\end{equation}
%where $\mathbf{e}_{\mathbf{R},u}$ is the radiation pattern at the $u$-th Rx, $\mathbf{H}^{(k)}_\mathbf{v}$ is the angular virtual channel at the $k$-th location, $\mathbf{e}^{}_\mathbf{T}(n)$ is the radiation pattern of the $n$-th PRBFN-FAS state. 

\section{Discussions}

In this section, we provide further discussions on several issues in our proposed PRBFN-FAS design. 

\subsection{Advantages of PRBFN over conventional BFN}

In this design, the PRBFN architecture is adopted due to its high reconfigurability, which is able to support the large number of reconfigurable states $N$ required by FAS. The core principle involves precisely controlling the outputs of the PRBFN to satisfy the specific beamforming criteria. However, existing BFN designs can also support the FAS as long as the amplitudes and phases meet the requirements. 

\begin{table}[!t]
	\vspace{-0.05cm}
	\caption{Comparison of the PRBFN and recent BFN designs}
	\label{Comparison_BFN} 
	\centering
	\resizebox{\linewidth}{!}{
	\begin{tabular}{cccccc}
		\toprule
		Ref. &  \cite{BFN1} & \cite{BFN2} & \cite{BFN4} & \cite{BFN5} & This\\
		\midrule
		$f_0$ (GHz) & 1 & 1.15$\sim$1.4 & 1$\sim$1.1 & 2.4 & 2.6  \\
		BW (\%) & 20 & 6.4$\sim$9.5 & 5 & 20 & 5 \\
		RL (dB) & $<$-10 & $<$-15 & $<$-10 & $<$-10 & $<$-10 \\
		IS (dB) & $>$20 & N.A. & $>$20 & $>$15 & $>$20 \\
		Phase ($^\circ$) & 0/180 & -180$\sim$180 & -180$\sim$180 & -180$\sim$180 &  -180$\sim$180 \\
		IL (dB) & 0.8$\sim$1.8 & 2.6$\sim$2.9  & 4.2$\sim$4.9 & 2.2 & 0.4$\sim$2.9 \\
		PDR (dB) & 0 & 0 & -5$\sim$5 & -9$\sim$9 & -10$\sim$10 \\
		size ($\lambda_0^2$) & 0.27$\times$0.18 & N.A. & 0.56$\times$0.4 & 0.43$\times$0.13 & 0.36$\times$0.36 \\
		RC & 4 VDs & 13 VDs & 22 VDs & 6 VDs & 32 diodes \\ 
		\bottomrule
	\end{tabular}}
	\begin{tablenotes}
		\footnotesize
		\item[1] \hspace{-0.4cm} $\bullet$ BW: bandwidth. RL: return loss. IS: isolation. IL: insertion loss. PDR: power division ratio. RC: reconfigurable components. VD: varactor diode. \\
	\end{tablenotes}
\end{table}

To highlight the merits of our PRBFN over conventional BFN designs, Table \ref{Comparison_BFN} compares several reconfigurable PDs (BFNs) from \cite{BFN1,BFN2,BFN4,BFN5}. Those BFNs rely on many varactor diodes (VDs), each requiring a high‑precision DC voltage supply or digital‑to‑analog converter (DAC) for accurate control, which is a heavy burden in practical systems. In contrast, our PRBFN uses digitally controlled diodes driven directly by digital signals, making state switching more straightforward. Furthermore, traditional BFNs are not optimized for FAS, i.e. their varactor values and control voltages would need to be calibrated. Our PRBFN, however, is specifically designed for FAS, with reconfigurable outputs that closely approximate the target $\hat{\mathbf{B}}$ while maintaining relatively low insertion loss. Together with competitive size and sufficient 5\% bandwidth, these advantages establish the PRBFN as a very useful alternative for FAS implementations.

\begin{figure}[!t]
	\vspace{-0.1cm}
	\centering
	\includegraphics[width=0.65\linewidth]{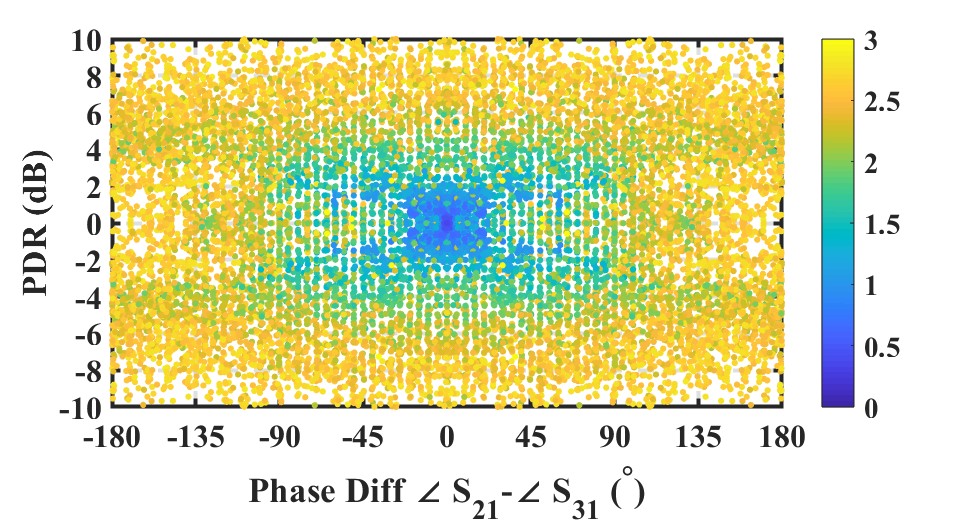}
	\vspace{-0.3cm}
	\caption{For the proposed PRBFN unit cell, each point represents a reconfigurable state satisfying $S_{11},S_{22},S_{33}<-10$ dB and $S_{23}<-20$ dB at 2.6 GHz. The color indicates the insertion loss.}
	\label{BF_Evaluation}
\end{figure}

It is also worth noting that the phase and amplitude of the reconfigurable power dividers (PDs) in \cite{BFN4,BFN5} can be controlled independently, whereas our PRBFN unit cell achieves output control through optimized pixel connections and thus amplitude and phase are not independently selected. However, by using specific pixel connections, it is indeed possible to mimic approximately independent amplitude and phase outputs. Fig. \ref{BF_Evaluation} shows selected reconfigurable states under both matching and isolation conditions, covering nearly the full PDR and phase difference range with fine resolution. By choosing appropriate 32‑bit diode states, our digitally controlled PRBFN unit cell achieves comparable function to conventional VD-based PDs.

%\subsection{Applications of PRBFN-FAS in Rx}
%
%In the PRBFN-FAS architecture illustrated in Fig. \ref{FAS_Architecture}, the low-power RF signal, after passing through the PRBFN, is amplified by PAs with minimal distortion, making the PRBFN-FAS well-suited for Tx applications. However, when employed at the Rx side, the insertion loss of the PRBFN can degrade the SNR, potentially limiting its performance. To overcome this, low-noise amplifiers (LNAs) must be deployed for PRBFN-FAS applications at Rx. $N_\text{A}$ LNAs are connected after $N_\text{A}$ antenna ports, and the LNA outputs feed the proposed PRBFN. 

\subsection{Comparison with previous FAS designs}

For practical FAS implementations, various hardware architectures have been proposed in previous works. This subsection reviews existing FAS hardware designs and compares them with our proposed PRBFN-FAS architecture, highlighting the advantages of our approach. A detailed comparison is provided in Table \ref{Comparison_FAS}, part of which incorporates the reference data from \cite{FAS_Compare}. Ideal FAS based-on physical movement \cite{FAS} with corresponding FAS port number $N$ and aperture $W\lambda$ is selected as the benchmark.

\begin{table*}[!t]
	\caption{Comparison of the proposed PRBFN-FAS and other early attempted FAS antennas in previous works}
	\label{Comparison_FAS} 
	\centering
	\resizebox{\linewidth}{!}{
	\begin{tabular}{cccccccccccc}
		\toprule
		Ref. & \makecell[c]{Method} & \makecell[c]{Freq.\\(GHz)} & \makecell[c]{Port Num.\\ $N$} & \makecell[c]{FAS Size\\ $W$($\lambda_0$)} & \makecell[c]{Port Density\\ $N/W$} & \makecell[c]{Bandwidth\\(\%)}& \makecell[c]{Radiation\\ Eff. (\%)} & \makecell[c]{Switching\\ Speed} & Scalability & Control & \makecell[c]{Linearity} \\
		\midrule
		\cite{MMA-FAS} & Motor & 3.5/27.5 & 600/100 & 6/5 & 100/20 & N.A. & N.A. & very slow & poor & motor & High \\
		\cite{liquid2} & Liquid Metal & 27 & continuous & 1 & $\geq$12 & 20 & 80 & slow & moderate & pump & High \\
		\cite{PRA-FAS} & PRA & 2.5 & 12 & 0.5 & 24 & 2 & 70 & $<1\mu$s & hard & FPGA & Low\\
		\cite{Meta-fluid} & Meta-fluid & 26.5 & 15 (8) & 6.6 (2.6) & 2.3 (3) & 3.8 & N.A. & $<1\mu$s & easy & FPGA & Low\\ 
		\cite{liquid3} & Liquid Metal & 3.55 & 64 & N.A. & N.A. & 120.9 & N.A. & slow & hard & pump & High\\
		\cite{RHS} & RHS & 26.2 & 48 (8) & 13.7 (8.6) & 3.5 (0.9) & N.A. & N.A. & $<1\mu$s & easy & FPGA & Low \\
		\cite{PRA_iotj} & PRA & 6.8 & 64 & N.A. & N.A. & 1.5 & N.A. & $<1\mu$s & hard & FPGA & Low\\
		\midrule
		This & PRBFN & 2.6 & 18 & 1.5 & 12 & 5 & 90 & $<1\mu$s & moderate & FPGA & High\\ 
		\bottomrule
	\end{tabular}}
	\begin{tablenotes}
		\footnotesize
		\item[1] \hspace{-0.4cm} $\bullet$ To support FAS with high performance, fast switching speed ($\leq 1$ ms) and a port density $N/W>10$ are required. \\
		\item[2] \hspace{-0.4cm} $\bullet$ The designs in \cite{Meta-fluid} and \cite{RHS} are 2D FAS. The value inside ($\cdot$) denotes the $y$-axis parameter, and the other denotes the $x$-axis parameter.\\
		\item[3] \hspace{-0.4cm} $\bullet$ Unlike conventional FAS concept, the design in \cite{liquid3} enhances array gain at specific angles by adjusting the 64 reconfigurable states per unit. 
		\item[4] \hspace{-0.4cm} $\bullet$ The antenna in \cite{PRA_iotj} exploits pattern diversity from $2^6=64$ reconfigurable states (6 diodes), rather than following the conventional FAS correlation. 
	\end{tablenotes}
\end{table*}

A comparative analysis reveals distinct trade-offs among existing FAS hardware implementations. Designs based on physical movement \cite{MMA-FAS,liquid2,liquid3} achieve nearly continuous port switching but are severely limited by low switching speeds. Alternatively, rapidly switchable architectures such as PRA-FAS \cite{PRA-FAS,PRA_iotj}, meta-fluid \cite{Meta-fluid}, and RHS FAS \cite{RHS} face challenges in Tx applications due to the nonlinearity of reconfigurable components \cite{pixel0}. These approaches also exhibit inherent limitations: the PRAs suffer from limited scalability of size, while the meta-fluid antenna and RHS antenna are constrained by low FAS port density, which ultimately restricts the performance of FAS.

In contrast, the PRBFN-FAS proposed in this work achieves a port density ($N/W$) exceeding 10, $\mu$s-level switching speed, and maintains signal integrity in Tx operation. Furthermore, the scaling methodology in Section \Rmnum{3} allows the deployment of the PRBFN-FAS applications with larger FAS size $W$. The comparative advantages of PRBFN-FAS are demonstrated through these performance metrics and architectural features.

\subsection{Implicit Beamforming Principle in PRA-FAS}

It should be noted that the PRA-FAS design in \cite{PRA-FAS} implicitly employs the beamforming principles mentioned in this work. However in \cite{PRA-FAS} it is masked by the intricate optimization process and it has not been explicitly discussed. Using the proposed theory from this work, the PRA-FAS operation can be more clearly understood. The PRA-FAS contains several orthogonal radiation components and this is where the identity matrix $\mathbf{K_M}$ arises. Optimizing the $N$ reconfigurable pixel connections corresponds to controlling the $N$ proportions of these components, and these correspond to the beamforming currents $\mathbf{B}$ in formula (\ref{eqn2-7}). Similarly, the settings of these $N$ reconfigurable pixels can be optimized to approximate Bessel function correlation $\mathbf{C}_\text{obj}$.  On the other hand, the PRA-FAS approach \cite{PRA-FAS} utilizes numerical optimization without introducing any explicit beamforming concept. In contrast, our proposed method offers a more intuitive and transparent interpretation of beamforming theory for enabling FAS.

\subsection{Compatibility with Existing Techniques}

\begin{figure}[!t]
	\vspace{-0.35cm}
	\centering
	\includegraphics[width=1\linewidth]{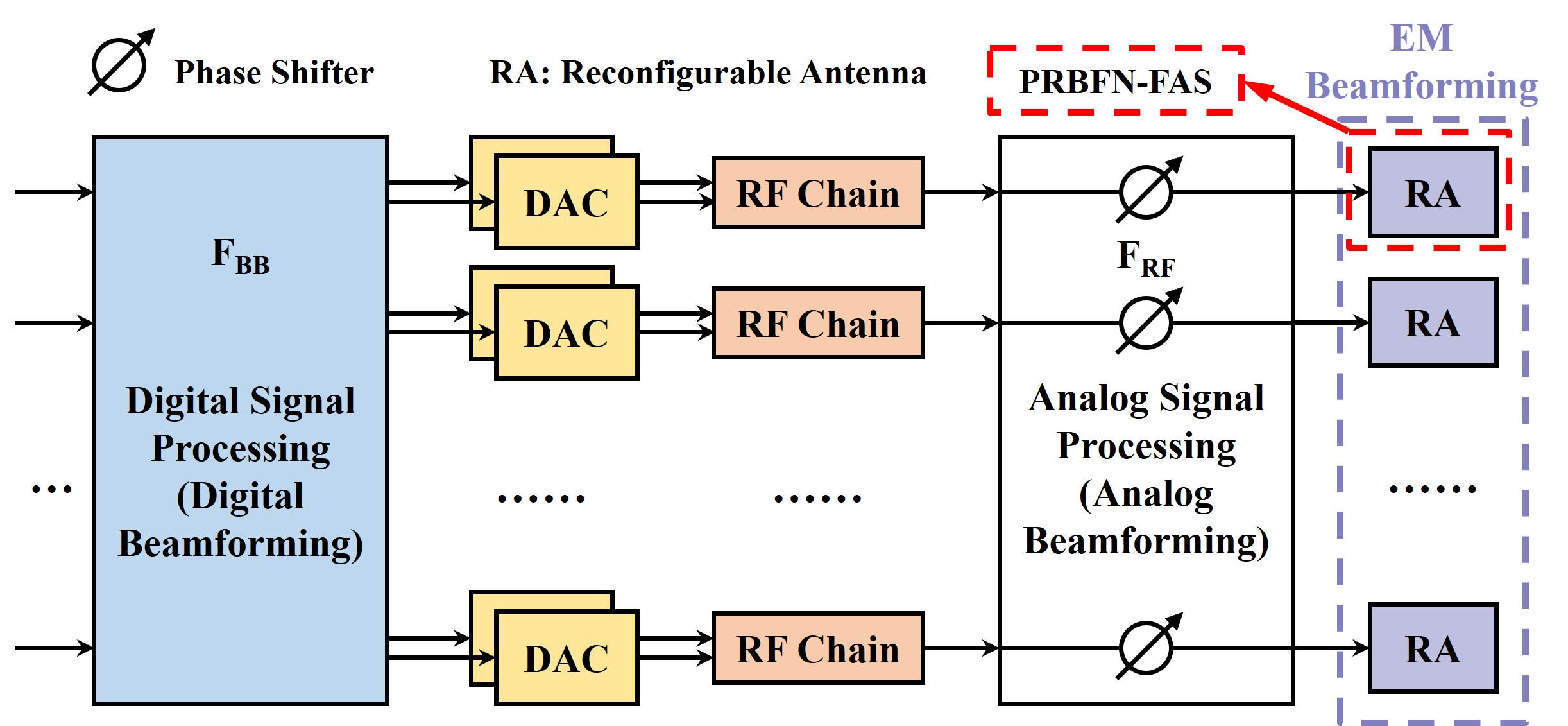}
	\vspace{-0.7cm}
	\caption{Tri-hybrid MIMO architecture integrated with digital beamforming, analog beamforming, and EM beamforming by reconfigurable antennas (RAs). Our proposed PRBFN-FAS design actually serves as RAs, which can cooperate with the traditional digital and analog beamforming.}
	\label{BF}
\end{figure}

\subsubsection{Beamforming in MIMO}

In the theoretical analysis presented in this paper, we establish that pattern-domain FAS can be interpreted through a beamforming framework. While recent research has explored various beamforming implementations for FAS \cite{FAS_BF1,FAS_BF2,FAS_BF3}, our method operates at the antenna level and remains fully compatible with conventional array-level beamforming techniques in MIMO.

To clarify this distinction, we consider the emerging tri-hybrid MIMO architecture \cite{Tri-Hybrid}, which integrates digital beamforming, analog beamforming, and electromagnetic (EM) beamforming implemented through reconfigurable antennas (RAs), as given in Fig. \ref{BF}. Traditional MIMO beamforming methods, such as digital beamforming and analog beamforming, operate at the array level and have been the focus of recent FAS-related studies \cite{FAS_BF1,FAS_BF2,FAS_BF3}. In contrast, the PRBFN-FAS proposed in this work serves as one RA element, which can achieve so-called EM beamforming. Therefore, the PRBFN-FAS be incorporated into larger array-level beamforming systems without conflict. This positions PRBFN-FAS as a complementary rather than competing technology in beamforming using FAS as well as MIMO \cite{MIMO-FAS}.

\subsubsection{Phased array}

The goal of a phased array is beamsteering, i.e., directing a main lobe toward a desired LoS direction. This requires precise phase shifters and amplitude controls, typically with analog or mixed‑signal circuits. In contrast, FAS (and thus PRBFN‑FAS) targets spatial diversity in NLoS environments. The objective is to generate a set of radiation patterns whose correlation follows a prescribed model (e.g., Clarke’s Bessel function), not to steer a single narrow beam.

Generating FAS correlation indeed can be achieved by phased arrays. For instance, a standard 4-element phased array (corresponding the $N_\text{A}=4$ in our PRBFN-FAS design) actually can generate required $N=18$ patterns if the excitation weight of every phased array element can be freely controlled by 4 independent reconfigurable phase shifters and attenuators. In contrast, both the aperture size and the RF chain complexity are substantially greater than those of our PRBFN‑FAS.

\section{Conclusion}

This paper has presented a novel PRBFN‑FAS architecture that enables high‑speed reconfiguration and high‑power operation, making it well suited for both Tx and Rx applications in FAS. Theoretically, we have shown that FAS can be interpreted as beamforming in the radiation pattern domain, where conventional FAS port switching corresponds to switching reconfigurable beamforming patterns in the PRBFN‑FAS framework. A systematic design methodology has been introduced, along with a scalable expansion approach to support FAS implementations with various parameter sets.

To validate the proposed design, one PRBFN-FAS prototype was developed targeting FAS configurations with parameters $W = 1.5, N = 18,N_\text{A}=4$. By controlling the states of integrated diodes, the PRBFN dynamically adjusts the pixel interconnections to achieve the desired amplitude and phase distribution across output ports. Measured results confirm that the PRBFN-FAS achieves a bandwidth exceeding 5\%, with S-parameters, radiation pattern correlation, and communication performance all aligning closely with theoretical expectations. Experimental deployments in real communication scenarios further verified its robust operation under practical scattering conditions, demonstrating expected correlation behavior, channel capacity, and enhancement of multiple access.

By establishing a connection between the traditional FAS spatial domain and the radiation pattern domain, the PRBFN-FAS offers a scalable, hardware-efficient pathway toward practical high-performance FAS implementations.

\section*{Data Availability}
The PRBFN-FAS E-field data is available at \href{URL}{https://github. com/ZhangJichen2001}. We encourage researchers in related fields of wireless communication to use these results. 

\bibliographystyle{IEEEtran}
\bibliography{IEEEabrv,IEEEreference}

\vfill

\end{document}